%% file: vorticesarxivv1.tex
\documentclass[12pt]{article}
\usepackage{jheppub}
\pdfoutput=1

\usepackage{amsmath,bbm,array,amsfonts,graphicx,wrapfig,lscape,float,mathtools,multirow,longtable}

\input{pref}
\newcommand{\V}{\mathcal{V}}
\newcommand{\wV}{\widetilde{\mathcal{V}}}
\newcommand{\wmaster}{\widetilde{\master}}
\newcommand{\Tr}{\text{Tr}}
\newcommand{\ra}{\rightarrow}
\newcommand{\cp}{\mathbb{C}\mathbb{P}}

\title{Hilbert Series and \\ Moduli Spaces of $k$ $U(N)$ Vortices}

\author[a]{Amihay Hanany}
\author[b]{and Rak-Kyeong Seong}

\affiliation[a]{
Theoretical Physics Group, The Blackett Laboratory,
Imperial College London, \\
Prince Consort Road, London SW7 2AZ, UK
}

\affiliation[b]{
School of Physics, Korea Institute for Advanced Study, \\
85 Hoegi-ro, Seoul 130-722, South Korea
}

\emailAdd{a.hanany@imperial.ac.uk}
\emailAdd{rkseong@kias.re.kr}

\preprint{
\begin{flushright}
Imperial/TP/13/AH/04\\
KIAS-P14016
\end{flushright}
}

\abstract{
We study the moduli spaces of $k$ $U(N)$ vortices which are realized by the Higgs branch of a $U(k)$ supersymmetric gauge theory. The theory has $4$ supercharges and lives on $k$ D1-branes in a $N$ D3- and NS5-brane background. We realize the vortex moduli space as a $\mathbb{C}^{*}$ projection of the vortex master space. The Hilbert series is calculated in order to characterize the algebraic structure of the vortex master space and to identify the precise $\mathbb{C}^{*}$ projection. As a result, we are able to fully classify the moduli spaces up to $3$ vortices.
}

\begin{document}

\maketitle

\section{Introduction}

The study of vortices has attracted much interest in the past since their first appearance in theoretical physics \cite{Abrikosov1957199,1973NuPhB..61...45N}. More recently, the work in \cite{Hanany:2003hp} presented a brane construction for a supersymmetric gauge theory with 4 supercharges describing $k$ $U(N)$ vortices. For simplicity, we will restrict to a $3d$ $\mathcal{N}=4$ theory in this work. The construction involves $k$ D1-branes in a $N$ D3-brane and NS5-brane background. Of our interest is the worldvolume theory of the D1-branes realized on the Higgs branch of the vortex theory. The D1-branes play the role of vortices and their worldvolume theory is effectively a $1d$ theory which is a dimensional reduction of a $2d$ $\mathcal{N}=(2,2)$ supersymmetric gauge theory. The vacuum moduli space of this theory is identified as the moduli space of $k$ $U(N)$ vortices on $\mathbb{C}$. The following work is interested in studying these moduli spaces using the construction in \cite{Hanany:2003hp}. 

Vortex moduli spaces are complex projective spaces \cite{Hanany:2003hp,Auzzi:2005gr,Hashimoto:2005hi,Eto:2006cx,Eto:2010aj,Hanany:2005bc,Hanany:2004ea,Eto:2007yv,Auzzi:2003fs,Eto:2005yh,Eto:2006pg}. In this work, we express the vortex moduli spaces as partial $\mathbb{C}^{*}$ projections of vortex \textit{master spaces}. These are spaces of mesonic and baryonic chiral operators which are invariant under the non-Abelian part of the gauge symmetry \cite{Hanany:2010zz,Forcella:2008bb}.\footnote{Strictly speaking, the master space is the space of invariants under the non-Abelian part of the gauge symmetry \textit{and} under the F-term constraints which are obtained by the superpotential of the theory. We will later see that the vortex theory does not have relevant a superpotential.} For the case of $k$ $U(N)$ vortices, this is the $SU(k)$ part of the $U(k)$ gauge symmetry. Master spaces have been studied in various setups in string theory, with a particular focus on $4d$ $\mathcal{N}=1$ supersymmetric quiver gauge theories which can be represented by brane tilings \cite{Zaffaroni:2008zz,Forcella:2009bv,Butti:2007jv}. The $\mathbb{C}^{*}$ projection of the vortex master space is along the remaining $U(1)$ gauge symmetry and leads in general to a partially weighted projective space. This is precisely the full vortex moduli space we want to study in this work.

We take inspiration from the recent fruitful studies of moduli spaces of instantons on $\mathbb{C}^2$ \cite{Benvenuti:2010pq,Hanany:2012dm,Dey:2013fea}. The \textit{ADHM construction} \cite{Atiyah1978185} for instanton moduli spaces arising from D3-D7 brane constructions resembles remarkably the construction of vortex moduli spaces. From this point of view, the vortex construction in \cite{Hanany:2003hp} is often referred to as a $\frac{1}{2}$-ADHM construction for vortices. As it has been used for the study of instanton moduli spaces, we make use of \textit{Hilbert series} \cite{Benvenuti:2006qr,Hanany:2006uc,Feng:2007ur} as a tool to analyze the algebraic structure of vortex moduli spaces. The use of Hilbert series combined with the use of vortex master spaces allows us to fully characterize the algebraic structure of vortex moduli spaces for up to 3 $U(N)$ vortices on $\mathbb{C}$. 

Hilbert series have been very successfully used to study vacuum moduli spaces of various supersymmetric gauge theories.  They are partition functions of gauge invariant chiral operators. Hilbert series have been used for instance to shed light on moduli spaces of SQCD with classical gauge groups \cite{Hanany:2008sb} and toric moduli spaces of brane tilings \cite{Hanany:2012hi,Hanany:2012vc,Cremonesi:2013aba}. 

We first compute the Hilbert series of the vortex master space by taking gauge invariance under the $SU(k)$ non-Abelian part of the gauge symmetry. The Hilbert series allows us to identify the full algebraic structure of the vortex master space, including information about its generators and quadratic relations satisfied by the generators. In general, we observe and verify that the $k$ $U(N)$ vortex master space is a non-compact singular Calabi-Yau cone of complex dimension $kN+1$. The algebraic variety of the vortex master space is weighted under the remaining $U(1)$ symmetry. These weights are part of the $\mathbb{C}^{*}$ projection which lifts the vortex master space to the full now partially compact vortex moduli space of complex dimension $kN$. This work for the first time uses Hilbert series to fully classify the vortex moduli spaces up to 3 $U(N)$ vortices and presents the Hilbert series for 4 $U(1)$ and $U(2)$ vortices.

The outline of the paper is as follows. Section \sref{snack} reviews the analysis of instanton moduli spaces from the ADHM construction and summarizes the similarities to the vortex construction in \cite{Hanany:2003hp}. The section introduces the computation for the Hilbert series of the vortex master space and explains its $\mathbb{C}^{*}$ projection into the full vortex moduli space. Using the techniques presented in section \sref{snack}, sections \sref{s1} to \sref{s3} present the a classification of vortex moduli spaces up to $3$ vortices for any $U(N)$. Section \sref{s4} illustrates with Hilbert series for $4$ $U(1)$ and $U(2)$ vortices that the classification scheme we introduce here with this paper is generalizeable to any number of $U(N)$ vortices. Finally, section \sref{apphs} summarizes the unrefined Hilbert series for our classification of vortices as well as uses a new compact form of presenting character expansions of Hilbert series. 
\\

\section{Background \label{snack}}

In this section, we review the theoretical background on vortices based on \cite{Hanany:2003hp}. An outline is given for the brane construction of vortices and the corresponding quiver diagram of the worldvolume theory. We are interested in the Higgs branch moduli space of the vortex theory which we will describe as a $\mathbb{C}^{*}$ projection of the master space. The Hilbert series is obtained in order to characterize the master space, the $\mathbb{C}^{*}$ projection and ultimately the vortex moduli space itself.

The quiver theory for $k$ $U(N)$ vortices is strikingly similar to the ADHM construction of $k$ $U(N)$ instantons. Given that the Higgs branch moduli space of $k$ $U(N)$ instantons has been extensively studied with the help of Hilbert series \cite{Benvenuti:2006qr,Hanany:2006uc,Feng:2007ur}, let us review the study of instanton moduli spaces as a warm-up for vortices.
\\

\subsection{Vortices from instantons \label{sback1}}

\paragraph{$k$ $U(N)$ instantons revisited.}
The moduli space of instantons on $\mathbb{C}^2$ is the Higgs branch of a $\mathcal{N}=2$ supersymmetric gauge theory in $3+1$ dimensions. It is the worldvolume theory of $k$ D3-branes in a $N$ D7-brane background. At the Higgs branch of the theory of $k$ $U(N)$ instantons, the $k$ D3-branes are on top of the $N$ D7-branes, the position of the D3-branes in the D7-branes being in $\mathbb{C}^2$.
  	
The worldvolume theory is a $4d$ $\mathcal{N}=2$ quiver gauge theory. Its quiver diagram consists of a $\mathcal{N}=2$ hypermultiplet and an $\mathcal{N}=2$ adjoint hypermultiplet. There is a $U(k)$ vector multiplet with a $U(N)$ global symmetry. The $U(1)$ of $U(N)$ can be absorbed into the local $U(k)$ giving us a two noded quiver where one node corresponds to the local $U(k)$ and the other to the global $SU(N)$.

The quiver and superpotential can be expressed in terms of $\mathcal{N}=1$ language by decomposing the $\mathcal{N}=2$ hyper and vector multiplets into $\mathcal{N}=1$ chiral and vector multiplets. The resulting quiver diagram is shown in \fref{tinstanton}.

\begin{table}[ht!]
\begin{center}
\begin{tabular}{|c|cc|c|cc|}
\hline 
\; & \multicolumn{2}{c|}{$U(k)_{gauge}$} & $U(N)_{global}$ & \multicolumn{2}{c|}{} \\
\; & $SU(k)_w$ & $U(1)_z$ & $SU(N)_x$ & $SU(2)_{\mathbb{C}^2}$ & $U(1)_r$
\\
\hline\hline
$\Phi$ & $[1,0,\dots,0,1]_w+1$ & 0 & $[0,\dots,0]_x$ & $[0]_q$ & 0\\
$\phi_1,\phi_2$ & $[1,0,\dots,0,1]_w+1$ & 0 & $[0,\dots,0]_x$ & $[1]_q$ & 1\\
$Q$ & $[1,0,\dots,0]_w$ & $+1$ & $[0,\dots,0,1]_x$ & $[0]_q$ & 1\\
$\widetilde{Q}$ & $[0,\dots,0,1]_w$ & $-1$ & $[1,0,\dots,0]_x$ & $[0]_q$ & 1\\
\hline
\end{tabular}
\caption{Charges carried by the fields of the $k$ $U(N)$ instanton quiver. The $U(1)$ of the global $U(N)$ has been absorbed into the local $U(k)$ for simplicity. The fugacity for the $SU(2)_R$ charge is $t$ and for the $SU(2)_{\mathbb{C}^2}$-charge is $q$. \label{tinstanton}}
\end{center}
\end{table}

\begin{figure}[ht!!]
\begin{center}
\resizebox{0.55\hsize}{!}{
\includegraphics[trim=0cm 0cm 0cm 0cm,totalheight=16 cm]{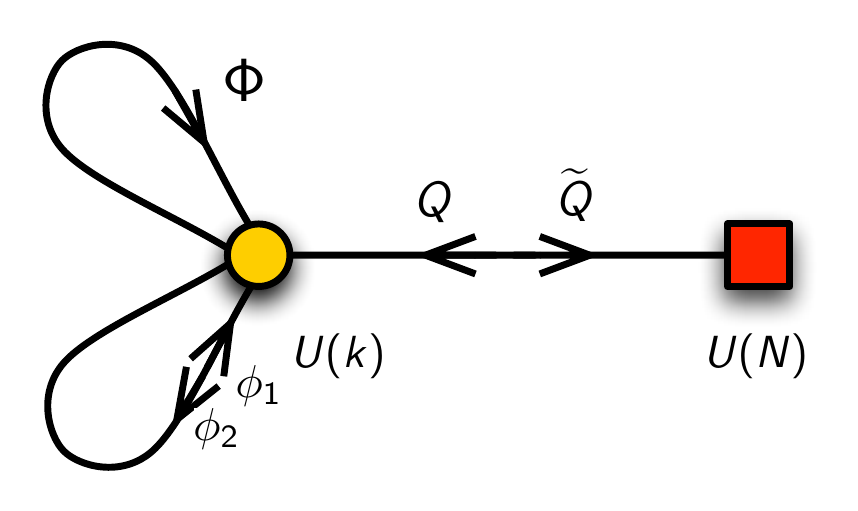}
}  \caption{$\mathcal{N}=1$ quiver for the theory describing $k$ $U(N)$ instantons in $\mathbb{C}^2$. \label{finstantonquiver}}
 \end{center}
 \end{figure}

The fields carry the gauge and global charges as shown in \tref{tinstanton}. Spacetime is broken to $\mathbb{R}^{3,1}\times \mathbb{C}^2\times \mathbb{C}$ where $\mathbb{C}^2$ has an isometry of $U(2) = SU(2)_{\mathbb{C}^2}\times U(1)_r$ where the $U(1)_r$ is the Cartan element of $SU(2)_R$. 

From the quiver in \fref{tinstanton}, the superpotential can be written as follows,
\beal{es000}
W =\Tr( Q \cdot \Phi \cdot \widetilde{Q}   + \phi_1 \cdot \Phi \cdot \phi_2 - \phi_2 \cdot \Phi \cdot \phi_1 )~~.
\eea
For the Higgs branch, the $k$ D3-branes are on top of the $N$ D7-branes and hence we have $\langle \Phi \rangle = 0$.

The ADHM data \cite{Atiyah1978185} is summarized by the quiver fields and the F-terms originating from the superpotential. Following the construction, we analyze the Higgs branch of the above supersymmetric quiver gauge theory as the moduli space of $k$ $U(N)$ instantons. The \textbf{Hilbert series} \cite{Benvenuti:2006qr,Hanany:2006uc,Feng:2007ur} characterizes the instanton moduli space. The quiver fields contribute to the computation of the Hilbert series the following plethystic exponential\footnote{For a multivariate function $f(t_1,\dots,t_n)$, the \textit{plethystic exponential} is defined as $\PE{f(t_1,\dots,t_n)}=\exp\left[\sum_{k=1}^{\infty}\frac{f(t_1^k,\dots,t_n^k)}{k!}\right]$. The PE plays the role of symmetrizing the function $f(t_1,\dots,t_n)$.}
\beal{es001}
f_1 &=& 
\PE
\Big[
[1,0,\dots,0,1]_{w} [1]_{q} t 
\nn\\
&& \hspace{1cm}
+ [1,0,\dots,0]_{w} z [0,\dots,0,1]_{x} t + [0,\dots,0,1]_{w} z^{-1} [1,0,\dots,0]_{x} t
\Big]~~.
\nn\\
\eea
The F-term takes the form
\beal{es002}
\mathcal{F}:= \partial_{\Phi} W = Q \cdot \widetilde{Q}  + \phi_1 \cdot \phi_2 - \phi_2 \cdot \phi_1 ~~.
\eea
The F-term carries the $U(1)_r$ charge of the superpotential and also transforms under the adjoint representation of the gauge group. As such, it contributes the following plethystic exponential to the Hilbert series computation,
\beal{es003}
f_2 = 
\PE\Big[
-([1,0,\dots,0,1]_w+1) t^2
\Big]~~.
\eea

Overall, the Hilbert series of the $k$ $U(N)$ instanton moduli space $\mathcal{M}_{k,N}^{inst}$ can be calculated when one integrates out the gauge group dependence of the above plethystic exponentials,
\beal{es004}
g(x,q,t;\mathcal{M}_{k,N}^{inst}) = 
\oint \ud\mu_{SU(N)} \ud\mu_{U(1)} 
~f_1 f_2~~,
\eea
where $\ud\mu_{SU(N)}$ and $\ud\mu_{U(1)}$ are the Haar measures of $SU(N)$ and $U(1)$ respectively.
\\

\paragraph{From the instanton to the vortex.} It has been outlined in \cite{Hanany:2003hp} and further evaluated in \cite{Eto:2006cx,Eto:2010aj} and consecutive papers that the instanton theory is related to the theory of $k$ $U(N)$ vortices. The worldvolume theory of the vortices is effectively a $1d$ theory which is a dimensional reduction of a $2d$ $\mathcal{N}=2$ superymmetric gauge theory. The construction of the vortex moduli space from this quiver theory resembles the ADHM construction of the instanton moduli space. In literature, the construction for vortices is also referred to as a $\frac{1}{2}$-ADHM construction. 

Considering the quiver for the ADHM construction of instantons, the construction for the vortex moduli space precisely requires half of the quiver field content: a single fundamental $Q$ between $U(k)$ and $U(N)$, and a single adjoint $\phi\equiv\phi_1$. The combination of $\{\phi, Q\}$ precisely forms the field content of the vortex theory. Let us consider first the brane construction for the $k$ $U(N)$ vortex in the following section.
\\

\subsection{The brane construction and the vortex moduli space \label{sback2}}

 \begin{table}[htt!]
 \begin{center}
 \begin{tabular}{r|cccccccccc}
\; & 0 & 1 & 2 & 3 & 4 & 5 & 6 & 7 & 8 & 9 
 \\
 \hline
 NS5 & $\times$ & $\times$ & $\times$ & $\times$ & $\times$ & $\times$ & & & & 
 \\
 $N$ D3 & $\times$ & $\times$ & $\times$ & & & & $\times$ & & &
 \\
 $k$ D1 & & & & & & & & & & $\times$
 \end{tabular}
 \caption{$k$ D1-branes and $N$ D3-branes which are suspended between NS5-branes. The theory for this brane construction has $4$ supercharges. For simplicity we restrict to a $3d$ $\mathcal{N}=4$ $U(N)$ theory in this work. \label{tvbranes}}
 \end{center}
 \end{table}

\begin{figure}[ht!!]
\begin{center}
\resizebox{0.6\hsize}{!}{
\includegraphics[trim=0cm 0cm 0cm 0cm,totalheight=16 cm]{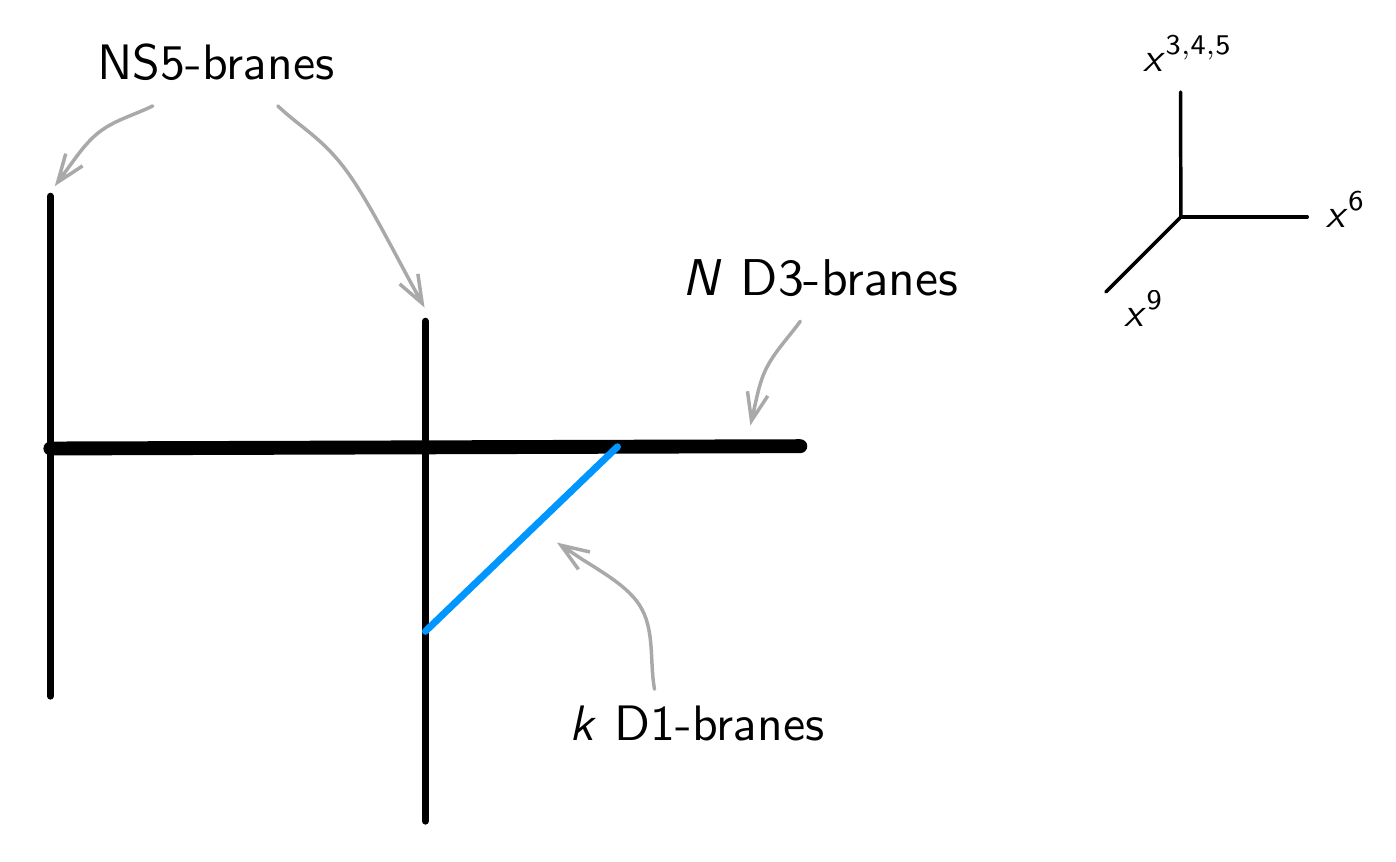}
}  \caption{Brane construction for $k$ $U(N)$ vortices. It represents the Higgs branch of a $3d$ $\mathcal{N}=4$ $U(N)$ Yang-Mills Higgs theory. \label{fvbranes}}
 \end{center}
 \end{figure}

Vortices in Type IIB string theory are realized by a construction of $k$ D1-branes and $N$ D3-branes which are suspended between NS5-branes \cite{Hanany:2003hp,Hanany:1996ie}. The theory for this brane construction is a theory with 4 supercharges. For simplicity, we take it here to be
a $3d$ $\mathcal{N}=4$ $U(N)$ Yang-Mills Higgs theory. \tref{fvbranes} shows the brane picture in 9+1 dimensions.

The BPS vortices are represented by the $k$ D1-branes. We are interested in the Higgs branch of the Yang-Mills Higgs theory where the FI parameter $\zeta$ is non-zero. The mass and size of the vortices scale respectively as $M_{v}\sim \zeta$ and $M_\gamma \sim \zeta^{-1/2}$. The finite FI parameter $\zeta$ relates to a decoupling of a NS5 brane from the rest of the construction. Between the decoupled NS5-brane and $N$ D3-branes are the $k$ suspended D1-branes as shown in \fref{fvbranes}.

In order to describe the vortex moduli space, we consider the worldvolume theory of the $k$ D1-branes. The worlvolume theory is effectively a $1d$ theory which is a dimensional reduction of a $2d$ $\mathcal{N}=(2,2)$ supersymmetric gauge theory as mentioned in section \sref{sback1}. The field content $\{\phi,Q\}$ of the theory consists of a complex scalar adjoint $\phi$ which parameterizes the position of the $k$ D1-branes in the $x_1$-$x_2$ plane. It also contains a fundamental $Q$ in the $U(k)$ gauge group arising from D1-D3 strings. The vector multiplet contains the gauge field and complex scalar fields parameterizing the degrees of freedom of the D1-branes in the $x^{3,4,5}$ directions.

The Higgs branch $\mathcal{M}_{k,N}^{vort}\equiv \mathcal{V}_{k,N}$ of the vortex theory is determined by a K\"ahler quotient of $\mathbb{C}^{kN+k^2}$ parameterized by the $k\times k$ matrices $\phi$ and $k\times N$ matrices $Q$. There is no superpotential and therefore no F-terms. The D-terms are given by
\beal{es101}
D^{m}_{~n} = \phi^{m}_{~i} (\phi^\dagger)^{i}_{~n} - [Q,Q^\dagger]^{m}_{~n} - r\delta^{m}_{~n}~~,
\eea
where $r$ is the finite non-zero FI parameter of the vortex theory for the Higgs branch $\mathcal{V}_{k,N}$. The D-terms impose $k^2$ constraints and with the $U(k)$ gauge group which results in a further $k^2$ reduction, the total real dimension of the Higgs branch reduces to
\beal{es102}
\text{dim}_{\mathbb{R}}\left(\mathcal{V}_{k,N}\right) =2 (k^2+ kN) - k^2 - k^2 = 2kN~~.
\eea
The complex dimension is $\text{dim}_{\mathbb{C}}=k N$, and is precisely half the dimension of the $k$ $U(N)$ instanton moduli space which we discussed in section \sref{sback1}.

We demand the D-term equations to be satisfied. For the vortex master space, gauge invariance is only taken for the non-Abelian $SU(k)$ part of the $U(k)$ symmetry. The remaining $U(1)$ has a FI-term in the D-term equations, and as such for non-zero FI-terms the D-term equations are set to be constant. This together with $U(1)$ gauge invariance amounts to complex rescaling of the $SU(k)$ invariant baryonic operators along the $U(1)$ direction. This is the partial $\mathbb{C}^{*}$ projection of the vortex master space in order to obtain the full moduli space of the vortices. In the following section, we review the $\frac{1}{2}$-ADHM construction similar to the construction of the instanton moduli space, and outline the method of using Hilbert series to study partial projective spaces as moduli spaces of $k$ $U(N)$ vortices.
\\

\subsection{The quiver and the Hilbert series}

\begin{figure}[ht!!]
\begin{center}
\resizebox{0.6\hsize}{!}{
\includegraphics[trim=0cm 0cm 0cm 0cm,totalheight=16 cm]{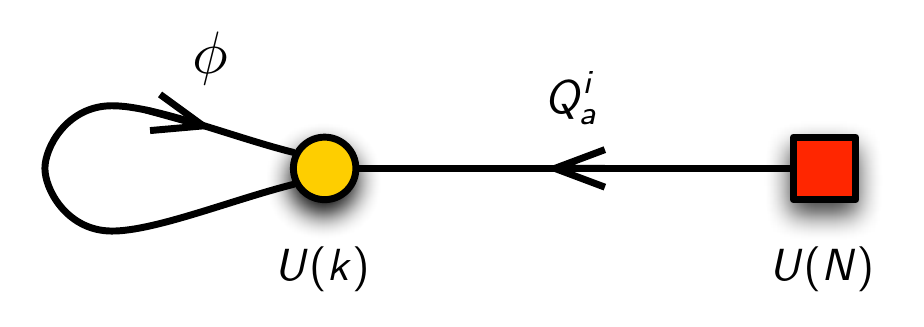}
}  \caption{Quiver diagram of the k $U(N)$ vortex theory. \label{fvquiver}}
 \end{center}
 \end{figure}

\paragraph{Vortex quiver.}
The field content of the vortex theory consisting of the adjoint scalar $\phi$ and the fundamental $Q$ in the gauge group $U(k)$ can be represented by a 2-noded quiver diagram as shown in \fref{fvquiver}. The two nodes of the quiver diagram represent the gauge group $U(k)$ and the global symmetry group $U(N)$.  The transformation laws of the fields are summarized in \tref{tvcharges}. 

There is no superpotential and hence no corresponding F-term equations. For gauge invariance, we demand invariance under the non-Abelian $SU(k)$ part of the $U(k)$ gauge symmetry. The remaining $U(1)$ symmetry has a FI-term in the D-term equations. They are set to be a constant due to the FI-term which combined with $U(1)$ gauge invariance amounts to complex rescaling of the $SU(k)$ invariant baryonic operators along the $U(1)$ direction. From the quiver in \fref{fvquiver} we observe that only the fundamental $Q$ transforms under the $U(1)$ parts of both the gauge $U(k)$ and global $U(N)$ symmetries. These transformations under the two $U(1)$'s are not independent and as such we can absorb the $U(1)$ of the local symmetry $U(k)$ into the $U(1)$ of the global $U(N)$ without any loss of generality.
\\
 
\begin{table}[t!]
\centering
\begin{tabular}{|c|cc|cc|c|c|}
\hline
\; & \multicolumn{2}{c|}{$U(k)_{gauge}$}
& \multicolumn{2}{c|}{$U(N)_{global}$} & &
\\
\; & $SU(k)_w$ & $U(1)_z$ & $SU(N)_x$ & $U(1)_q$ & $U(1)_r$ & $U(1)_s$\\
\hline\hline
$\phi$ & $[1,0,\dots,0,1]_w$+1 & 0 & $[0,\dots,0]_{x}$ & 0 & 1 & 1 \\
$Q_a^i$ & $[0,\dots,0,1]_{w}$ & +1 & $[1,0,\dots,0]_{x}$ & $-1$ & 1 & 0 \\
\hline
\end{tabular}
\caption{Quiver fields of the vortex theory and their transformation properties. \label{tvcharges}}
\end{table} 
 
\paragraph{Vortex moduli space.} The vortex moduli space $\V_{k,N}$ for $k$ $U(N)$ vortices is a partially weighted projective space originating from a partial $\mathbb{C}^{*}$ projection of the vortex master space $\master_{k,N}$. We denote this relationship as follows,
\beal{es901}
\V_{k,N} \equiv \mathbb{WP}_{U(1)}[\master_{k,N}]~~.
\eea
The projection of the master space $\master_{k,N}$ is along the $U(1)$ part of the $U(k)$ gauge symmetry.

The vortex master space $\master_{k,N}$ is a space of gauge invariant operators which are invariant under the $SU(k)$ part of the gauge symmetry. We make use of the Hilbert series to identify the algebraic structure of the master space. By identifying the $U(1)$ gauge charges carried by the generators of $\master_{k,N}$, we can specify the projection into the full vortex moduli space $\V_{k,N}$ as a partially weighted projected space. 

We make use of the following notation to describe the projection of $\master_{k,N}$ along the $U(1)$ gauge charges,
\beal{es902}
\V_{k,N} = \master_{k,N} / \{
\alpha_1 \simeq \lambda^{w_1} \alpha_1 ~,~
\dots
~,~
\alpha_n \simeq \lambda^{w_n} \alpha_n 
\}
~,~
\eea
where $\alpha_i$ are the generators of $\master_{k,N}$, $w_i$ are the $U(1)$ gauge charges, and $\lambda$ is the $\mathbb{C}^{*}$ parameter. The master space is determined by the following symplectic quotient
\beal{es903}
\master_{k,N} = \mathbb{C}^c // SU(k)~,~
\eea
where $c$ is the dimension of the freely generated space of all quiver fields. Note that there is no superpotential, and hence the master space here is determined without an ideal made of F-term constraints.

The dimension of the master space $\master_{k,N}$ for the $k$ $U(N)$ vortex theory is 
\beal{es904}
\dim_{\mathbb{C}} \master_{k,N} = k N + 1 ~~,
\eea
which reduces via the $\mathbb{C}^{*}$ projection to the dimension of the vortex moduli space
\beal{es905}
\dim_{\mathbb{C}} \V_{k,N} = k N ~~.
\eea
\\

\begin{figure}[ht!!]
\begin{center}
\resizebox{0.9\hsize}{!}{
\includegraphics[trim=0cm 0cm 0cm 0cm,totalheight=16 cm]{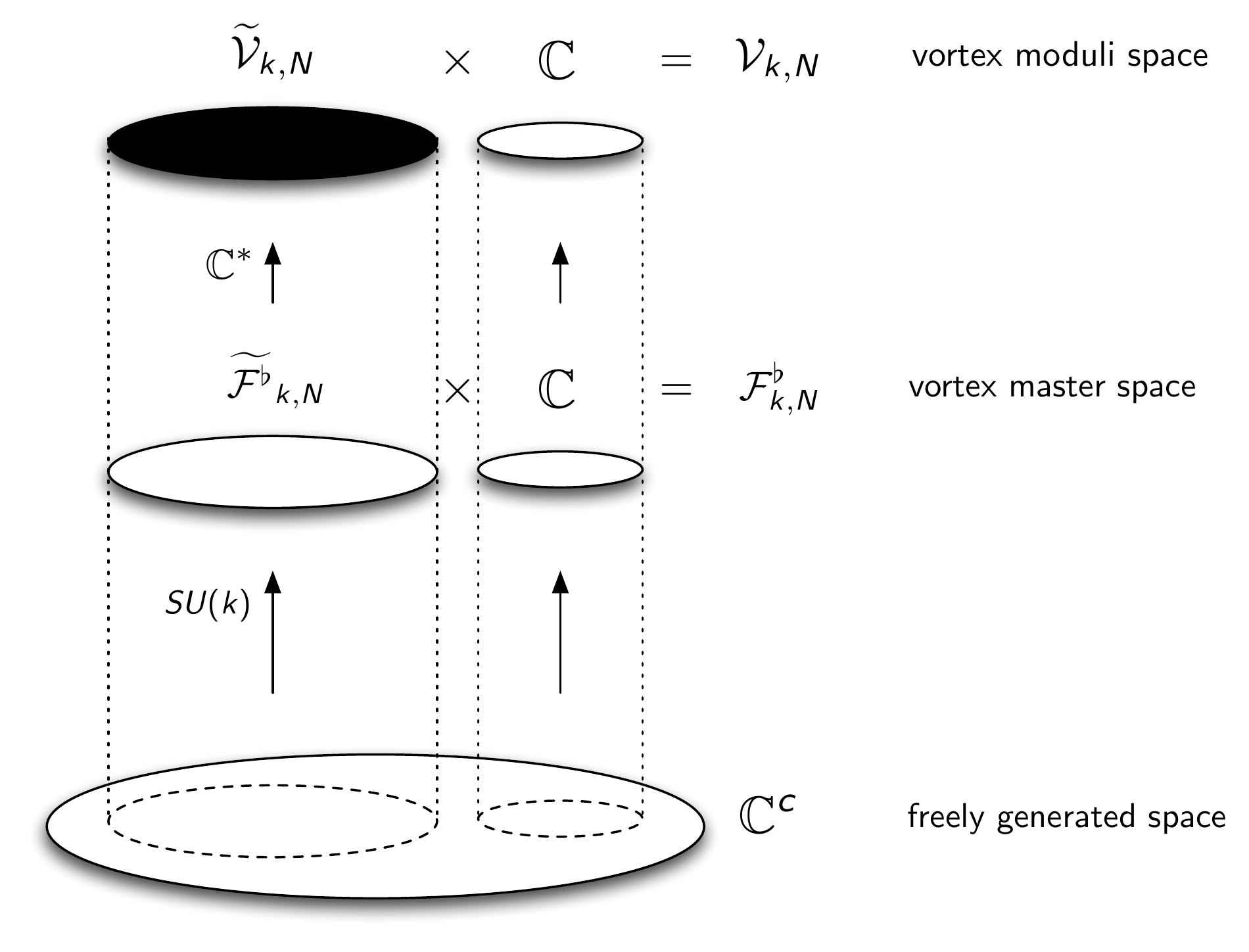}
}  \caption{\textit{Vortex moduli space.}  The freely generated space of quiver fields is lifted to the vortex master space $\master_{k,N}$ by quotienting out the $SU(k)$ gauge charges. The full vortex master space $\master_{k,N}$ contains the irreducible part of the master space $\wmaster_{k,N}$ and the center of mass position factor $\mathbb{C}$. The $\mathbb{C}^{*}$ projection leads to the vortex moduli space $\V_{k,N}$. \label{fspacedia}}
 \end{center}
 \end{figure}

\paragraph{Vortex Hilbert series.}
For the Hilbert series computation, we have the following plethystic exponentials that contribute to the Hilbert series. For the fundamental $Q_{a}^{i}$ we have the contribution
\beal{es300}
\PE\Big[
[1,0,\dots,0]_{x} [0,\dots,0,1]_w 
z
~
t
\Big]
= \frac{1}{\prod_{i=1}^{k-1} \prod_{j=1}^{N-1} (1- \frac{x_{i}}{x_{i-1}} \frac{w_{j-1}}{w_{j}} ~z~t)}~~,
\eea
where $x_0=x_k=1$ and $w_0=w_N=1$. The contribution from the adjoint $\phi$ is the plethystic exponential 
\beal{es301}
\PE\Big[
([1,0,\dots,0,1]_{w}+1) 
s
\Big]
=
\frac{1}{1-s}
\PE\Big[
[1,0,\dots,0,1]_{w} 
~s
\Big]~~.
\eea

The Molien integral for the Hilbert series of the master space of $k$ $U(N)$ vortices is
\beal{e1}
g_{k,U(N)} (s,t,x; \master_{k,N})
&=&
\oint
\ud\mu_{SU(k)}
\PE
\Big[
[1,0,\dots,0]_{x}
[0,\dots,0,1]_{w}
~
z
~
t
\nn\\
&& \hspace{5cm}
+
([1,0,\dots,0,1]_{w}+1) 
~
s
\Big]
\nn\\
&=&
\frac{1}{1-s}
\oint
\ud\mu_{SU(k)}
\PE
\Big[
[1,0,\dots,0]_{x}
[0,\dots,0,1]_{w}
~
z~
t
\nn\\
&& \hspace{5cm}
+
[1,0,\dots,0,1]_{w} 
~
s
\Big] ~~,
\eea
where $\ud\mu_{SU(k)}$ is the Haar measure for $SU(k)$.
Note that the $\frac{1}{1-s}$ prefactor corresponds to a $\mathbb{C}$ factor of the vortex moduli space which parameterizes the centre of mass of the vortices. 
\\

\section{$1$ $U(N)$ vortex on $\mathbb{C}$ \label{s1}}

As a pedagogical introduction, we go through the example of $1$ $U(N)$ vortices and re-derive known results from \cite{Hanany:2003hp} by making use of Hilbert series and the notion of vortex master spaces.

\begin{figure}[H]
\begin{center}
\resizebox{0.6\hsize}{!}{
\includegraphics[trim=0cm 0cm 0cm 0cm,totalheight=16 cm]{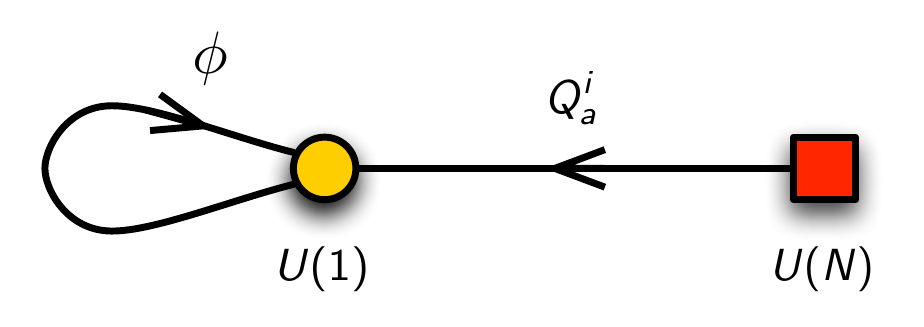}
}  \caption{Quiver diagram of the 1 $U(N)$ vortex theory. \label{fquiverk1}}
 \end{center}
 \end{figure}

\paragraph{The moduli space.} The Hilbert series of the $1$ $U(N)$ vortex master space $\master_{1,N}$ does not require a Molien integral. The $U(1)$ gauge charges are kept in the Hilbert series in order to identify the $\mathbb{C}^{*}$ projection weights for the generators of the master space. Recall that the generators of the master space $x_1,\dots,x_d$ play the role of the projection coordinates, where the $\mathbb{C}^{*}$ projection is given by
\beal{es20aa1}
(x_1,\dots,x_d) \simeq (\lambda^{w_1} x_1, \dots,\lambda^{w_d} x_d) ~.~
\eea
 $\lambda$ is the $\mathbb{C}^{*}$ parameter and $w_1,\dots,w_d$ are the projection weights along the corresponding projection coordinates $x_1,\dots,x_d$. These $U(1)$ weights $w_1,\dots,w_d$ are used to partially project the master space $\master_{1,N}$ into the vortex moduli space $\V_{1,N}$. The vortex moduli space then takes the following form in analogy to the above projection,
\beal{es20aa2}
\V_{1,N} = \master_{1,N} /\{ 
x_1 \simeq \lambda^{w_1} x_1 , 
\dots,
x_d \simeq \lambda^{w_d} x_d
\}~,~
\eea
where $\master_{1,N}$ is parameterized by its generators $x_1,\dots,x_d$. 
\\

\paragraph{The Molien integral and Hilbert series.}  The $U(1)$ gauge charge is carried by the fugacity $t$ corresponding to $Q$. Accordingly, in terms of the $U(1)$ gauge and $SU(N)$ global symmetries of the vortex theory, the Hilbert series for the $1$ $U(N)$ vortex master space can be written as,
\beal{es20}
g(t,s,x;\master_{1,N}) 
&=& 
\PE\Big[
[1,0,\dots,0]_x t  + s
\Big]
=
\frac{1}{1-s}
\PE\Big[
[1,0,\dots,0]_x t 
\Big]~~,
\eea
where $[1,0,\dots,0]_x$ is the fundamental representation of $SU(N)$.
\\

\paragraph{Center of mass position.} Note that the overall factor $\frac{1}{1-s}$ in \eref{es20} corresponds to the centre of mass position of the vortex in $\mathbb{C}$. We can ignore the overall position of the vortex and obtain the Hilbert series of the \textit{reduced} master space $\wmaster_{1,N}$ excluding the centre of mass position. The Hilbert series of the reduced master space is given by
\beal{es20b}
g(t,s,x;\wmaster_{1,N}) = (1-s) \times g(t,s,x;\master_{1,N})~.~
\eea 
Note that the centre of mass is not involved in the projection of the master space into the vortex moduli space because it is not charged under the $U(1)$ gauge symmetry. Accordingly, we will later make use of this fact and project the reduced master space $\wmaster_{1,N}$ into the reduced vortex moduli space $\V_{1,N} = \wV_{1,N}\times\mathbb{C}$. In the following discussion, we, for simplicity, refer interchangeably to $\wV_{1,N}$ and $\V_{1,N}$ as the moduli space of the $1$ $U(N)$ vortex on $\mathbb{C}$.
\\

\subsection{$1$ $U(N)$ vortex on $\mathbb{C}$ \label{s1b}}

The first few examples of the refined Hilbert series for the master space of the $1$ $U(N)$ vortex theory are as follows,
\beal{es30}
g(t,s,x;\wmaster_{1,1}) &= &
\frac{1}{1-t}
~~,
\nn\\
g(t,s,x;\wmaster_{1,2}) &= &
\frac{1}{(1- x t)(1-\frac{1}{x}t)} 
~~,
\nn\\
g(t,s,x;\wmaster_{1,3}) &= &
\frac{1}{(1-x_1 t)(1-\frac{x_2}{x_1} t)(1-\frac{1}{x_2}t)}~~,
\eea
We observe that the master spaces $\master_{1,N}$ are freely generated spaces of dimension $N+1$, 
\beal{es30b}
\master_{1,N} = \mathbb{C}^{N+1}~.~
\eea

As a character expansion in terms of characters of irreducible representations of the global $SU(N)$ symmetry, the Hilbert series is
\beal{es33}
g(t,s,x;\wmaster_{1,N}) =&&
\sum_{n=0}^{\infty} [n,0,\dots,0]_x t^{n}~~,
\eea
where the $\PE$ in \eref{es20} acts as a function generating symmetric products of the fundamental representation of $SU(N)$. 

The plethystic logarithm is the inverse function of the plethystic exponential. Given that the Hilbert series of the full master space $\master_{1,N}=\mathbb{C}\times\wmaster_{1,N}$ was generated via $\phi$ counted by $s$ and $Q$ counted by $t$ in the vortex quiver diagram, the plethystic logarithm is simply,
\beal{es32}
\PL\Big[g(t,s,x;\master_{1,N})\Big]
=
\PL\Big[g(t,s,x;\wmaster_{1,N})\Big] + s 
=
[1,0,\dots,0]_x t + s ~~.
\eea
We summarize the generators of $\wmaster_{1,N}$ as follows,
\beal{es34}
~[1,0,\dots,0]_x t 
&\ra& 
Q^{i} ~~.
\eea

We expect that the $1$ $U(N)$ vortex moduli space $\V_{1,N}$ has complex dimension $N$. The $\mathbb{C}^{*}$ projection reduces the dimension of the master space $\master_{1,N}$ by 1. According to the Hilbert series in \eref{es32}, the  $Q^i$ are the only objects carrying a $U(1)$ charge, which is interpreted as $w_i=1$ for all $Q^i$. Therefore, the $\mathbb{C}^{*}$ projection of the reduced master space $\wmaster_{1,N}$ is given by
\beal{es34b}
\wV_{1,N} =
\wmaster_{1,N} / \{
Q^i \simeq \lambda Q^i
\}~,~
\eea 
which implies that the $1$ $U(N)$ vortex moduli space is
\beal{es34b3}
\wV_{1,N}
= \cp^{N-1}~,~
\eea
in other words, as the complex projective space of dimension $N-1$. This result is known for example from \cite{Hanany:2003hp}.
\\

\section{$2$ $U(N)$ vortices on $\mathbb{C}$ \label{s2}}

We proceed here to the case of $2$ $U(N)$ vortices on $\mathbb{C}$ by generalizing the previous analysis on $1$ $U(N)$ vortices with Hilbert series and vortex master spaces.

\begin{figure}[H]
\begin{center}
\resizebox{0.6\hsize}{!}{
\includegraphics[trim=0cm 0cm 0cm 0cm,totalheight=16 cm]{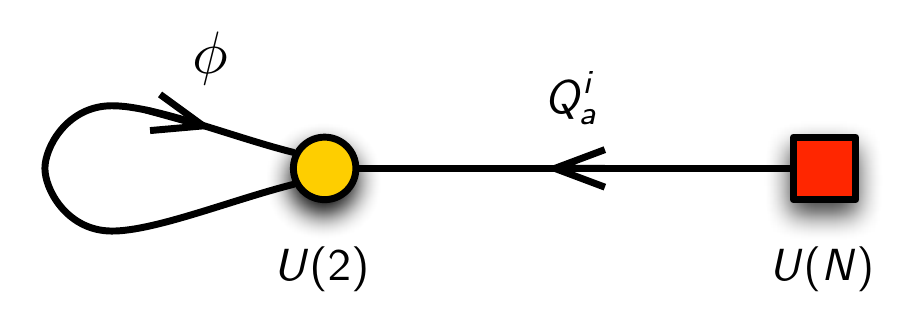}
}  \caption{Quiver diagram of the 2 $U(N)$ vortex theory. \label{fquiverk2}}
 \end{center}
 \end{figure}

\paragraph{The moduli space.} We recall that the $2$ $U(N)$ vortex moduli space $\V_{2,N}$ is a partial $\mathbb{C}^{*}$ projection of the corresponding master space $\master_{2,N}$. The generators $x_1,\dots,x_d$ of $\master_{2,N}$ are projection coordinates with corresponding projection weights $w_1,\dots,w_d$. Under the $\mathbb{C}^{*}$ projection, one has
\beal{es50beg1}
(x_1,\dots,x_d) \simeq 
(\lambda^{w_1} x_1,\dots,\lambda^{w_d} x_d) ~,~
\eea
where $\lambda$ is the $\mathbb{C}^{*}$ parameter. Analogously, the master space is $\mathbb{C}^{*}$ projected as follows to give the vortex moduli space
\beal{es50beg2}
\V_{2,N} = \master_{2,N} / \{
x_1 \simeq \lambda^{w_1} x_1 , \dots , 
x_d \simeq \lambda^{w_d} x_d 
\} ~,~
\eea
where the generators $x_1,\dots,x_d$ parameterize the master space $\master_{2.N}$.
\\

\paragraph{The Molien integral and Hilbert series.} The computation for the Hilbert series of the $2$ $U(N)$ vortex master space $\master_{2,N}$ requires an integral over the $SU(2)$ gauge charges of the vortex theory. The Molien integral is
\beal{es50}
g(t,s,x;\mathcal{V}_{2,N}) = 
\oint \ud\mu_{SU(2)}~
\PE\Big[
[1,0,\dots,0]_x [1]_w t + [2]_w s
\Big]~~,
\eea
where $\ud\mu_{SU(2)}$ is the Haar measure of $SU(2)$. The entries in the plethystic exponential correspond to $Q_\alpha^i$ transforming as $[1,0\dots,0]_x [1]_w t$, and $\phi$ transforming in $[2]_w s$.
\\

\paragraph{Center of mass position.} Recall that the character for the adjoint representation of $SU(2)$ is $[2]_w= w^2 + 1 + w^{-2}$. Accordingly, the integrand in \eref{es50} can be rewritten as
\beal{es51aa1}
\PE\Big[[1,0,\dots,0]_x [1]_w t + [2]_w s\Big] =
\frac{1}{1-s}\PE\Big[[1,0,\dots,0]_x [1]_w t + (w^{2}+w^{-2}) s\Big]
~.~
\eea
The $\frac{1}{1-s}$ factor above completely decouples from the Molien integral and corresponds to the $\mathbb{C}$ factor of the master space and vortex moduli space. It relates to the center of mass position of the 2 vortices and can be ignored in the following discussion of moduli spaces. We call the vortex master and moduli spaces reduced when we factor out the center of mass position contribution and denote the spaces respectively by $\wmaster_{2,N}$ and $\wV_{2,N}$. The partial $\mathbb{C}^{*}$ quotient of the master space does not involve the center of mass $\mathbb{C}$ factor. The Hilbert series of the reduced moduli space $\wV_{2,N}$ is obtained therefore simply from the full moduli space $\V_{2,N}$ Hilbert series
\beal{es51aa2}
g(t,s,x;\wV_{2,N}) = (1-s) \times g(t,s,x;\V_{2,N}) ~.~
\eea
\\

\paragraph{The Hilbert series as a rational function.} The first few examples of the refined Hilbert series for the $2$ $U(N)$ vortex master spaces are
\beal{es51} 
g(t,s,x;\wmaster_{2,1}) &= &
\frac{1}{(1 - s^2) (1 - s t^2)}
~~,
\nn\\
g(t,s,x;\wmaster_{2,2}) &= &
\frac{1-s^2 t^4}{
(1-s^2) (1-t^2) 
(1 -  x^2 s t^2) (1-s t^2)
(1-x^{-2} s t^2)
}
~~,
\nn\\
g(t,s,x;\wmaster_{2,3}) &= &
\nn\\
&&
\hspace{-2.5cm}
\frac{
1 
+ [0,1]_x s t^2
- [1,0]_x s t^4
- s^2 t^6 
}{
(1 - s^2) (1 - x_1^{-1} t^2) (1 - x_1^2 s t^2 ) (1- x_1 x_2^{-1} t^2) (1 - x_2 t^2 ) (1 - x_2^{-2} s t^2) (1 - x_1^{-2} x_2^2 s t^2 )
}~.~
\nn\\
\eea
\\

\paragraph{Quiver fields and convention.} $SU(N)_x$ flavor indices are given by small $i_1,i_2=1,\dots,N$ and $SU(k)_w$ color indices are given by $\alpha,\beta=1,2$. The antisymmetric tensor $\epsilon^{ab}$ is used as a raising and lowering operator for gauge indices. The Casimir operators are given by,
\beal{es992}
u_1 &=& \Tr(\phi) = 0 \nn\\
u_2 &=& \Tr(\phi^2) =  2 \phi_{12}^2 - 2 \phi_{11} \phi_{22} ~~.
\eea
\\

\subsection{$2$ $U(1)$ vortices on $\mathbb{C}$ \label{s2b1}}

The Hilbert series of the master space of 2 $U(1)$ vortices is given by the following Molien integral,
\beal{es1000}
g(t,s,x;\master_{2,1}) =  \oint \ud\mu_{SU(2)} \PE\Big[
[1]_w t + [2]_w s
\Big]~~,
\eea
where $\ud\mu_{SU(2)}$ is the $SU(2)$ Haar measure. The Hilbert series is
\beal{es1001}
g(t,s,x;\wmaster_{2,1}) = 
\frac{1}{(1 - s^2) (1- s t^2)} ~~.
\eea
The plethystic logarithm is
\beal{es1002}
\PL\Big[
g(t,s,x;\wmaster_{2,1})
\Big]
= s^2 + s t^2~~.
\eea
We note that $\wmaster_{2,1}$ has dimension $2$. The generators are
\beal{es1003}
s^2 &\ra&u_2 = \Tr(\phi^2) \nn\\
st^2 &\ra& A =  \epsilon^{\alpha_1 \alpha_2} \epsilon^{\beta_1 \beta_2} Q_{\alpha_1} \phi_{\alpha_2 \beta_1} Q_{\beta_2} ~~.
\eea
$\wmaster_{2,1}$ is a freely generated space where the generators do not form any relations.
\\

\paragraph{Vortex moduli space:} The $\mathbb{C}^{*}$ projection of the master space gives
\beal{es1003c}
\wV_{2,1} =
\wmaster_{2,1} / \{A \simeq \lambda^2 A\} = \mathbb{C}
~.~
\eea
For the general case of $k$ $U(1)$ vortices, the moduli space is simply
\beal{es1003cc}
\wV_{k,1} =
\wmaster_{k,1} / \{A \simeq \lambda^2 A\} = \mathbb{C}^{k-1}
~.~
\eea
The metric for the $2$ $U(1)$ vortex moduli space was studied in \cite{Hanany:2005bc,Kim:2002qj}. 
\\

\subsection{$2$ $U(2)$ vortices on $\mathbb{C}$\label{s2b2}}

 The $2$ $U(2)$ vortex master space has the following Molien integral for the Hilbert series,
\beal{es1010}
g(t,s,x;\master_{2,2})
&=&  \oint \ud\mu_{SU(2)} \PE\Big[
[1]_w [1]_x t + [2]_w s
\Big]~~,
\eea
where $[1]_x$ is the character of the fundamental representation of $SU(2)$. By removing the contribution from the center of mass position of the 2 vortices, the Hilbert series is
\beal{es1011}
g(t,s,x;\wmaster_{2,2}) &=&
\frac{
1 - s^2 t^4 
}{
(1 - s^2) (1 - t^2) (1- x^2 s t^2 ) (1 - s t^2) (1 - x^{-2} s t^2 )
}~~.\nn\\
\eea
We observe that $\wmaster_{2,2}$ is a complete intersection of dimension $4$. The character expansion of the Hilbert series is\beal{es1012}
g(t,s,x;\wmaster_{2,2}) = \frac{1}{1-s^2} \sum_{n_0,n_1=0}^{\infty} [2n_1]_x s^{n_1} t^{2n_0+2n_1}~~.
\eea

The plethystic logarithm of the Hilbert series takes the form
\beal{es1013}
\PL\Big[
g(t,s,x;\wmaster_{2,2})
\Big]
=
s^2+t^2+[2]_x s t^2 - s^2 t^4
~~.
\eea
The generators of the master space are
\beal{es1014}
s^2 &\ra& u_2=\Tr(\phi^2) \nn\\
t^2 &\ra& M = \epsilon^{\alpha_1 \alpha_2} \epsilon_{ij} Q_{\alpha_1}^{i} Q_{\alpha_2}^{j} 
\nn\\ 
~[2]_x s t^2 &\ra& A^{ij} = \epsilon^{\alpha_1 \alpha_2} \epsilon^{\beta_1 \beta_2} Q_{\alpha_1}^{i} \phi_{\alpha_2 \beta_1} Q_{\beta_2}^{j}
~~,
\eea
where we note that 
\beal{es1014b}
A^{ij} = A^{ji} ~~.
\eea
\\

\paragraph{Quadratic relations:} Let us consider the symmetric product
\beal{es1015b}
\text{Sym}^2 ~[2]_x = [4]_x + [0]_x ~~,
\eea
and the term in the plethystic logarithm corresponding to the quadratic relation,
\beal{es1015c}
- s^2 t^4 ~~.
\eea
The quadratic relation can be constructed as follows, 
\beal{es1015}
-s^2 t^4 &\ra&
R = \det{A} - \frac{1}{8} u_2 M^2= 0
~~.
\eea
Note that $-s^2 t^4$ is the corresponding contribution in the plethystic logarithm of the Hilbert series shown in \eref{es1013}.
\\

\paragraph{Vortex moduli space:} The $\mathbb{C}^{*}$ projection of $\wmaster_{2,2}$ depends on the $U(1)$ gauge charges encoded in the $Q_\alpha^i$ fugacity $t$ in the Hilbert series. The vortex moduli space is
\beal{es1016}
\wV_{2,2} =
\wmaster_{2,2} /\{ M\simeq \lambda^2 M, A^{ij}\simeq \lambda^2 A^{ij}\}~~,
\eea
where
\beal{es1016b}
\wmaster_{2,2} = \mathbb{C}[u_2, M, A^{ij} ]/ \{ R = 0 \}~~.
\eea
The dimension of $\wV_{2,2}$ is $3$. This reproduces the result in \cite{Eto:2010aj,Eto:2006cx}.
\\

\subsection{$2$ $U(3)$ vortices on $\mathbb{C}$ \label{s2b3}}

The Hilbert series for the $2$ $U(3)$ vortex master space is given by the Molien integral
\beal{es1020}
g(t,s,x;\master_{2,3}) = 
\oint \ud\mu_{SU(2)} 
\PE\Big[
[1]_w [1,0]_x +[2]_w s 
\Big]~~.
\eea
When the contribution from the center of mass of the vortices is removed, the integral gives 
\beal{es1021}
&&
g(t,s,x;\wmaster_{2,3}) = 
\nn\\
&&
\hspace{0.5cm}
\frac{
1 
+ [0,1]_x s t^2
- [1,0]_x s t^4
- s^2 t^6 
}{
(1 - s^2) (1 - x_1^{-1} t^2) (1 - x_1^2 s t^2 ) (1- x_1 x_2^{-1} t^2) (1 - x_2 t^2 ) (1 - x_2^{-2} s t^2) (1 - x_1^{-2} x_2^2 s t^2 )
}~~,
\nn\\
\eea
where we notice that $\wmaster_{2,3}$ is a Calabi-Yau space of dimension $6$. The character expansion is
\beal{es1022}
g(t,s,x;\wmaster_{2,3}) =
\frac{1}{1-s^2}
\sum_{n_0,n_1=0}^{\infty}
[2n_1,n_0]_{x} s^{n_1} t^{2n_0+2n_1}
~~.
\eea

The plethystic logarithm of the Hilbert series has the following expansion
\beal{es1023}
\PL\Big[
g(t,s,x;\wmaster_{2,3}) 
\Big]
&=&
s^2 + [0,1]_x t^2 
+ [2,0]_x s t^2 
- [1,0]_x s t^4
- [0,2]_x s^2 t^4 
+ \dots ~~.
\nn\\
\eea
The generators can be identified from the above plethystic logarithm as
\beal{es1024}
s^2 &\ra&u_2=\Tr(\phi^2)
\nn\\
~[0,1]_x t^2 &\ra&  
M^{ij} = \epsilon^{\alpha_1 \alpha_2} Q_{\alpha_1}^{i} Q_{\alpha_2}^{j}
\nn\\
~[2,0]_x s t^2 &\ra&
A^{ij} = \epsilon^{\alpha_1 \alpha_2} \epsilon^{\beta_1 \beta_2} Q_{\alpha_1}^{i} \phi_{\alpha_2 \beta_1} Q_{\beta_2}^{j}
~~,
\eea
where we have
\beal{es1024b}
M^{ij} = - M^{ji}~~,~~
A^{ij} = A^{ji} ~~.
\eea
\\

\paragraph{Quadratic relations:} For the quadratic relations, we consider the following products of representations,
\beal{es1025a1}
~
[0,1]_x \times [2,0]_x 
&=&
 [2,1]_x + [1,0]_x ~~,
\nn\\
\text{Sym}^2 ~ [2,0]_x 
&=& 
[4,0]_x  + [0,2]_x~~.
\eea
From the plethystic logarithm in \eref{es1023}, we identify the terms 
\beal{es1025a2}
-[1,0]_x s t^4- [0,2]_x s^2 t^4
\eea
as corresponding to the following quadratic relations respectively,
\beal{es1025rel}
-[1,0]_x s t^4 &\ra&
R_i = \frac{1}{2} \epsilon^{jkl} A_{ij} M_{kl} = 0
\nn\\
-[0,2]_x s^2 t^4 &\ra&
S_{ij} = 
A_{ii} A_{jj} - A_{ij}^2 + \frac{1}{2} u_2 M_{ij} M_{ji} = 0 
~~.
\eea
where 
\beal{es1025a2b}
S_{[i j]}=0~~.
\eea
\\

\paragraph{Vortex moduli space:} The vortex moduli space is
\beal{es1025}
\wV_{2,3} =
\wmaster_{2,3} /\{ M_k \simeq \lambda^2 M_k , A^{ij} \simeq \lambda^2 A^{ij}\}
~,~
\eea
where
\beal{es1025b}
\wmaster_{2,3}
= \mathbb{C}[u_2,M_k, A^{ij}] / \{ R_i = 0 , S_{ij} = 0\} ~~.
\eea
The dimension of the above reduced vortex moduli space is $5$. The $2$ $U(2)$ vortex moduli space has been studied in \cite{Hashimoto:2005hi} in the context of cosmic $U(2)$ strings.
\\

\subsection{$2$ $U(4)$ vortices on $\mathbb{C}$ \label{s2b4}}

The master space of the 2 $U(4)$ vortex theory has the Molien integral for the Hilbert series,
\beal{es2000}
g(t,s,x;\master_{2,4})= 
\oint \ud\mu_{SU(2)}~
\PE\Big[
[1]_w [1,0,0]_x t + [2]_w s
\Big] ~~,
\eea
where $[1,0,0]_x$ is the character for the fundamental representation of $SU(4)_x$.

The character expansion is
\beal{es2001}
g(t,s,x;\wmaster_{2,4})= \frac{1}{1-s^2} \sum_{n_0,n_1=0}^{\infty} [2n_1,n_0,0]_x s^{n_1} t^{2n_0+2n_1} ~.~
\eea
 The corresponding plethystic logarithm is given by
\beal{es2002}
\PL\Big[
g(t,s,x;\wmaster_{2,4})
\Big]
&=&
s^2 + [0,1,0]_x t^2 + [2,0,0]_x s t^2
- t^4 - [1,0,1]_x s t^4 
\nn\\
&&
- [0,2,0]_x s^2 t^4 + \dots ~~.
\eea
The plethystic logarithm has an infinite expansion. Accordingly, the reduced master space is a non-complete intersection Calabi-Yau space of dimension $8$.

The generators are
\beal{es2003}
s^2 &\ra& u_2 = \Tr(\phi^2)
\nn\\
~[0,1,0]_x t^2 &\ra&
M^{ij} = \epsilon^{\alpha_1 \alpha_2} Q_{\alpha_1}^{i} Q_{\alpha_2}^{j}
\nn\\
~[2,0,0]_x s t^2 &\ra&
A^{ij} =  Q_{\alpha_1}^{i} \phi^{\alpha_1 \alpha_2} Q_{\alpha_2}^j
~~,
\eea
where 
\beal{es2004}
M^{ij} = - M^{ji}
~~,~~ 
A^{ij} = A^{ji} ~~.
\eea
\\

\paragraph{Quadratic relations:} The terms in the plethystic logarithm in \eref{es2002} corresponding to the quadratic relations are
\beal{es2005}
- t^4 - [1,0,1]_x s t^4 - [0,2,0]_x s^2 t^4~~.
\eea

Let us consider the following products of $SU(4)$ representations
\beal{es2006}
\text{Sym}^2 ~ [0,1,0]_x 
&=&
[0,2,0]_x  + [0,0,0]_x
~~, 
\nn\\
~[2,0,0]_x \times [0,1,0]_x 
&=&
[2,1,0]_x + [1,0,1]_x
~~,
\nn\\
\text{Sym}^2 ~ [2,0,0]_x 
&=&
[4,0,0]_x + [0,2,0]_x
~~.
\eea
Using the representation products, we construct the quadratic relations as follows
\beal{es2010}
-t^4 
&\ra&
R = 
\frac{1}{8} \epsilon^{ijkl} M_{ij} M_{kl} = 0  
\nn\\
-[1,0,1]_x s t^4 
&\ra&
S^{i}_{~j} =\frac{1}{2} \epsilon^{iklm} A_{jk} M_{lm} = 0
\nn\\
- [0,2,0]_x s^2 t^4 
&\ra&
T_{ijkl} = 
 A_{ik} A_{jl} - A_{il} A_{jk} + \frac{1}{2} u_2 M_{ij} M_{kl}
=0
~~.
\eea
The relations exhibit the following properties,
\beal{es2010b}
&
S^{i}_{~i} = 0 
&
\nn\\
&
T_{ijkl} = T_{jilk}
~,~
T_{ijkl} = - T_{jikl}
~,~
T_{ijkl} = - T_{ijlk}
&
~.~
\eea
\\

\paragraph{Vortex moduli space:} We can express the vortex moduli space as the following $\mathbb{C}^*$ projection
\beal{es2011}
\wV_{2,4}
= \wmaster_{2,4} /
\{
M^{ij} \simeq \lambda^2 M^{ij},
A^{ij} \simeq \lambda^2 A^{ij}
\}
~~,
\eea
where $\lambda$ is the $\mathbb{C}^{*}$ parameter.
The master space is given by
\beal{es2012}
\wmaster_{2,4}
=
\mathbb{C}[u_2, M^{ij}, A^{ij}]
/ \{
R = 0,
S^{i}_{~j} = 0,
T_{ijkl} = 0
\}~~.
\nn\\
\eea 
The dimension of the above reduced vortex moduli space is $7$.
\\

\subsection{$2$ $U(5)$ vortices on $\mathbb{C}$ \label{s2b5}}

 Let us consider the master space of 2 $U(5)$ vortices whose Hilbert series can be computed via the following Molien integral,
\beal{es2100}
g(t,s,x;\master_{2,5}) 
= 
\oint \ud\mu_{SU(2)} \PE\Big[
[1]_w [1,0,0,0]_x + [2]_w s
\Big]~~.
\eea

The character expansion of the Hilbert series is
\beal{es2101}
g(t,s,x;\wmaster_{2,5})
=
\frac{1}{1-s^2} \sum_{n_0,n_1=1}^{\infty} [2n_1,n_0,0,0]_x s^{n_1} t^{2n_0 + 2n_1}~.~
\eea
The plethystic logarithm of the Hilbert series is
\beal{es2012}
\PL\Big[
g(t,s,x;\wmaster_{2,3})
\Big]
&=&
 s^2 + [0,1,0,0]_x t^2 + [2,0,0,0]_x s t^2
- [0,0,0,1]_x t^4
\nn\\
&&
- [1,0,1,0]_x s t^4
- [0,2,0,0]_x s^2 t^4
+ \dots ~~.
\eea
The reduced master space is a non-complete intersection Calabi-Yau and has dimension $10$.

The generators of the master space are
\beal{es2013}
s^2 &\ra& u_2 = \Tr(\phi^2)
\nn\\
~[0,1,0,0]_x t^2 
&\ra&  
M^{ij} = \epsilon^{\alpha_1 \alpha_2} Q_{\alpha_1}^{i} Q_{\alpha_2}^{j}
\nn\\
~[2,0,0,0]_x s t^2 
&\ra&
A^{ij} = Q_{\alpha_1}^{i} \phi^{\alpha_1 \alpha_2} Q_{\alpha_2}^{j}
~~,
\eea
where we have 
\beal{es2013b}
M^{ij} = - M^{ji} ~~,~~
A^{ij} = A^{ji} ~~.
\eea
\\

\paragraph{Quadratic relations:} The plethystic logarithm of the Hilbert series in \eref{es2012} has the following terms corresponding to the quadratic relations between generators,
\beal{es2015}
- [0,0,0,1]_x t^4
- [1,0,1,0]_x s t^4
- [0,2,0,0]_x s^2 t^4
~~.
\eea
Let us consider the following representation products
\beal{es2016}
\text{Sym}^2 [0,1,0,0]_x 
&=&
[0,2,0,0]_x + [0,0,0,1]_x
~~,
\nn\\
~[2,0,0,0]_x  \times [0,1,0,0]_x 
&=&
[2,1,0,0]_x + [1,0,1,0]_x
~~,
\nn\\
\text{Sym}^2 [2,0,0,0]_x 
&= &
[4,0,0,0]_x  + [0,2,0,0]_x
~~.
\eea

Using the representation products, we construct the quadratic relations as follows,
\beal{es2017}
-[0,0,0,1]_x t^4 
&\ra&
R^{i} = \frac{1}{8} \epsilon^{ijklm} M_{jk} M_{lm} = 0
\nn\\
-[1,0,1,0]_x s t^4
&\ra&
S^{ij}_{~~k} = \frac{1}{2} \epsilon^{ijlmn} A_{kl} M_{mn} = 0
\nn\\
-[0,2,0,0]_x s ^2 t^4
&\ra&
T_{ijkl} = A_{ik} A_{jl} - A_{il} A_{jk} + \frac{1}{2} u_2 M_{ij} M_{kl} = 0 
~~,
\eea
where the relations satisfy
\beal{es2017b}
&
S^{ik}_{~~k} = 0
&
~,~
\nn\\
&
T_{ijkl} = T_{jilk}
~,~
T_{ijkl} = - T_{jikl}
~,~
T_{ijkl} = - T_{ijlk}
&
~.~
\eea
\\

\paragraph{Vortex moduli space:} Given the generators and the quadratic relations of the master space, we can express the vortex moduli space as the following $\mathbb{C}^{*}$ projection,
\beal{es2018}
\wV_{2,5} = 
\wmaster_{2,5} /\{
M^{ij} \simeq \lambda^{2} M^{ij},
A^{ij} \simeq \lambda^2 A^{ij}
\}
~~,
\eea
where $\lambda$ is the $\mathbb{C}^{*}$ parameter. The master space is
\beal{es2019}
\wmaster_{2,5} = \mathbb{C}[u_2,M^{ij},A^{ij}] / \{
R^{i} = 0 ,
S^{ij}_{~~k} = 0 ,
T_{ijkl} = 0 
\}~~.
\eea
The dimension of $\wmaster_{2,5}$ is $9$.
\\

\subsection{$2$ $U(N)$ vortices on $\mathbb{C}$ \label{s2bN}}

For the general case of $2$ $U(N)$ vortices, the Molien integral for the Hilbert series of the master space is
\beal{es1030}
g(t,s,x;\master_{2,N}) 
= 
\oint \ud\mu_{SU(2)}
\PE\Big[
[1]_w [1,0,\dots,0]_x + [2]_w s 
\Big]~~,
\eea
where $[1,0,\dots,0]_x$ is the character for the fundamental representation of $SU(N)_x$. When unrefined by setting the fugacities for the $SU(N)_x$ characters to $x_i=1$, the Hilbert series for the first few values of $N$ are 
\beal{es1031}
g(t,s;\wmaster_{2,1}) &=& 
\frac{1}{(1-s^2)(1-s t^2)}
~,
\nn\\
g(t,s;\wmaster_{2,2}) &=& 
 \frac{1 - s^2 t^4}{(1-s^2)(1- t^2) (1- s t^2)^2}
~,
\nn\\
g(t,s;\wmaster_{2,3}) &=&  
\frac{1 + 3 s t^2 - 3 s t^4 - s^2 t^6}{(1-s^2)(1 - t^2)^3 (1 - s t^2)^3}
~,
\nn\\
g(t,s;\wmaster_{2,4}) &=& 
\frac{
1 + t^2 + 6 s t^2 - 9 s t^4 + s^2 t^4 + s t^6 - 9 s^2 t^6 + 6 s^2 t^8 + s^3 t^8 + s^3 t^{10}
}{
(1-s^2)(1 - t^2)^5 (1 - s t^2)^4
}
~,
\nn\\
g(t,s;\wmaster_{2,5}) &=& 
\frac{
1
}{
(1-s^2)(1 - t^2)^7 (1 - s t^2)^5
}
\times
(
1 + 3 t^2 + 10 s t^2 + t^4 - 15 s t^4 
\nn\\
&& 
+ 5 s^2 t^4  - 5 s t^6 - 40 s^2 t^6 + 40 s^2 t^8 + 5 s^3 t^8 - 5 s^2 t^{10} + 15 s^3 t^{10} 
\nn\\
&&
- s^4 t^{10} - 10 s^3 t^{12}  - 3 s^4 t^{12} - s^4 t^{14}
)
~,
\nn\\
g(t,s;\wmaster_{2,6}) &=& 
\frac{
1
}{
(1-s^2 )(1 - t^2)^9 (1 - s t^2)^6
}
\times
(
1 + 6 t^2 + 15 s t^2 + 6 t^4 - 15 s t^4 
\nn\\
&&
+ 15 s^2 t^4  + t^6 - 36 s t^6 - 120 s^2 t^6 + s^3 t^6 - 6 s t^8 + 126 s^2 t^8  + 6 s^3 t^8
\nn\\
&&
+ 6 s^2 t^{10}  + 126 s^3 t^{10}  - 6 s^4 t^{10} + s^2 t^{12} - 120 s^3 t^{12} - 36 s^4 t^{12} + s^5 t^{12}
\nn\\
&&
+ 15 s^3 t^{14}  - 15 s^4 t^{14} + 6 s^5 t^{14} + 15 s^4 t^{16} + 6 s^5 t^{16} + s^5 t^{18}
)
~.
\nn\\
g(t,s;\wmaster_{2,7}) &=& 
\frac{
1
}{
(1-s^2) (1 - t^2)^{11} (1 - s t^2)^7
}\times
(1 + 10 t^2 + 21 s t^2 + 20 t^4 
\nn\\
&&
+ 35 s^2 t^4 + 10 t^6 - 112 s t^6 - 280 s^2 t^6 + 7 s^3 t^6 + t^8 - 70 s t^8 + 224 s^2 t^8 
\nn\\
&&
- 35 s^3 t^8 - 7 s t^{10} + 210 s^2 t^{10} + 658 s^3 t^{10} - 21 s^4 t^{10} + 21 s^2 t^{12} - 658 s^3 t^{12} 
\nn\\
&&
- 210 s^4 t^{12} + 7 s^5 t^{12} + 35 s^3 t^{14} - 224 s^4 t^{14} + 70 s^5 t^{14} - s^6 t^{14} - 7 s^3 t^{16}
\nn\\
&&
 + 280 s^4 t^{16} + 112 s^5 t^{16} - 10 s^6 t^{16} - 35 s^4 t^{18} - 20 s^6 t^{18} - 21 s^5 t^{20} 
 \nn\\
 &&
 - 10 s^6 t^{20} - s^6 t^{22}
)
~,~
\eea
where we have removed the contribution from the centre of mass position of the 2 vortices.

We observe that the numerators in \eref{es1031} are all palindromic. This indicates that the vortex master space is a Calabi-Yau manifold. Refined and as a character expansion, the Hilbert series for the reduced  $2$ $U(N)$ vortex master space takes the following general form
\beal{es1033}
g(t,s,x;\wmaster_{2,N}) &=&
\frac{1}{1-s^2}
\sum_{n_0=0}^{\infty} \sum_{n_1=0}^{\infty} [2 n_1,n_0,0,\dots,0]_x s^{n_1} t^{2(n_0+n_1)} ~~.
\eea

The plethystic logarithm of the Hilbert series is
\beal{es1034}
\PL\Big[
g(t,s,x;\wmaster_{2,N})
\Big]
&=&
s+ s^2 
+ [0,1,0,\dots,0]_x t^2
+ [2,0,\dots,0]_x s t^2
\nn\\
&&
- [0,0,0,1,0,\dots,0]_x t^4
- [1,0,1,0,\dots,0]_x s t^4
\nn\\
&&
- [0,2,0,\dots,0]_x s^2 t^4
+ \dots ~~.
\eea
The first positive terms of the plethystic logarithm correspond to the generators
\beal{es1035}
s^2 
&\ra&
u_2=\Tr(\phi^2) 
\nn\\
~[0,1,0,\dots,0]_x t^2 
&\ra& 
M^{ij} = \epsilon^{\alpha_1 \alpha_2} Q_{\alpha_1}^{i} Q_{\alpha_2}^{j}
\nn\\
~[2,0,\dots,0]_x s t^2 
&\ra& 
A^{ij} = Q_{\alpha_1}^{i} \phi^{\alpha_1 \alpha_2} Q_{\alpha_2}^{j}
~~.
\eea
The generators satisfy the following,
\beal{es1035r1}
M^{ij} = - M^{ji} ~,~
A^{ij} = A^{ij} ~.~
\eea
\\

\paragraph{Quadratic relations:}
The plethystic logarithm of the Hilbert series of the master space exhibits the following terms corresponding quadratic relations between the generators of the master space,
\beal{es1035r2}
-[0,0,0,1,0,\dots,0]_x t^4 -[1,0,1,0\dots,0]_x s t^2 -[0,2,0,\dots,0]_x s^2 t^4~~.
 \eea
 which respectively correspond to the following quadratic relations between generators,
\beal{es1036r1}
-[0,0,0,1,0,\dots,0]_x t^4
&\ra& 
R^{i_1\dots i_{N-4}}
= \frac{1}{8} \epsilon^{i_1\dots i_{N-4} jklm} M_{jk} M_{lm} = 0 
\nn\\
-[1,0,1,0,\dots,0]_x s t^2 
&\ra&
S^{i_1\dots i_{N-3}}_{~~~~~~~~~j}
= 
\frac{1}{2} \epsilon^{i_1\dots i_{N-3}klm}
A_{jk} M_{lm}
= 0
\nn\\
-[0,2,0,\dots,0]_x s^2 t^4
&\ra&
T_{i j k l} = 
A_{ik} A_{jl} - A_{il}A_{jk} + \frac{1}{2} u_2 M_{ij} M_{kl} = 0 ~~.
\eea
The above relations satisfy
\beal{es1036r2}
&
S^{i_1\dots i_{N-3}}_{~~~~~~~~~i_{N-3}} = 0 
&
~,~
\nn\\
&
T_{ijkl} = T_{jilk}
~,~
T_{ijkl} = - T_{jikl}
~,~
T_{ijkl} = - T_{ijlk}
&
~.~
\eea
\\

\paragraph{Vortex moduli space:} The moduli space of 2 $U(N)$ vortices is expressed as the following $\mathbb{C}^{*}$ projection,
\beal{es1037}
\wV_{2,N} = \wmaster_{2,N} / \{
M^{ij} \simeq \lambda^2 M^{ij},
A^{ij} \simeq \lambda^2 A^{ij}
\}~~,
\eea
where the master space is given by
\beal{es1038}
\wmaster_{2,N} = 
\mathbb{C}[u_2,M^{ij},A^{ij}] / \{
R^{i_1\dots i_{N-4}} = 0,
S^{i_1\dots i_{N-3}}_{~~~~~~~~~j} = 0,
T_{ijkl} = 0
\}~~.
\eea
The dimension of the reduced $2$ $U(N)$ vortex moduli space is $2N-1$.
\\

\section{$3$ $U(N)$ vortices on $\mathbb{C}$ \label{s3}}

\begin{figure}[H]
\begin{center}
\resizebox{0.6\hsize}{!}{
\includegraphics[trim=0cm 0cm 0cm 0cm,totalheight=16 cm]{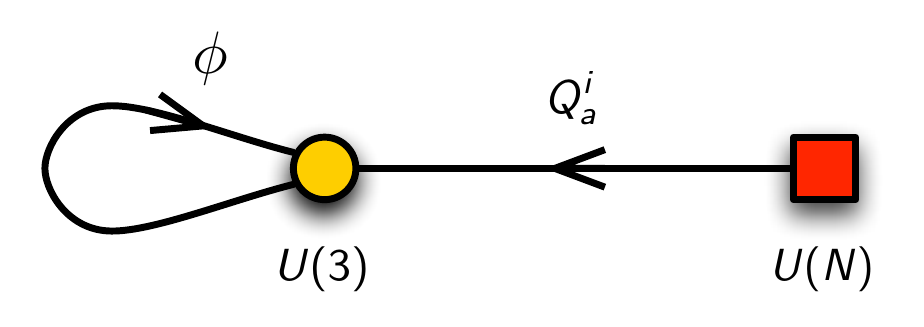}
}  \caption{Quiver diagram of the 3 $U(N)$ vortex theory. \label{fquiverk3}}
 \end{center}
 \end{figure}

\paragraph{The moduli space.} The moduli space of $3$ $U(N)$ vortices $\V_{3,N}$ can be expressed as a $\mathbb{C}^{*}$ projection of the master space $\master_{3,N}$ of the vortex theory. Given the generators of the master spaces, $x_1,\dots,x_d$, the $\mathbb{C}^{*}$ projection acts as follows on the master space coordinates
\beal{es1100aa1}
(x_1,\dots,x_d) \simeq (\lambda^{w_1} x_1, \dots, \lambda^{w_d} x_d) ~.~
\eea
$\lambda$ is the $\mathbb{C}^{*}$ parameter and $w_1,\dots,w_d$ are the $U(1)$ weights for the $\mathbb{C}^{*}$ projection of the master space. Following the coordinate identification in \eref{es1100aa1}, the master space is $\mathbb{C}^{*}$ projected as follows,
\beal{es1100aa2}
\V_{3,N} = \master_{3,N}/\{
x_1 \simeq \lambda^{w_1} x_1 , \dots, x_d \simeq \lambda^{w_d} x_d 
\}~.~
\eea
\\

\paragraph{The Molien integral and Hilbert series.} The Hilbert series of the master space of $3$ $U(N)$ vortices is given by the following Molien integral 
\beal{es1100aa5}
g(t,s,x;\master_{3,N}) = \oint \ud\mu_{SU(3)} ~
\PE\Big[
[1,0,\dots,0]_x [0,1]_w t + [1,1]_w s
\Big]~,~
\eea
where $\ud\mu_{SU(3)}$ is the Haar measure for $SU(3)$. The entries of the plethystic exponential correspond as expected to  $Q_{\alpha}^{i}$ transforming in $[1,0,\dots,0]_x [0,1]_w t$ and $\phi$ transforming in $[1,1]_w s$.
\\

\paragraph{Center of mass position.} The integrand in the Molien integral for the Hilbert series of the vortex master space can be rewritten as follows,
\beal{es1100aa6}
&&
\PE\Big[
[1,0,\dots,0]_x [0,1]_w t + [1,1]_w s
\Big]
= 
\frac{1}{1-s} \PE\Big[
[1,0,\dots,0]_x [0,1]_w t 
\nn\\
&& \hspace{2cm}
+ 
(w_1 w_2 + w_1^2 w_2^{-1} + w_1^{-1} w_2^2 + 1 + w_1 w_2^{-2} + w_1^{-2} w_2 + w_1^{-1} w_2^{-1} )
\Big]~,~
\nn\\
\eea 
where the character of the adjoint of $SU(3)$ is given by
\beal{es1100aa7}
[1,1]_w = w_1 w_2 + w_1^2 w_2^{-1} + w_1^{-1} w_2^2 + 2 + w_1 w_2^{-2} + w_1^{-2} w_2 + w_1^{-1} w_2^{-1} ~.~
\eea
The prefactor $\frac{1}{1-s}$ in \eref{es1100aa6} refers to the center of mass position of the 3 vortices. The center of mass position contributes in the vortex master space $\master_{3,N}$ and moduli space $\V_{3,N}$ with a $\mathbb{C}$ factor. This contribution does not interact with the $\mathbb{C}^{*}$ projection of the master space and therefore can be safely taken out from the following discussion. The reduced vortex master and moduli spaces are denoted respectively by $\wmaster_{3,N}$ and $\wV_{3,N}$.
\\

\paragraph{Quiver fields and convention.} The $SU(N)$ flavor indices are given by $i_1,i_2,i_3=1,\dots,N$ and $SU(k)$ color indices are given by $\alpha,\beta=1,2,3$. The components for the adjoint $\phi_{\alpha\beta}$ satisfy $\phi_{33}= -\phi_{11}-\phi_{22}$ such that
\beal{es1100a2}
\phi_{\alpha}^{~\alpha} = 0 ~.~
\eea
Note that for $SU(3)$ we use $\delta_{\alpha\beta}$ as a raising or lowering operator.
Using the choice of adjoint above, we obtain the following traces which we use in the discussion of vortex moduli spaces,
\beal{es1100a3}
u_1 &=& \Tr(\phi) = 0
\nn\\
u_2 &=& \Tr(\phi^2) = 
2 \phi_{11}^2 + 2 \phi_{12} \phi_{21} + 2 \phi_{11} \phi_{22} + 2 \phi_{22}^2 + 2 \phi_{13} \phi_{31} + 2 \phi_{23} \phi_{32}
\nn\\
u_3 &=& \Tr(\phi^3) = 
3 \phi_{11} \phi_{12} \phi_{21} - 3 \phi_{11}^2 \phi_{22} + 3 \phi_{12} \phi_{21} \phi_{22} - 3 \phi_{11} \phi_{22}^2 - 3 \phi_{13} \phi_{22} \phi_{31} + 3 \phi_{12} \phi_{23} \phi_{31}
\nn\\
&&
\hspace{2cm}
 + 3 \phi_{13} \phi_{21} \phi_{32} - 3 \phi_{11} \phi_{23} \phi_{32}
~.~
\eea
\\

\subsection{$3$ $U(1)$ vortices on $\mathbb{C}$ \label{s3b1}}

The Hilbert series of the $3$ $U(1)$ vortex master space is
\beal{es1102}
g(t,s,x;\wmaster_{3,1}) =
\frac{1}{(1-s^2)(1-s^3)(1-s^3 t^3)}
~~.
\eea
The master space generators are 
\beal{es1104}
s^2 &\ra& u_2 = \Tr(\phi^2)
\nn\\
s^3 &\ra& u_3 = \Tr(\phi^3) 
\nn\\
s^3 t^3 &\ra& 
A_{012}
= \epsilon^{\alpha_1 \alpha_2 \alpha_3} Q_{\alpha_1} \phi_{\alpha_2}^{\beta_1} Q_{\beta_1} \phi_{\alpha_3}^{\beta_2} \phi_{\beta_ 2}^{\beta_3} Q_{\beta_3}
~,~
\eea
and  the vortex moduli space under the $\mathbb{C}^{*}$ projection becomes
\beal{es1105}
\wV_{3,1} = \wmaster_{3,1} / \{
A_{012} \simeq \lambda^3 A_{012}
\} = \mathbb{C}^2 ~.~
\eea
This agrees with the generalization for $k$ $U(1)$ vortices.
\\

\subsection{$3$ $U(2)$ vortices on $\mathbb{C}$ \label{s3b2}}

 The Hilbert series of the $3$ $U(2)$ vortex master space can be obtained by solving the following Molien integral
\beal{es1110}
g(t,s,x;\wmaster_{3,2}) =
\oint \ud\mu_{SU(3)} 
\PE\Big[
[0,1]_w [1]_x t + [1,1]_w s
\Big]~.~
\eea
 The solution to the integral is
\beal{es1111}
g(t,s,x;\wmaster_{3,2}) =
\frac{
1+ [1]_x s^2 t^3 + [1]_x s^3 t^3 + s^5 t^6
}{
(1-s^2) (1-s^3) 
(1-x^3 s^3 t^3)
(1-x s t^3)
(1-x^{-1} s t^3)
(1-x^{-3} s^3 t^3)
}~~.
\nn\\
\eea
$\wmaster_{3,2}$ is a non-complete intersection Calabi-Yau space of dimension $7$.

As a character expansion, the Hilbert series is
\beal{es1112}
&&
g(t,s,x;\wmaster_{3,2}) = 
\frac{1}{(1-s^2)(1-s^3)} \times
\nn\\
&&
\hspace{1cm}
\sum_{n_1=0}^{\infty} \sum_{n_2=0}^{\infty} \sum_{n_3=0}^{\infty} 
\Bigg[
[n_1+n_2+3n_3]_x s^{n_1+2n_2+3n_3} t^{3n_1+3n_2+3n_3}
\nn\\
&&
\hspace{4cm}
+
[n_1+n_2]_x s^{n_1+2n_2+3n_3+3} t^{3n_1+3n_2+6n_3+6}
\Bigg]~.~
\eea

The plethystic logarithm of the Hilbert series is
\beal{es1113}
\PL[g(t,s,x;\master_{3,2})] 
&=&
s + s^2 + s^3
+ [1]_x s t^3
+ [1]_x s^2 t^3
+ [3]_x s^3 t^3 
\nn\\
&&
- [2]_x s^4 t^6
- [2]_x s^5 t^6
- [2]_x s^6 t^6
+ \dots ~~.
\eea
The generators of the master space are identified from the plethystic logarithm as follows
\beal{es1114}
s^2 &\ra& u_2=\Tr(\phi^2) 
\nn\\
s^3 &\ra& u_3=\Tr(\phi^3) 
\nn\\
~[1]_x s t^3 &\ra& 
A_{001}^{i} =
\epsilon^{\alpha_1 \alpha_2 \alpha_3} \epsilon_{j k}
Q_{\alpha_1}^{j} Q_{\alpha_2}^{k} \phi_{\alpha_3}^{\beta_1} Q_{\beta_1}^{i_3}
\nn\\
~[1]_x s^2 t^3 &\ra& 
\left\{\ba{l}
A_{002}^{i} =
\epsilon^{\alpha_1 \alpha_2 \alpha_3} \epsilon_{j k}
Q_{\alpha_1}^{j} Q_{\alpha_2}^{k} \phi_{\alpha_3}^{\beta_1} \phi_{\beta_1}^{\beta_2} Q_{\beta_2}^{i}
\nn\\
A_{011}^{i} =
\epsilon^{\alpha_1 \alpha_2 \alpha_3} \epsilon_{j k}
Q_{\alpha_1}^{i} \phi_{\alpha_2}^{\beta_1} Q_{\beta_1}^{j} \phi_{\alpha_3}^{\beta_2} Q_{\beta_2}^{k}
\nn\\
\ra A_{002}^{i} = - A_{011}^{i}
\ea\right.
\nn\\
~[3]_x s^3 t^3 &\ra& 
\left\{
\ba{l}
A_{012}^{i j k}
= \epsilon^{\alpha_1 \alpha_2 \alpha_3} Q_{\alpha_1}^{i} \phi_{\alpha_2}^{\beta_1} Q_{\beta_1}^{j} \phi_{\alpha_3}^{\beta_2} \phi_{\beta_ 2}^{\beta_3} Q_{\beta_3}^{k}
\nn\\
\ra S_{012}^{i j k} = A_{012}^{i j k} + A_{012}^{j k i} + A_{012}^{k i j}
\ea\right.
~~.
\eea
The indices of the generators $A_{012}^{i_1 i_2 i_3}$ are symmetrized to obtain $S_{012}^{i j k}$.
\\

\paragraph{Quadratic relations:} The plethystic logarithm of the Hilbert series of the vortex master space encodes the quadratic relations formed by the generators. The following terms of the plethystic logarithm correspond to the quadratic relations,
\beal{es1114s12b1}
- [2]_x s^4 t^6 - [2]_x s^5 t^6 - [2]_x s^6 t^6 ~.~
\eea
From the above discussion of generators, we recall that the generators are
\beal{es1114b1}
u_2 ~,~
u_3 ~,~
A_{002}^{i_1 i_2 i_3} ~,~
S_{012}^{i_1 i_2 i_3} ~.~
\eea
Let us go through each of the relations separately as follows:
\begin{itemize}
\item \underline{$-[2]_x s^4 t^6$ relations:} We note that 
\beal{es1114b2}
\text{Sym}^2 [1]_x &=& [2]_x ~,~
\nn\\
~[1]_x \times [3]_x &=& [4]_x + [2]_x ~.~ 
\eea
Accordingly, the relations which transform in $[2]_x s^4 t^6$ can be obtained via the following products of generators which also transform in $[2]_x s^4 t^6$,
\beal{es1114b3}
{r_{(I)}}^{ij} &=& u_2 A_{001}^{i} A_{001}^{j} ~,~
\nn\\
{r_{(II)}}^{ij} &=& A_{002}^{i} A_{002}^{j} ~,~
\nn\\
{r_{(III)}}^{ij} &=& \epsilon_{p q} A_{001}^{p} S_{012}^{q ij} ~.~
\eea
Using the above product expressions, the relation at order $[2]_x s^4 t^6$ can be identified as
\beal{es1114b4}
- [2]_x s^4 t^6 &\ra&
R^{ij} = {r_{(I)}}^{ij} - 6 {r_{(II)}}^{ij} - 4 {r_{(III)}}^{ij} = 0 ~.~
\eea
Note that the relation above is symmetric in its indices,
\beal{es1114b5} 
R^{ij}= R^{ji}
~.~
\eea

\item \underline{$-[2]_x s^5 t^6$ relations:} The products that could contribute to a relation at order $[2]_x s^5 t^6$ are as follows,
\beal{es1115b10}
{p_{(I)}}^{ij} &=& u_3 A_{001}^{i} A_{001}^{j} ~,~
\nn\\
{p_{(II)}}^{ij} &=& u_2 A_{001}^{i} A_{002}^{j} ~,~
\nn\\
{p_{(III)}}^{ij} &=& \epsilon_{p q} A_{002}^{p} S_{012}^{q ij} ~.~
\eea
Given the above products, we are able to construct the following relation corresponding to the order $[2]_x s^5 t^6$,
\beal{es1115b11}
- [2]_x s^5 t^6
&\ra&
P^{ij} = 2 {p_{(I)}}^{ij} - {p_{(II)}}^{ij} - {p_{(II)}}^{ji} - 4 {p_{(III)}}^{ij} = 0 ~.~
\eea
The above relation is symmetric in its indices,
\beal{es1115b12}
P^{ij} = P^{ji} ~.~
\eea

\item \underline{$- [2]_x s^6 t^6$ relations:} Let us first write down the products of generators that correspond to order  $[2]_x s^6 t^6$ as follows,
\beal{es1115b11}
{o_{(I)}}^{ij} &=& u_2^2 A_{001}^{i} A_{001}^{j} ~,~
\nn\\
{o_{(II)}}^{ij} &=& u_2 A_{002}^{i} A_{002}^{j} ~,~
\nn\\
{o_{(III)}}^{ij} &=& u_3 A_{001}^{i} A_{002}^{j} ~,~
\nn\\
{o_{(IV)}}^{ij} &=& \epsilon_{p q} A_{001}^{p} S_{012}^{q i j} ~,~
\nn\\
{o_{(V)}}^{ij} &=& \epsilon_{p q} \epsilon_{r s} S_{012}^{p r i} S_{012}^{q s j}
~.~
\eea
From the above, the products ${o_{(I)}}^{ij}, {o_{(II)}}^{ij}, {o_{(IV)}}^{ij}$ can be used to form the following relation at order $[2]_x s^6 t^6$,
\beal{es1115b12}
- [2]_x s^6 t^6
& \ra &
O^{ij} = {o_{(I)}}^{ij} - 6 {o_{(II)}}^{ij} - 4 {o_{(IV)}}^{ij} = 0 ~.~
\eea
The relation above is symmetric in its indices,
\beal{es1115b13}
O^{ij} = O^{ji} ~.~
\eea

\end{itemize}

\paragraph{Vortex moduli space:} The vortex moduli space can be expressed as a partial $\mathbb{C}^{*}$ projection of the vortex master space $\wmaster_{3,2}$. The projection is given as follows,
\beal{es1115}
&&
\wV_{3,2} = \wmaster_{3,2} / \{
{A_{001}}^{i} = \lambda^3 {A_{001}}^{i},
{A_{002}}^{i} = \lambda^3 {A_{002}}^{i},
{S_{012}}^{i j k} = \lambda^3 {S_{012}}^{i j k}
\}
~,~
\eea
where $\lambda$ is the $\mathbb{C}^{*}$ parameter. The vortex moduli space is as expected $5$ dimensional. The vortex master space is given by the following quotient
\beal{es1115b}
\wmaster_{3,2} = \mathbb{C}[u_2, u_3, {A_{001}}^{i}, {A_{002}}^{i},{S_{012}}^{i j k}] /
\{
R^{ij} = 0,
P^{ij} = 0,
O^{ij} = 0
\}
~.~
\eea
\\

\subsection{$3$ $U(3)$ vortices on $\mathbb{C}$ \label{s3b3}}

 For $3$ $U(3)$ vortices, the Hilbert series of the master space is given by the following Molien integral
\beal{es1120}
g(t,s,x;\wmaster_{3,3}) = \oint \ud\mu_{SU(3)}
\PE\Big[
[0,1]_w [1,0]_x t + [1,1]_w s
\Big]~~.
\eea
The Hilbert series for $\wmaster_{3,3}$ is
\beal{es1121}
g(t,s;\wmaster_{3,3}) 
&=& 
\frac{
1
 }{
 (1 - s^2) (1 - s^3) (1 - t^3) (1 - s t^3)^4 (1 - s^3 t^3)^3
 } 
 \times
 \nn\\
 &&
  (1 + 4 s t^3 + 8 s^2 t^3 
 + 7 s^3 t^3 + s^2 t^6 + 5 s^3 t^6 + 
 10 s^4 t^6 + 11 s^5 t^6 + s^6 t^6 
 \nn\\
 &&
 - s^4 t^9 - 11 s^5 t^9 
 - 10 s^6 t^9 - 5 s^7 t^9 - s^8 t^9 - 7 s^7 t^{12} - 8 s^8 t^{12} - 
 4 s^9 t^{12}
 \nn\\
 &&
  - s^{10} t^{15})
~~,
\eea
where for simplicity we have set the global $SU(3)$ fugacities to $x_1=x_2=1$. $\wmaster_{3,3}$ is a non-complete intersection Calabi-Yau space of dimension $9$. The character expansion of the Hilbert series is 
\beal{es1122}
&&
g(t,s,x;\wmaster_{3,3}) 
=
\frac{1}{(1-s^2)(1-s^3)} 
\times
\nn\\
&&
\hspace{1cm}
\sum_{n_0=0}^{\infty}
\sum_{n_1=0}^{\infty}
\sum_{n_2=0}^{\infty}
\sum_{n_3=0}^{\infty}
\Bigg[
[n_1+n_2+3n_3,n_1+n_2]_x 
s^{n_1 + 2n_2 + 3n_3}
t^{3n_0 + 3n_1+3n_2+3n_3}
\nn\\
&&
\hspace{1.5cm}
+
[n_1+n_2,n_1+n_2+3n_3+3]_x 
s^{n_1+2n_2+3n_3+3}
t^{3n_0+3n_1+3n_2+6n_3+6}
\Bigg]
~~.
\nn\\
&&
\eea

The plethystic logarithm is
\beal{es1123}
\PL\Big[
g(t,s,x;\wmaster_{3,3})
\Big]
&=&
s^2+s^3
+ t^3
+ [1,1]_x s t^3
+ [1,1]_x s^2 t^3
+ [3,0]_x s^3 t^3
\nn\\
&&
- (1+ [1,1]_x ) s^2 t^6
- (1+ 2 [1,1]_x + [3,0]_x ) s^3 t^6
\nn\\
&&
- (1+ 2 [1,1]_x + [2,2]_x + [3,0]_x) s^4 t^6
+ \dots
~~.
\eea
The generators of $\wmaster_{3,3}$ are as follows
\beal{es1124}
s^2 &\ra& u_2 = \Tr(\phi^2)
\nn\\
s^3 &\ra& u_3 = \Tr(\phi^3)
\nn\\
t^3 &\ra& 
\left\{\ba{l}
B^{i j k} = \epsilon^{\alpha_1 \alpha_2 \alpha_3} Q_{\alpha_1}^{i} Q_{\alpha_2}^{j} Q_{\alpha_3}^{k}
\nn\\
\ra B = B^{123} + B^{231} + B^{312} 
\ea\right.
\nn\\
~[1,1]_x s t^3 &\ra& 
\left\{
\ba{l}
{A_{001}}_{i}^{~j} =
\epsilon^{\alpha_1 \alpha_2 \alpha_3}
\epsilon_{i k_1 k_2}
Q_{\alpha_1}^{k_1} Q_{\alpha_2}^{k_2} \phi_{\alpha_3}^{\beta} Q_{\beta}^{j}
\nn\\
{A_{001}}_{i}^{~i} = 0
\ea
\right.
\nn\\
~[1,1]_x s^2 t^3 &\ra& 
\left\{\ba{l}
{A_{002}}_{i}^{~j} =
\epsilon^{\alpha_1 \alpha_2 \alpha_3}
\epsilon_{i k_1 k_2}
Q_{\alpha_1}^{k_1} Q_{\alpha_2}^{k_2} \phi_{\alpha_3}^{\beta_1} \phi_{\beta_1}^{\beta_2} Q_{\beta_2}^{j}
\nn\\
{A_{011}}_{i}^{~j} =
\epsilon^{\alpha_1 \alpha_2 \alpha_3} \epsilon_{i k_1 k_2}
Q_{\alpha_1}^{j} \phi_{\alpha_2}^{\beta_1} Q_{\beta_1}^{k_1} \phi_{\alpha_3}^{\beta_2} Q_{\beta_2}^{k_2}
\nn\\
\ra
{A_{002}}_{i}^{~j} = 
- {A_{011}}_{i}^{~j}
\ea
\right.
\nn\\
~[3,0]_x s^3 t^3 &\ra& 
\left\{\ba{l}
{A_{012}}^{i j k}
= \epsilon^{\alpha_1 \alpha_2 \alpha_3} Q_{\alpha_1}^{i} \phi_{\alpha_2}^{\beta_1} Q_{\beta_1}^{j} \phi_{\alpha_3}^{\beta_2} \phi^{\beta_3}_{\beta_2} Q_{\beta_3}^{k}
\nn\\
{S_{012}}^{ijk}
= {A_{012}}^{i j k} + {A_{012}}^{jki} + {A_{012}}^{kij}
\ea
\right.
~~.
\eea
The generators for $[3,0]_x s^3 t^4$ come only from the partition $012$, giving $A_{012}^{i_1 i_2 i_3}$. This in turn is symmetrized to give $S_{012}^{i_1 i_2 i_3}$.
\\

\paragraph{Quadratic relations:} The terms in the plethystic logarithm corresponding to quadratic relations are
\beal{es1114f1}
&
-s^2 t^6 - [1,1]_x s^2 t^6
&
\nn\\
&
- s^3 t^6
- 2 [1,1]_x s^3 t^6
- [3,0]_x s^3 t^6
&
\nn\\
&
- s^4 t^6
- 2 [1,1]_x s^4 t^6
- [2,2]_x s^4 t^6
- [3,0]_x s^4 t^6
&
\nn\\
&
- [1,1]_x s^5 t^6
- [2,2]_x s^5 t^6
- [3,0]_x s^5 t^6
&
\nn\\
&
- [2,2]_x s^6 t^6
&
~.~
\eea
We recall, the generators are
\beal{es1114x1}
u_2 ~,~ 
u_3 ~,~
B ~,~
{A_{001}}_{i}^{~j} ~,~
{A_{002}}_{i}^{~j} ~,~
{S_{012}}^{i j k} ~.~
\eea
Let us go through each of the quadratic relations that are given by the terms in \eref{es1114f1} in the plethystic logarithm.

The first set of quadratic relations at orders of $s^2 t^6$ are as follows:
\begin{itemize}

\item \underline{$-s^2 t^6$ relation:} We see that this quadratic relation transforms as a singlet in $SU(3)_x$. Considering the following symmetric product which contains the singlet $[0,0]_x$,
\beal{es1114x5}
\text{Sym}^2 ~[1,1]_x = [2,2]_x + [1,1]_x + [0,0]_x ~,~
\eea
we identify the following generator products in order to construct the desired quadratic relation,
\beal{es1114x6}
{r_{(I)}}_{i}^{~j} &=&
B {A_{002}}_{i}^{~j}
\nn\\
{r_{(II)}}_{i}^{~j} &=&
{A_{001}}_{i}^{~k} {A_{001}}_{k}^{~j}
\nn\\
r_{(III)} &=& u_3 B B ~,~
\eea
which all transform in $-s^2 t^6$.
Using the above products correctly, we construct the relation
\beal{es1114x6}
- s^2 t^6 \ra
R_{(I)} = 4 r_{(III)} - 9   ({r_{(II)}}_{1}^{~1}+{r_{(II)}}_{2}^{~2}+{r_{(II)}}_{3}^{~3})= 0~.~  
\eea
\\

\item \underline{$- [1,1]_x s^2 t^6$ relations:}  Given that at this order the relations need to transform in $[1,1]_x s^2 t^6$, the following relations can be identified using the products in \eref{es1114x6}
\beal{es1114x7}
-[1,1]_x s^2 t^6 \ra
{R_{(II)}}_{i}^{~j} = 3 {r_{(II)}}_{i}^{~j} - 2 {r_{(I)}}_{i}^{~j} = 0 ~.~
\eea
\\

\end{itemize}

The next set of quadratic relations are at orders of $s^3 t^6$. In order to construct the quadratic relations, we consider the following representation products,
\beal{es1114x7b}
\text{Sym}^{2} [1,1]_x &=&
[2,2]_x + [1,1]_x + [0,0]_x ~,~
\nn\\
~[1,1]_x \times [1,1]_x &=&
[2,2]_x + [3,0]_x + [0,3]_x + 2 [1,1]_x + [0,0]_x 
~.~
\eea
The quadratic relations are as follows:
\begin{itemize}
\item\underline{$- s^3 t^6$ relation:} For this order, we first consider the following generator products,
\beal{es1114x8}
{p_{(I)}}_{i}^{~j} &=& {A_{001}}_{i}^{~k}  ~ {A_{002}}_{k}^{~j}
~,~
\nn\\
p_{(II)} &=& u_3 B B
~.~
\eea
The quadratic relation for this order is
\beal{es1114x9}
P_{(I)} &=&
4 p_{(II)} - 9 ({p_{(I)}}_{1}^{~1}+{p_{(I)}}_{2}^{~2}+{p_{(I)}}_{3}^{~3}) = 0
~.~
\eea
\\

\item\underline{$-[1,1]_x s^3 t^6$ relations:} For the quadratic relations at this order, we have to consider the following products of generators,
\beal{es1114xa10}
{p_{(III)}}^{i}_{~j} &=&
{A_{001}}_{p}^{~s} {A_{002}}_{q}^{~i} \epsilon^{pqr} \epsilon_{sri}
~,~
\nn\\
{p_{(IV)}}^{i}_{~j} &=&
{A_{002}}_{p}^{~s} {A_{001}}_{q}^{~i} \epsilon^{pqr} \epsilon_{sri}
~.~
\eea
The above generator products transform in the correct representation of $SU(4)$ for this order, and satisfy the following quadratic relation,
\beal{es1114xa11}
{P_{(II)}}^{i}_{~j} = 
{p_{(III)}}^{i}_{~j} + {p_{(IV)}}^{i}_{~j}
= 0 ~.~
\eea
The above is the correct relation for this order of the plethystic logarithm.
\\

\end{itemize}

The next set of quadratic relations are of the order $s^4 t^6$. In order to construct the quadratic relations at this order, we consider first the following $SU(3)$ representation products,
\beal{es1114xa12}
~[3,0]_x \times [1,1]_x &=&
[4,1]_x + [2,2]_x + [3,0]_x + [1,1]_x
~,~
\nn\\
\text{Sym}^{2} [1,1]_x &=&
[2,2]_x + [1,1]_x + [0,0]_x ~,~
\nn\\
~[1,1]_x \times [1,1]_x &=&
[2,2]_x + [3,0]_x + [0,3]_x + 2 [1,1]_x + [0,0]_x 
~.~
\eea
We make use of the above information to construct the quadratic relations at this order as follows:
\begin{itemize}
\item\underline{$-s^4 t^6$ relations:} For the quadratic relations at this order, we need to consider the following generator products,
\beal{es1114xa14}
u_{(I)} &=& {A_{002}}_{i}^{~j} {A_{002}}_{j}^{~i} ~,~
\nn\\
u_{(II)} &=&
(u_2 B)^2 ~,~
\eea
which transform as a singlet of $SU(3)$. The above generator products form the following single unique quadratic relation,
\beal{es1114xa15}
U_{(I)} = u_{(I)} - \frac{2}{9} u_{(II)} = 0
~,~
\eea
which is the relation we are looking for at this order.
\\

\item\underline{$-2[1,1]_x s^4 t^6$ relations:} The quadratic relations at this order can be constructed from the following generator products,
\beal{es1114xa16}
{u_{(III)}}_{i}^{~j} &=&
{A_{002}}_{i}^{~k} {A_{002}}_{k}^{~j} 
~,~
\nn\\
{u_{(IV)}}_{i}^{~j} &=&
{A_{002}}_{i}^{~j} {A_{002}}_{k}^{~k}
~,~
\nn\\
{u_{(V)}}_{i}^{~j} &=&
u_3 {A_{001}}_{i}^{~j} B
~,~
\nn\\
{u_{(VI)}}_{i}^{~j} &=&
u_2 {A_{002}}_{i}^{~j} B
~.~
\eea
All the above generator products transform in the adjoint of $SU(3)$. They satisfy the following quadratic relations,
\beal{es1114xa17}
{U_{(II)}}_{i}^{~j} &=& {u_{(IV)}}_{i}^{~j} - \frac{2}{3} {u_{(VI)}}_{i}^{~j} = 0
~,~
\nn\\
{U_{(III)}}_{i}^{~j} &=& {u_{(III)}}_{i}^{~j}  -  \frac{1}{3} {u_{(VI)}}_{i}^{~j} - \frac{2}{9} {u_{(V)}}_{i}^{~j} 
= 0 ~.~
\eea
The above are precisely the two quadratic relations expected for this order.
\\

\item\underline{$-[2,2]_x s^4 t^6$ relations:} For this order, the quadratic relation can be constructed by considering the following generator product and its index symmetrisations and anti-symmetrizations. The generator product to be considered is as follows,
\beal{es1114xa18}
{u_{(VII)}}^{ijklmn} = {A_{001}}_{p}^{~k} {S_{012}}^{lmn} ~,~
\eea
which we anti-symmetrize in its indices $[kl]$ and $[mn]$,
\beal{es1114xa19}
{u_{(VIII)}}^{ijklmn} = {u_{(VII)}}^{ijklmn} - {u_{(VII)}}^{ijlkmn} - {u_{(VII)}}^{ijklnm} + {u_{(VII)}}^{ijlknm} ~.~
\eea
We further symmetries the above in the pair of indices $[kl]$ and $[mn]$ as follows,
\beal{es1114xa20}
{U_{(IV)}}^{ijklmn} = {u_{(VIII)}}^{ijklmn} + {u_{(VIII)}}^{ijmnkl} = 0
~,~
\eea
which vanishes exactly. This is precisely the quadratic relations at this order.
\\

\item\underline{$-[3,0]_x s^4 t^6$ relations:} The quadratic relation at this order can be obtained by considering the following product of generators and its symmetrisation of indices. The generator product to consider is,
\beal{es1114xa21}
{u_{(IX)}}^{ijk} = {A_{001}}_{s}^{~q} {S_{012}}^{rjk} \epsilon^{sip} \epsilon_{qrp} ~,~
\eea
where we symmetrize on the indices $ijk$ to obtain,
\beal{es1114xa22}
{U_{(V)}}^{ijk} = {u_{(IX)}}^{ijk} + {u_{(IX)}}^{jki} + {u_{(IX)}}^{kij} = 0 ~.~
\eea
The above vanishes exactly and corresponds precisely to the quadratic relation at this order.
\\
\end{itemize}

The following set of quadratic relations involves the order $s^5 t^6$. In order to construct the relations at this order, we consider the following $SU(3)$ representation products,
\beal{es1114xa50}
~[3,0]_x \times [1,1]_x &=&
[4,1]_x + [2,2]_x + [3,0]_x + [1,1]_x
~,~
\nn\\
\text{Sym}^{2} [1,1]_x &=&
[2,2]_x + [1,1]_x + [0,0]_x ~,~
\nn\\
~[1,1]_x \times [1,1]_x &=&
[2,2]_x + [3,0]_x + [0,3]_x + 2 [1,1]_x + [0,0]_x 
~.~
\eea
The above representation products help us in finding the following quadratic relations between moduli space generators:
\begin{itemize}
\item\underline{$- [1,1]_x s^5 t^6$ relations:} For the quadratic relation at this order, we consider first the following generator products,
\beal{es1114xa51}
{v_{(I)}}_{i}^{~j} &=&
{A_{002}}_{i}^{~p} {S_{012}}^{qrj} \epsilon_{pqr}
~,~
\nn\\
{v_{(II)}}_{i}^{~j} &=&
u_3 {A_{001}}_{i}^{~p} {A_{001}}_{p}^{~j}
~.~
\eea
The above products transform as desired in the adjoint of $SU(3)$. Furthermore, the products satisfy the following quadratic relation,
\beal{es1114xa52}
{V_{(I)}}_{i}^{~j} = {v_{(I)}}_{i}^{~j} + \frac{1}{2} {v_{(II)}}_{i}^{~j} = 0 ~,~
\eea
which is precisely the relation we are looking for at this order.
\\

\item\underline{$- [2,2]_x s^5 t^6$ relations:} For the quadratic relation at this order, we have to consider the following generator products with their index symmetrization and anti-symmetrizations. The first generator product to consider is the following,
\beal{es1114xa53}
{v_{(III)}}^{ijklmn} = {A_{011}}_{p}^{~k} {S_{012}}^{lmn} \epsilon_{pij} ~,~
\eea
which we anti-symmetrize in the indices $[kl]$ and $[mn]$ to get
\beal{es1114xa54}
{v_{(IV)}}^{ijklmn} = {v_{(III)}}^{ijklmn} - {v_{(III)}}^{ijlkmn} - {v_{(III)}}^{ijklnm} + {v_{(III)}}^{ijlknm}  ~.~
\eea
We further anti-symmetrize the above in the pairs of indices $[ij]$ and $[mn]$ as follows 
\beal{es1114xa55}
{v_{(V)}}^{ijklmn} = {v_{(IV)}}^{ijklmn} - {v_{(IV)}}^{mnklij} ~.~
\eea
The second generator product to consider is as follows
\beal{es1114xa56}
{v_{(VI)}}^{ijklmn} = u_3 {A_{001}}_{p}^{~k} {A_{001}}_{r}^{~n} \epsilon_{pij} \epsilon_{rlm} ~,~
\eea
which we anti-symmetrize in the indices $[kl]$ and $[mn]$ to get
\beal{es1114xa57}
{v_{(VII)}}^{ijklmn} = {v_{(VI)}}^{ijklmn} - {v_{(VI)}}^{ijlkmn} - {v_{(VI)}}^{ijklnm} + {v_{(VI)}}^{ijlknm} ~.~
\eea
We symmetrize the above product in the pairs of indices $[ij]$ and $[kl]$ to obtain
\beal{es1114xa58}
{v_{(VIII)}}^{ijklmn} = {v_{(VII)}}^{ijklmn} + {v_{(VII)}}^{klijmn} ~.~
\eea
The above two generator products form the following quadratic relation,
\beal{es1114xa59}
{V_{(II)}}^{ijklmn} = {v_{(V)}}^{ijklmn} + \frac{1}{6} {v_{(VIII)}}^{ijmnkl} = 0 ~,~
\eea
which is precisely the relation we are looking for at this order.
\\

\item\underline{$- [3,0]_x s^5 t^6$ relations:} The quadratic relation at this order can be found by considering the following generator products. The products to consider are 
\beal{es1114xa60}
{v_{(IX)}}^{ijk} &=& {A_{002}}_{p}^{~p} {S_{012}}^{ijk} ~,~
\nn\\
{v_{(X)}}^{ijk} &=& u_2 {S_{012}}^{ijk} B ~.~
\eea
The above two products satisfy the following quadratic relation,
\beal{es1114xa63}
{V_{(III)}}^{ijk} = {v_{(IX)}}^{ijk} - \frac{2}{3} {v_{(X)}}^{ijk} = 0 ~,~
\eea
which is precisely the relation we are looking for at this order.
\\
\end{itemize}

The final quadratic relation is at order $s^6 t^6$. We consider the following $SU(3)$ representation products for the construction of this relation,
\beal{es1114xa100}
\text{Sym}^{2} [3,0]_x &=& [6,0]_x + [2,2]_x 
~,~
\nn\\
~[3,0]_x \times [1,1]_x &=&
[4,1]_x + [2,2]_x + [3,0]_x + [1,1]_x 
~,~
\nn\\
\text{Sym}^{2} [1,1]_x &=&
[2,2]_x + [1,1]_x + [0,0]_x ~,~
\nn\\
~[1,1]_x \times [1,1]_x &=&
[2,2]_x + [3,0]_x + [0,3]_x + 2 [1,1]_x + [0,0]_x 
~.~
\eea
The above products help us in finding the correct generator product which leads to the following final relation between generators:
\begin{itemize}
\item\underline{$-[2,2]_x s^6 t^6$ relations:} The final quadratic relations can be obtained by considering the following generator product,
\beal{es1114xa101}
{z_{(I)}}^{ijklmn} = {S_{012}}^{ijk} {S_{012}}^{lmn} ~,~
\eea
which we anti-symmetrize in the indices $[kl]$ and $[mn]$ as follows
\beal{es1114xa102}
{z_{(II)}}^{ijklmn} = {z_{(I)}}^{ijklmn} - {z_{(I)}}^{ijlkmn} - {z_{(I)}}^{ijklnm} + {z_{(I)}}^{ijlknm} ~.~
\eea
Another symmetrization of the pairs of indices $[kl]$ and $[mn]$ gives
\beal{es1114xa103}
Z^{ijklmn} = {z_{(II)}}^{ijklmn} + {z_{(II)}}^{ijmnkl} = 0 ~,~
\eea
which vanishes exactly and represents precisely the quadratic relation at this order.
\\
\end{itemize}

\paragraph{Vortex moduli space:} The vortex moduli space can be expressed as a $\mathbb{C}^*$ projection of the master space. The $\mathbb{C}^*$ projection gives as the vortex moduli space the following,
\beal{es1114g1}
\wV_{3,3} = \wmaster_{3,3}/\{
B_{i} \simeq \lambda^3 B ~,~
{A_{001}}_{i}^{~j} \simeq \lambda^3 {A_{001}}_{i}^{~j}~,~
{A_{002}}_{i}^{~j} \simeq \lambda^3 {A_{002}}_{i}^{~j}~,~
{S_{012}}^{ijk} \simeq \lambda^3 {S_{012}}^{ijk}
\}
~.~
\nn\\
\eea
The master space is expressed as follows,
\beal{es1114g2}
&&
\wmaster_{3,3} = \mathbb{C}[u_2, u_3,B, {A_{001}}_{i}^{~j},{A_{002}}_{i}^{~j},{S_{012}}^{ijk}] /\{
\nn\\
&&
\hspace{2cm}
R_{(I)}=0~,~
{R_{(II)}}_{i}^{~j}=0~,~
\nn\\
&&
\hspace{2cm}
P_{(I)}=0~,~
{P_{(II)}}^{i}_{~j}=0~,~
\nn\\
&&
\hspace{2cm}
U_{(I)} = 0 ~,~
{U_{(II)}}_{i}^{~j} = 0 ~,~
{U_{(III)}}_{i}^{~j} = 0 ~,~
{U_{(IV)}}^{ijklmn} = 0 ~,~
{U_{(V)}}^{ijk} = 0 ~,~
\nn\\
&&
\hspace{2cm}
{V_{(I)}}_{i}^{~j} = 0 ~,~
{V_{(II)}}^{ijklmn} = 0 ~,~
{V_{(I)}}^{ijk} = 0 ~,~
\nn\\
&&
\hspace{2cm}
Z^{ijklmn} = 0 
~~~
\}
~,~
\eea
where the quotient is taken by all quadratic relations of the master space generators.
\\

\subsection{$3$ $U(4)$ vortices on $\mathbb{C}$ \label{s3b4}}

The Hilbert series for the $3$ $U(4)$ vortex master space is obtained from the following Molien integral
\beal{es1130}
g(t,s,x;\wmaster_{3,4}) = \oint \ud\mu_{SU(3)} 
\PE\Big[
[0,1]_w [1,0,0]_{x} t + [1,1]_w s
\Big]
~~,
\eea
where $[1,0,0]_x$ is the character of the fundamental representation of the global $SU(4)$. The integral gives the following Hilbert series 
\beal{es1131}
g(t,s;\wmaster_{3,4}) 
&=&
\frac{1}{
(1 - s^2) (1 - s^3) (1 - t^3)^4 (1 - s t^3) (1 - s^3 t^3)^4
}
\times
\nn\\
&&
(1 + 14 s t^3 + 20 s^2 t^3 + 16 s^3 t^3 - 16 s t^6 + 5 s^2 t^6 + 46 s^3 t^6 + 90 s^4 t^6 
\nn\\
&&
+ 60 s^5 t^6 + 10 s^6 t^6 + 4 s t^9 - 44 s^2 t^9 - 104 s^3 t^9 - 156 s^4 t^9 - 156 s^5 t^9 
\nn\\
&&
- 60 s^6 t^9 - 24 s^7 t^9 + 21 s^2 t^{12} + 25 s^3 t^{12} + 23 s^4 t^{12} + 65 s^5 t^{12} - 73 s^6 t^{12}
\nn\\
&& 
- 195 s^7 t^{12} - 207 s^8 t^{12} - 81 s^9 t^{12} + s^{10} t^{12} + s^{11} t^{12} + 14 s^3 t^{15} 
\nn\\
&&
+ 34 s^4 t^{15} + 22 s^5 t^{15} + 206 s^6 t^{15} + 438 s^7 t^{15} + 438 s^8 t^{15} + 206 s^9 t^{15} 
\nn\\
&&
+ 22 s^{10} t^{15} + 34 s^{11} t^{15} + 14 s^{12} t^{15} + s^4 t^{18} + s^5 t^{18} - 81 s^6 t^{18}  - 207 s^7 t^{18}
\nn\\
&& 
- 195 s^8 t^{18} - 73 s^9 t^{18} + 65 s^{10} t^{18} + 23 s^{11} t^{18} + 25 s^{12} t^{18} + 21 s^{13} t^{18}
\nn\\
&&
 - 24 s^8 t^{21} - 60 s^9 t^{21} - 156 s^{10} t^{21} - 156 s^{11} t^{21} - 104 s^{12} t^{21} - 44 s^{13} t^{21}
 \nn\\
 &&
 +  4 s^{14} t^{21} + 10 s^9 t^{24} + 60 s^{10} t^{24} + 90 s^{11} t^{24} + 46 s^{12} t^{24} + 5 s^{13} t^{24}
 \nn\\
 && 
 - 16 s^{14} t^{24} + 16 s^{12} t^{27} + 20 s^{13} t^{27} + 14 s^{14} t^{27} + s^{15} t^{30})
~~,
\eea
where for simplicity we set the global $SU(4)$ fugacities to $x_1=x_2=x_3=1$. Accordingly, the master space of the vortex theory is a non-complete intersection Calabi-Yau space of dimension $13$. 

As a character expansion, the Hilbert series is
\beal{es1132}
&&
g(t,s,x;\wmaster_{3,4}) 
=
\frac{1}{(1-s^2)(1-s^3)} 
\times
\nn\\
&&
\hspace{1cm}
\sum_{n_0=0}^{\infty}
\sum_{n_1=0}^{\infty}
\sum_{n_2=0}^{\infty}
\sum_{n_3=0}^{\infty}
\Bigg[
[n_1+n_2+3n_3,n_1+n_2,n_0]_x 
s^{n_1 + 2n_2 + 3n_3}
t^{3n_0 + 3n_1+3n_2+3n_3}
\nn\\
&&
\hspace{1.5cm}
+
[n_1+n_2,n_1+n_2+3n_3+3,n_0]_x 
s^{n_1+2n_2+3n_3+3}
t^{3n_0+3n_1+3n_2+6n_3+6}
\Bigg]
~~.
\nn\\
&&
\eea
The corresponding plethystic logarithm is
\beal{es1133}
\PL\Big[
g(t,s,x;\wmaster_{3,4}) 
\Big]
&=&
s^2+s^3
+ [0,0,1]_x t^3
+ [1,1,0]_x s t^3
+ [1,1,0]_x s^2 t^3
+ [3,0,0]_x s^3 t^3
\nn\\
&&
- ([0,1,0]_x + [2,0,0]_x) s t^6
- ([0,1,0]_x + 2 [2,0,0]_x + [0,0,2]_x 
\nn\\
&&
+ [1,1,1]_x) s^2 t^6
- ([0,1,0]_x +[0,0,2]_x + 2[2,0,0]_x + 2[1,1,1]_x 
\nn\\
&&
+[3,0,1]_x) s^3 t^6 
- ([2,0,0]_x + [0,0,2]_x + [1,1,1]_x + [2,2,0]_x 
\nn\\
&&
+ [3,0,1]_x) s^4 t^6
+ \dots~~.
\eea
The generators of the vortex master space are encoded in the Hilbert series above. They are as follows,
\beal{es1134}
s^2 &\ra& u_2=\Tr(\phi^2)
\nn\\
s^3 &\ra& u_3=\Tr(\phi^3)
\nn\\
~[0,0,1]_x t^3 &\ra&
{B}_{i}= \epsilon^{\alpha_1 \alpha_2 \alpha_3} \epsilon_{k_1 k_2 k_3 i} Q_{\alpha_1}^{k_1} Q_{\alpha_2}^{k_2} Q_{\alpha_3}^{k_3}
\nn\\
~[1,1,0]_x s t^3 &\ra&
\left\{
\ba{l}
{A_{001}}_{i j}^{~~k} =
\epsilon^{\alpha_1 \alpha_2 \alpha_3}
\epsilon_{k_1 k_2 i j}
Q_{\alpha_1}^{k_1} Q_{\alpha_2}^{k_2} \phi_{\alpha_3}^{\beta} Q_{\beta}^{k}
\nn\\
{A_{001}}_{i k}^{~~k} = 0
\ea
\right.
\nn\\
~[1,1,0]_x s^2 t^3 &\ra&
\left\{
\ba{l}
{A_{002}}_{ij}^{~~k} =
\epsilon^{\alpha_1 \alpha_2 \alpha_3}
\epsilon_{k_1 k_2 i j}
Q_{\alpha_1}^{k_1} Q_{\alpha_2}^{k_2} \phi_{\alpha_3}^{\beta_1}\phi_{\beta_1}^{\beta_2} Q_{\beta_2}^{k}
\nn\\
3 {A_{002~}}_{ik}^{~~k} = - u_2 {B}_{i}
\nn\\
{A_{011}}_{ij}^{~~k}=
\epsilon^{\alpha_1 \alpha_2 \alpha_3}
\epsilon_{k_1 k_2 i j}
Q_{\alpha_1}^{k} \phi_{\alpha_2}^{\beta_1} Q_{\beta_1}^{k_1} \phi_{\alpha_3}^{\beta_2} Q_{\beta_2}^{k_2}
\nn\\
6 {A_{011}}_{ik}^{~~k}= u_2 {B}_{i}
\nn\\
\ra {A_{002}}_{ijk} = {A_{011~}}_{ijk} 
\nn\\
~~~~ {A_{002}}_{ikk} = {A_{011~}}_{ikk}  - \frac{1}{6} u_2 {B}_{i}
\ea\right.
\nn\\
~[3,0,0]_x s^3 t^3 &\ra&
\left\{\ba{l}
{A_{012}}^{ijk}
= 
\epsilon^{\alpha_1 \alpha_2 \alpha_3} 
Q_{\alpha_1}^{i}
\phi_{\alpha_2}^{\beta_1} Q_{\beta_1}^{j}
\phi_{\alpha_3}^{\beta_2} \phi_{\beta_2}^{\beta_3} Q_{\beta_3}^{k}
\nn\\
{S_{012}}^{ijk} =
{A_{012}}^{ijk}
+ {A_{012}}^{jki}
+ {A_{012}}^{kij}
\ea\right.
~~.
\eea
\\

\paragraph{Quadratic relations:} The terms in the plethystic logarithm corresponding to the quadratic relations between generators are as follows,
\beal{es1134y1}
&
- [0,1,0]_x s t^6 - [2,0,0]_x s t^6
&
\nn\\
&
- [0,1,0]_x s^2 t^6 
- 2 [2,0,0]_x s^2 t^6
- [0,0,2]_x s^2 t^6
- [1,1,1]_x s^2 t^6
&
\nn\\
&
- [0,1,0]_x s^3 t^6
- [0,0,2]_x s^3 t^6
- 2 [2,0,0]_x s^3 t^6
- 2 [1,1,1]_x s^3 t^6
- [3,0,1]_x s^3 t^6
&
\nn\\
&
- [2,0,0]_x s^4 t^6
- [0,0,2]_x s^4 t^6
- [1,1,1]_x s^4 t^6
- [2,2,0]_x s^4 t^6
- [3,0,1]_x s^4 t^6
&
\nn\\
&
-[1,1,1]_x s^5 t^6
-[2,2,0]_x s^5 t^6
-[3,0,1]_x s^5 t^6
&
\nn\\
&
-[2,2,0]_x s^6 t^6
&
~~.
\nn\\
\eea
From the discussion above regarding generators of the master space, we note that the quadratic relations can only be formed by the following generators
\beal{es1134rel0}
u_2 ~,~
u_3 ~,~
{B}_{i} ~,~
{A_{001}}_{ij}^{~~k} ~,~
{A_{002}}_{ij}^{~~k} ~,~
{S_{012}}^{ijk} ~.~
\eea
Let us go through the quadratic relations corresponding to terms in \eref{es1134y1} one by one.
\\

The first relations to consider are at orders at $st^6$. For these relations, we consider the following product of $SU(4)$ representations,
\beal{es1134aa1}
[1,1,0]_x \times [0,0,1]_x = 
[1,1,1]_x + [2,0,0]_x + [0,1,0]_x ~.~
\eea
Using the above products of representations, we can easily construct the quadratic relations at this order. These are as follows:
\begin{itemize}
\item \underline{$-[0,1,0]_x s t^6$ relations:} The only non-trivial product of generators of the master space corresponding to the order $- [0,1,0]_x s t^6$ is as follows,
\beal{es1134rel1}
{R_{(I)}}_{ij}={A_{001}}_{ij}^{~~k}~ {B}_{k} = 0 ~,~
\eea
where it turns out that the products vanishes. Given that
\beal{es1134rel1b}
{R_{(I)}}_{ij} = - {R_{(I)}}_{ji} ~,~
\eea
this is the quadratic relation for $-[0,1,0]_x s t^6$.
\\

\item \underline{$-[2,0,0]_x s t^6$ relations:} Another generator product which vanishes is
\beal{es1134rel2}
{A_{001}}_{i k}^{~~k} ~ {B}_{j} =0 ~.~
\eea
This product can be symmetrised as follows
\beal{es1134rel2b}
{R_{(II)}}_{ij} = {A_{001}}_{i k}^{~~k} ~ {B}_{j} + {A_{001}}_{j k}^{~~k} ~ {B}_{i}  = 0 ~,~
\eea
such that 
\beal{es1134rel2c}
{R_{(II)}}_{ij} = {R_{(II)}}_{ji} ~,~
\eea
and it corresponds to quadratic relations of the order $-[2,0,0]_x s t^6$.
\end{itemize}

The second set of quadratic relations are at orders of $s^2 t^6$. For these quadratic relations, we consider the following products of $SU(4)$ representations,
\beal{es1134bb1}
\text{Sym}^{2} [1,1,0]_x 
&=&
[2,2,0]_x + [1,1,1]_x + [2,0,0]_x + [0,0,2]_x ~,~
\nn\\
\text{Sym}^{2} [0,0,1]_x &=&
[0,0,2]_x ~,~
\nn\\
~[1,1,0]_x \times [0,0,1]_x 
&=& 
[1,1,1]_x + [2,0,0]_x + [0,1,0]_x ~.~
\eea
The above product decompositions of $SU(4)$ representations help us to construct the quadratic relations at this order as follows:
\begin{itemize}
\item \underline{$- [0,1,0]_x s^2 t^6$ relations:} The only product of master space generators that can correspond to order $-[0,1,0]_x s^2 t^6$ is the following,
\beal{es1134rel3}
{O_{(I)}}_{ij} = {A_{002}}_{ij}^{~~k} ~ {B}_{k} = 0 ~,~
\eea
which vanishes exactly. The above satisfies
\beal{es1134rel3b}
{O_{(I)}}_{ij} = -{O_{(I)}}_{ji} ~,~
\eea
such that this is the quadratic relation at order $-[0,1,0]_x s^2 t^6$.
\\

\item \underline{$- 2 [2,0,0]_x s^2 t^6$ relations:} The following generator products relate to the o                                                                                                                                                                                                                                                                                                                                                                                                                                                                                                                                                                                                                                                                                                                                                                                                                                                                                                                                                                                                                                                                                                                                                                                                                                                                                                                                                                                                                                                                                                                                                                                                                                                                                                                                                                                                                                                                                                                                                                                                                                                                                                                                                                                                                                                                                                                                                                                                                                                                                                                                                                                                                                                                                                                                                                                                                                                                                                                                                                rder $- [2,0,0]_x s^2 t^6$,
\beal{es1134rel5}
{o_{(I)}}_{ij} &=& {A_{002}}_{ik}^{~~k} {B}_{j} + {A_{002}}_{jk}^{~~k} {B}_{i} ~,~
\nn\\
{o_{(II)}}_{ij} &=& u_2 {B}_{i} {B}_{j} ~,~
\nn\\
{o_{(III)}}_{ij} &=& {A_{001}}_{ik}^{~~m} {A_{001}}_{jm}^{~~k} ~,~
\eea
where for ${o_{(I)}}_{ij}$ we have symmetrized the product
\beal{es1134rel5b}
{A_{002}}_{ik}^{~~k} {B}_{j} ~.~
\eea
We have
\beal{es1134rel5c}
{o_{(I)}}_{ij} = {o_{(I)}}_{ji}
~,~
{o_{(II)}}_{ij} = {o_{(II)}}_{ji}
~,~
{o_{(III)}}_{ij} = {o_{(III)}}_{ji} ~.~
\eea
From the above products, we can identify the following independent quadratic relations,
\beal{es1134rel6}
{O_{(II)}}_{ij} &=& {o_{(I)}}_{ij} + 6 {o_{(III)}}_{ij} = 0 
\nn\\
{O_{(III)}}_{ij} &=& {o_{(II)}}_{ij} - 9 {o_{(III)}}_{ij} = 0 
~,~
\eea
which correspond to the order $- 2 [2,0,0]_x s^2 t^6$ of the plethystic logarithm.
\\

\item \underline{$- [0,0,2]_x s^2 t^6$ relations:}  The following generator products correspond to the order $- [0,0,2]_ s^2 t^6$
\beal{es1134rel10}
{o_{(IV)}}_{p q ij lm} &=&
\epsilon_{pqrs}{A_{001}}_{ij}^{~~ r} {A_{001}}_{lm}^{~~s} 
\nn\\
{o_{(V)}}_{p q ij lm} &=&
\epsilon_{p q r m} {A_{002}}_{ij}^{~~r} {B}_{l} 
~.~
\eea
The above products satisfy the following the quadratic relation,
\beal{es1134rel11}
{O_{(IV)}}_{p q ij lm} = 3 {o_{(IV)}}_{p q ij lm} - {o_{(V)}}_{p q ij lm} = 0 ~,~
\eea
which corresponds to the order $-[0,0,2]_x s^2 t^6$.
\\

\item \underline{$- [1,1,1]_x s^2 t^6$ relations:} The following generator product is the only one which corresponds to the order $-[1,1,1]_x s^2 t^6$ of the plethystic logarithm of the master space Hilbert series,
\beal{es1135rel15}
{O_{(V)}}_{ij~~l}^{~~~k} = 
{A_{001}}_{ij}^{~~k} {A_{001}}_{lm}^{~~m} = 0 ~.~
\eea
\end{itemize}

The third set of quadratic relation of generators at orders of $s^3 t^6$ involve the following representation products,
\beal{es1135cc1}
\text{Sym}^{2} [0,0,1]_x &=& [0,0,2]_x
~,~
\nn\\
~[1,1,0]_x \times [1,1,0]_x &=& [2,2,0]_x + [3,0,1]_x + [0,3,0]_x + [1,1,1]_x + [2,0,0]_x 
\nn\\
&&
+ [0,0,2]_x + [0,1,0]_x
~,~
\nn\\
~[1,1,0]_x \times [0,0,1]_x &=&
[1,1,1]_x + [2,0,0]_x + [0,1,0]_x
~,~
\nn\\
~[3,0,0]_x \times [0,0,1]_x &=&
[3,0,1]_x + [2,0,0]_x 
~.~
\eea
The above $SU(4)$ representation products are used to identify the quadratic relations at this order as follows:
\begin{itemize}
\item\underline{$-[0,1,0]_x s^3 t^6$ relations:} At this order, there is only the following generator product,
\beal{es1114}
{P_{(I)}}_{ij} = {A_{001}}_{ij}^{~~k} ~ {B}_{k} = 0 ~,~
\eea
which exactly vanishes. The relation satisfies,
\beal{es1114b}
{P_{(I)}}_{ij} = - {P_{(I)}}_{ji} ~,~
\eea
such that the quadratic relation corresponds to the order $-[0,1,0]_x s^3 t^6$.
\\

\item\underline{$-[0,0,2]_x s^3 t^6$ relations:} For the quadratic relations at this order, we have to consider the following generator products,
\beal{es1114cc1}
{p_{(I)}}_{ij} &=& 
{A_{001}}_{ip}^{~~q} {A_{002}}_{jq}^{~~p} 
~,~
\nn\\
{p_{(II)}}_{ij} &=& u_3 B_{i} B_{j} ~.~
\eea
The above products transform in $[0,0,2]_x$ and satisfy the following quadratic relation
\beal{es1114cc2}
{P_{(II)}}_{ij} = {p_{(I)}}_{ij} - \frac{1}{9} {p_{(II)}}_{ij} = 0
~,~
\eea
which is precisely the relation we are looking for at this order.
\\

\item\underline{$-2 [2,0,0]_x s^3 t^6$ relations:} One of the generator products which corresponds to the order $-[2,0,0]_x s^2 t^6$ is as follows,
\beal{es1115}
{P_{(III)}}^{ij} = {S_{012}}^{ijk} {B}_{k} = 0 ~,~
\eea
which vanishes exactly. The above satisfies
\beal{es1115b}
{P_{(III)}}^{ij} = {P_{(III)}}^{ji} ~,~
\eea
and is the first quadratic relation corresponding to $-[2,0,0]_x s^2 t^6$. Another set of generator products which correspond to this order is as follows,
\beal{es1116}
{p_{(III)}}_{ij} &=&
{A_{002}}_{ik}^{~~m} {A_{001}}_{im}^{~~k} 
\nn\\
{p_{(IV)}}_{ij} &=&
u_3 {B}_{i} {B}_{j} ~.~
\eea
These products satisfy the following quadratic relation,
\beal{es1117}
{P_{(IV)}}_{ij} = 
9 {p_{(III)}}_{ij} - {p_{(IV)}}_{ij} = 0 
~,~
\eea
which is the second relation corresponding to the order $-[2,0,0]_x s^2 t^6$.
\\

\item\underline{$- 2[1,1,1]_x s^3 t^6$ relations:} The following generator products transform in $-[1,1,1]_x s^3 t^6$,
\beal{es1120}
{p_{(V)}}_{jm~i}^{~~~l} &=& u_2 {A_{001}}_{jm}^{~~l} {B}_{i} ~,~
\nn\\
{p_{(VI)}}_{jm~i}^{~~~l} &=& {A_{002}}_{jm}^{~~k} {A_{001}}_{k i}^{~~l} ~,~
\nn\\
{p_{(VII)}}_{jm~i}^{~~~l} &=& {A_{001}}_{jm}^{~~k} {A_{002}}_{k i}^{~~l} ~.~
\eea
They satisfy the following quadratic relations,
\beal{es1121}
{P_{(V)}}_{jm~i}^{~~~l} &=&  {p_{(V)}}_{jm~i}^{~~~l} - 6 {p_{(VI)}}_{jm~i}^{~~~l} ~,~
\nn\\
{P_{(VI)}}_{jm~i}^{~~~l} &=&  {p_{(V)}}_{jm~i}^{~~~l} - 6 {p_{(VII)}}_{jm~i}^{~~~l} ~,~
\eea
which satisfy
\beal{es1122}
{P_{(V)}}_{jm~i}^{~~~l} = - {P_{(V)}}_{jm~i}^{~~~l} ~,~
{P_{(VI)}}_{jm~i}^{~~~l} = - {P_{(VI)}}_{jm~i}^{~~~l} ~,~
\eea
and hence correspond to the two quadratic relations at order $-[1,1,1]_x s^3 t^6$.
\\

\item\underline{$-[3,0,1]_x s^3 t^6$ relations:} For the quadratic relation at this order, we consider the following generator products. The first one to consider is,
\beal{es1122aa1}
{p_{(VIII)}}^{ijk}_{~~~m} = {A_{001}}_{pq}^{~~i} {A_{002}}_{rn}^{~~j} \epsilon^{pqrk}
~,~
\eea
which we symmetrize as follows,
\beal{es1122aa2}
{p_{(IX)}}^{ijk}_{~~~m} = {p_{(VIII)}}^{ijk}_{~~~m} + {p_{(VIII)}}^{jki}_{~~~m} + {p_{(VIII)}}^{kij}_{~~~m} ~.~
\eea
The second product to consider is
\beal{es1122aa3}
{p_{(X)}}^{ijk}_{~~~m} = {S_{012}}^{ijk} B_{m} ~.~
\eea
The above generator products satisfy the following quadratic relation
\beal{es1122aa3}
{P_{(VII)}}^{ijk}_{~~~m} = {p_{(IX)}}^{ijk}_{~~~m} - {p_{(X)}}^{ijk}_{~~~m} = 0 ~,~
\eea
which is the relation we are looking for at this order.
\\
\end{itemize}

The next set of quadratic relations at orders of $s^4 t^6$ can be identified by considering the following products of representations,
\beal{es1120a1}
\text{Sym}^{2} [1,1,0]_x &=&
[2,2,0]_x + [1,1,1]_x + [2,0,0]_x + [0,0,2]_x 
~,~
\nn\\
\text{Sym}^{2} [0,0,1]_x &=&
[0,0,2]_x 
~,~
\nn\\
~[3,0,0]_x \times [1,1,0]_x &=&
[3,1,1]_x + [2,0,2]_x + [2,1,0]_x + [1,0,1]_x 
~,~
\nn\\
~[1,1,0]_x \times [0,0,1]_x &=&
[1,1,1]_x + [2,0,0]_x + [0,1,0]_x
~.~
\eea
We make use of the above products to construct the quadratic relations of this order as follows:
\begin{itemize}
\item\underline{$-[2,0,0]_x s^4 t^6$ relations:} For order $-[2,0,0]_x s^4 t^6$, the following generator product applies,
\beal{es1200}
{U_{(I)}}^{ij} = \epsilon^{pqrs} {A_{002}}_{pq}^{~~i} {A_{002}}_{rs}^{~~j} = 0 ~,~
\eea
which vanishes exactly and satisfies,
\beal{es1201}
{U_{(I)}}^{ij} = {U_{(I)}}^{ji} ~.~
\eea
This is exactly the quadratic relation corresponding to the order $-[2,0,0]_x s^4 t^6$.
\\

\item\underline{$-[0,0,2]_x s^4 t^6$ relations:} For the quadratic relation at this order, we need to first consider the following products of generators,
\beal{es1202aa1}
{u_{(I)}}^{ij} &=& {A_{002}}_{ip}^{~~q} {A_{002}}_{jq}^{~~p}
~,~
\nn\\
{u_{(II)}}^{ij} &=& (u_2)^2 B_{i} B_{j} ~.~
\eea
The above products satisfy the following quadratic relation of this order,
\beal{es1202aa2}
{U_{(II)}}^{ij} = {u_{(I)}}^{ij} - \frac{1}{18} {u_{(II)}}^{ij} = 0~,~
\eea
which is the one we are looking for.
\\

\item\underline{$-[1,1,1]_x s^4 t^6$ relations:} The following generator products correspond to this order of the plethystic logarithm,
\beal{es1210}
{u_{(III)}}_{ij~l}^{~~k} &=&
{A_{002}}_{ij}^{~~k} {A_{002}}_{lm}^{~~m} ~,~
\nn\\
{u_{(IV)}}_{ij~l}^{~~k} &=& 
u_3 {A_{001}}_{ij}^{~~k} {B}_{l} ~,~
\nn\\
{u_{(V)}}_{ij~l}^{~~k} &=& 
u_2 {A_{002}}_{ij}^{~~k}  {B}_{l} ~.~
\eea
The above products satisfy the following quadratic relation,8
\beal{es1211}
{U_{(III)}}_{ij~l}^{~~k} = {u_{(III)}}_{ij~l}^{~~k} + 13 {u_{(IV)}}_{ij~l}^{~~k} - {u_{(V)}}_{ij~l}^{~~k} = 0 ~,~
\eea
which corresponds to the order $-[1,1,1]_x s^4 t^6$.
\\

\item\underline{$-[2,2,0]_x s^4 t^6$ relations:} For this order, the quadratic relations can be constructed by looking at generator products and their index symmetrizations and anti-symmetrizations. We consider
\beal{es1212cc1}
{u_{(VI)}}^{ijklmn} = \epsilon^{ijpq} {A_{001}}_{pq}^{~~k} {S_{012}}^{lmn} ~.
\eea
which we anti-symmetrize in the indices $[kl]$ and $[mn]$ as follows,
\beal{es1212cc2}
{u_{(VII)}}^{ijklmn} = {u_{(VI)}}^{ijklmn} - {u_{(VI)}}^{ijlkmn} - {u_{(VI)}}^{ijklnm} + {u_{(VI)}}^{ijlknm} ~.~
\eea
We follow with a further symmetrization of the pairs of indices $[kl]$ and $[mn]$
\beal{es1212cc3}
{U_{(IV)}}^{ijklmn} =  {u_{(VII)}}^{ijklmn} + {u_{(VII)}}^{ijmnkl} = 0 ~.~
\eea
The above vanishes exactly and forms the required quadratic relation at this order.
\\

\item\underline{$-[3,0,1]_x s^4 t^6$ relations:} For the quadratic relations at this order, we consider first the following product of generators of the vortex master space,
\beal{es1212cc5}
{u_{(VIII)}}^{ijk}_{~~~m} = {A_{001}}_{wv}^{~~q} {S_{012}}^{rjk} \epsilon^{wvip} \epsilon_{pqrm} ~.~
\eea
We symmetrize the above product as follows,
\beal{es1212cc6}
{U_{(V)}}^{ijk}_{~~~m} = {u_{(VIII)}}^{ijk}_{~~~m}+{u_{(VIII)}}^{jki}_{~~~m}+{u_{(VIII)}}^{kij}_{~~~m} = 0
~,~
\eea
where we see that the symmetrization vanishes non-trivially. This is precisely the relation related to the order $-[3,0,1]_x s^4 t^6$ of the plethystic logarithm of the Hilbert series.
\\
\end{itemize}

Let us move on to the next set of quadratic relations at orders of $s^5 t^6$. For these relations, we consider the following products of $SU(4)$ representations,
\beal{es1210a1}
\text{Sym}^{2} [1,1,0]_x &=&
[2,2,0]_x + [1,1,1]_x + [2,0,0]_x + [0,0,2]_x 
~,~
\nn\\
\text{Sym}^{2} [0,0,1]_x &=&
[0,0,2]_x 
~,~
\nn\\
~[3,0,0]_x \times [1,1,0]_x &=&
[3,1,1]_x + [2,0,2]_x + [2,1,0]_x + [1,0,1]_x 
~,~
\nn\\
~[1,1,0]_x \times [0,0,1]_x &=&
[1,1,1]_x + [2,0,0]_x + [0,1,0]_x
~,~
\nn\\
~[1,1,0]_x \times [1,1,0]_x &=&
[2,2,0]_x + [3,0,1]_x + [0,3,0]_x + 2 [1,1,1]_x 
\nn\\
&&
+ [2,0,0]_x + [0,0,2]_x 
~.~
\eea
We use the above products of representations in order to construct the quadratic relations of generators at this order as follows:
\begin{itemize}
\item \underline{$-[1,1,1]_x s^5 t^6$ relations:} For the quadratic relations at this order, we need to consider the following products of generators,
\beal{es1210cc1}
{v_{(I)}}_{ij~m}^{~~k} &=&
{A_{002}}_{ij}^{~~p} {S_{012}}^{qrk} \epsilon_{pqrm} ~,~
\nn\\ 
{v_{(II)}}_{ij~m}^{~~k} &=&
u_3 {A_{001}}_{ij}^{~~p} {A_{001}}_{pm}^{~~k} ~.~
\eea
These transform in the correct $SU(4)$ representation of this order. We find that they satisfy the following quadratic relations,
\beal{es1210cc2}
{V_{(I)}}_{ij~m}^{~~k} &=&
{v_{(I)}}_{ij~m}^{~~k} + \frac{1}{2} {v_{(II)}}_{ij~m}^{~~k} ~.~
\eea
This is precisely the relation of this order.
\\

\item \underline{$-[2,2,0]_x s^5 t^6$ relations:} The quadratic relation at this order is constructed by considering the following products of master space generators, with their index symmetrizations and anti-symmetrizations. The first generator product to consider is,
\beal{es1210cc3}
{v_{(III)}}^{ijklmn} = {A_{011}}_{pq}^{~~k} {S_{012}}^{lmn} \epsilon^{pqij} ~,~
\eea
which we anti-symmetrize on the indices $[kl]$ and $[mn]$ as follows,
\beal{es1210cc4}
{v_{(IV)}}^{ijklmn} = {v_{(III)}}^{ijklmn} - {v_{(III)}}^{ijlkmn} - {v_{(III)}}^{ijklnm} + {v_{(III)}}^{ijlknm} ~.~
\eea
A further anti-symmetrization on the pair of indices $[ij]$ and $[mn]$ gives
\beal{es1210cc5}
{v_{(V)}}^{ijklmn} = {v_{(IV)}}^{ijklmn} - {v_{(IV)}}^{mnklij} ~.~
\eea
The second generator product to consider is as follows,
\beal{es1210cc6}
{v_{(VI)}}^{ijklmn} = u_3 {A_{001}}_{pq}^{~~k} {A_{001}}_{rs}^{~~n} \epsilon^{pqij} \epsilon^{rslm} ~.~
\eea
We anti-symmetrize the above product in the indices $[kl]$ and $[mn]$,
\beal{es1210cc7}
{v_{(VII)}}^{ijklmn} = {v_{(VI)}}^{ijklmn} - {v_{(VI)}}^{ijlkmn} - {v_{(VI)}}^{ijklnm} + {v_{(VI)}}^{ijlknm} ~,~
\eea
and further symmetrize the product in the pairs of indices $[ij]$ and $[kl]$ as follows
\beal{es1210cc8}
{v_{(VIII)}}^{ijklmn} = {v_{(VII)}}^{ijklmn} + {v_{(VII)}}^{klijmn} ~.~
\eea
Using the above generator products, we identify the following quadratic relation,
\beal{es1210cc9}
{V_{(II)}}^{ijklmn} =  {v_{(V)}}^{ijklmn} + \frac{1}{4} {v_{(VIII)}}^{ijmnkl} ~,~
\eea
which is the relation we are looking for at this order.
\\

\item \underline{$-[3,0,1]_x s^5 t^6$ relations:} For the quadratic relations at this order, we have to consider the following generator products with their symmetrizations of indices. The first product to consider is as follows,
\beal{es1210cc10}
{v_{(IX)}}^{ijk}_{~~~m} = {A_{002}}_{vw}^{~~q} {S_{012}}^{rjk} \epsilon^{vwip} \epsilon_{pqrm} ~,~
\eea
which we symmetrize in the indices $ijk$ such that
\beal{es1210cc11}
{v_{(X)}}^{ijk}_{~~~m} = {v_{(IX)}}^{ijk}_{~~~m} + {v_{(IX)}}^{jki}_{~~~m} + {v_{(IX)}}^{kij}_{~~~m} ~.~
\eea
The second generator product to consider is
\beal{es1210cc12}
{v_{(XI)}}^{ijk}_{~~~m} = u_2 {S_{012}}^{ijk} B_{m} ~.~
\eea
With the first one, the above product satisfies the following quadratic relation,
\beal{es1210cc13}
{V_{(III)}}^{ijk}_{~~~m} = {v_{(X)}}^{ijk}_{~~~m} - \frac{4}{3} {v_{(XI)}}^{ijk}_{~~~m} = 0 ~,~
\eea
which is precisely the relation we are looking for at this order of the plethystic logarithm. 
\\
\end{itemize}

For the final quadratic relation at the order of $s^6 t^6$, we construct candidates for which the following products of $SU(4)$ representations are useful,
\beal{es1210a2}
\text{Sym}^{2} [3,0,0]_x &=& [6,0,0]_x + [2,2,0]_x 
~,~
\nn\\
\text{Sym}^{2} [1,1,0]_x &=& [2,2,0]_x + [1,1,1]_x + [2,0,0]_x + [0,0,2]_x 
~,~
\nn\\
~[3,0,0]_x \times [1,1,0]_x &=&
[4,1,0]_x + [2,2,0]_x + [3,0,1]_x + [1,1,1]_x 
~,~
\nn\\
~[1,1,0]_x \times [1,1,0]_x &=&
[2,2,0]_x + [3,0,1]_x + [0,3,0]_x + [1,1,1]_x 
\nn\\
&&
+ [2,0,0]_x + [0,0,2]_x + [0,1,0]_x
~.~
\eea
The above representation products can be used to identify the correct generator products and to construct the following quadratic relation at this order:
\begin{itemize}
\item \underline{$-[2,2,0]_x s^6 t^6$ relations:} The quadratic relation at this order can be obtained by considering the following generator product,
\beal{es1211cc1}
{z_{(I)}}^{ijklmn} = {S_{012}}^{ijk} {S_{012}}^{lmn} ~.~
\eea
We first anti-symmetrize the indices $[kl]$ and $[mn]$ as follows,
\beal{es1211cc2}
{z_{(II)}}^{ijklmn} = {z_{(I)}}^{ijklmn} - {z_{(I)}}^{ijlkmn} - {z_{(I)}}^{ijklnm} + {z_{(I)}}^{ijlknm}  ~,~
\eea
and then symmetrize the pairs of indices $[kl]$ and $[mn]$ to obtain
\beal{es1211cc3}
Z^{ijklmn} = {z_{(II)}}^{ijklmn} + {z_{(II)}}^{ijmnkl} = 0 ~,~
\eea
which vanishes non-trivially. As a result, we identify the above as the desired quadratic relation at this order of the plethystic logarithm.
\\
\end{itemize}

\paragraph{Vortex moduli space:} The vortex moduli space for $3$ $U(4)$ vortices can be expressed as a partial $\mathbb{C}^{*}$ projection of the vortex master space. The projection is given as follows,
\beal{es1135a1}
&&
\V_{3,4} = \master_{3,4} / \{ 
B_i \simeq \lambda^{3} B_i ,
{A_{001}}_{ij}^{~~k} \simeq \lambda^{3} {A_{001}}_{ij}^{~~k} ,
\nn\\
&&
\hspace{3cm}
{A_{002}}_{ij}^{~~k} \simeq \lambda^{3} {A_{002}}_{ij}^{~~k} ,
{S_{012}}^{i j k} \simeq \lambda^{3} {S_{012}}^{i j k} 
\}~~,
\eea
where $\lambda$ is the $\mathbb{C}^{*}$ parameter. The dimension of the partially projected space representing the vortex moduli space is as expected $12$. The master space is expressed as the following quotient,
\beal{es1135a2}
&&
\master_{3,4} 
= \mathbb{C}[u_2, u_3, B_{i}, {A_{001}}_{ij}^{~~k}, {A_{002}}_{ij}^{~~k}, {S_{012}}^{i j k}] / \{
\nn\\
&&
\hspace{2cm}
{R_{(I)}}_{ij}=0 ~,~
{R_{(II)}}_{ij}=0 ~,~
\nn\\
&&
\hspace{2cm}
{O_{(I)}}_{ij} = 0 ~,~
{O_{(II)}}_{ij} = 0 ~,~
{O_{(III)}}_{ij} = 0 ~,~
{O_{(IV)}}_{pqijlm} = 0 ~,~
{O_{(V)}}_{ij~l}^{~~k} = 0 ~,~
\nn\\
&&
\hspace{2cm}
{P_{(I)}}_{ij} = 0 ~,~
{P_{(II)}}_{ij} = 0 ~,~
{P_{(III)}}^{ij} = 0 ~,~
{P_{(IV)}}_{ij} = 0 ~,~
{P_{(V)}}_{jm~i}^{~~l} = 0 ~,~
\nn\\
&&
\hspace{2cm}
{P_{(VI)}}_{jm~i}^{~~l} = 0 ~,~
{P_{(VII)}}^{ijk}_{~~~m} = 0 ~,~
\nn\\
&&
\hspace{2cm}
{U_{(I)}}^{ij} = 0 ~,~
{U_{(II)}}^{ij} = 0 ~,~
{U_{(III)}}_{ij~l}^{~~k} = 0 ~,~
{U_{(IV)}}^{ijklmn} = 0 ~,~
{U_{(V)}}^{ijk}_{~~~m} = 0 ~,~
\nn\\
&&
\hspace{2cm}
{V_{(I)}}_{ij~m}^{~~k} = 0 ~,~
{V_{(II)}}^{ijklmn} = 0 ~,~
{V_{(III)}}^{ijk}_{~~~m} = 0 ~,~
\nn\\
&&
\hspace{2cm}
Z^{ijklmn} = 0
~~~
\}
~.~
\eea
\\

\subsection{$3$ $U(5)$ vortices on $\mathbb{C}$ \label{s3b5}}

The Hilbert series for the $3$ $U(5)$ reduced vortex master space is obtained from the following Molien integral,
\beal{es1150}
g(t,s,x;\wmaster_{3,5}) = \oint \ud\mu_{SU(3)} \PE\Big[
[0,1]_w [1,0,0,0]_x t + [1,1]_w s
\Big]~,~
\eea
where $[1,0,0,0]_x$ is the character of the fundamental representation of the global $SU(5)$. The integral leads to the following character expansion of the Hilbert series
\beal{es1152}
&&
g(t,s,x;\wmaster_{3,5}) = 
\frac{1}{(1-s^2)(1-s^3)} \times
\nn\\
&&
\sum_{n_0=0}^{\infty}
\sum_{n_1=0}^{\infty}
\sum_{n_2=0}^{\infty}
\sum_{n_3=0}^{\infty}
\Big[
[n_1+n_2+3n_3,n_1+n_2,n_0,0]_x s^{n_1+2n_2+3n_3} t^{3n_0+3n_1+3n_2 +3n_3}
\nn\\
&&
\hspace{1cm}
+ [n_1+n_2,n_1+n_2+3n_3+3,n_0,0]_x s^{n_1+2n_2+3n_3+3} t^{3n_0+3n_1+3n_2+6n_3+6}
\Big]
~,~\nn\\
\eea
where $[m_1,m_2,m_3,m_4]_x$ is a character of a $SU(5)$ irreducible representation.

The plethystic logarithm of the Hilbert series is
\beal{es1153}
\PL\Big[
g(t,s,x;\wmaster_{3,5})
\Big] 
&=&
s^2 + s^3 + [0,0,1,0]_x t^3
+[1,1,0,0]_x s t^3
+[1,1,0,0]_x s^2 t^3
\nn\\
&&
+[3,0,0,0]_x s^3 t^3
-[1,0,0,0]_x t^6
- (
[2,0,0,1]_x 
+ [0,1,0,1]_x
\nn\\
&&
+ [1,0,0,0]_x 
) s t^6
- (
[1,1,1,0]_x
+ [0,0,2,0]_x
+ 2 [2,0,0,1]_x
\nn\\
&&
+ [0,1,0,1]_x
+ [1,0,0,0]_x
) s^2 t^6
- (
[0,3,0,0]_x 
+ [3,0,1,0]_x 
\nn\\
&&
+ 2 [1,1,1,0]_x
+ [0,0,2,0]_x
+ 2 [2,0,0,1]_x
+ [0,1,0,1]_x
) s^3 t^6
\nn\\
&&
- (
 [2,2,0,0]_x
 + [3,0,1,0]_x
 + [0,0,2,0]_x
 + [2,0,0,1]_x
 \nn\\
 &&
 + 2 [1,1,1,0]_x
) s^4 t^6
+ \dots 
\nn\\
&&
- ([1,1,1,0]_x + [2,2,0,0]_x  + [3,0,1,0]_x ) s^5 t^6
+ \dots
\nn\\
&&
- [2,2,0,0]_x s^6 t^6 
+\dots
~~.
\eea
The generators of the vortex moduli space are indicated by the above plethystic logarithm. They are as follows,
\beal{es1154}
s^2 &\ra& u_2=\Tr(\phi^2)
\nn\\
s^3 &\ra& u_3=\Tr(\phi^3)
\nn\\
~[0,0,1,0]_x t^3 &\ra&
{B}_{ij}= \epsilon^{\alpha_1 \alpha_2 \alpha_3} \epsilon_{k_1 k_2 k_3 i j} Q_{\alpha_1}^{k_1} Q_{\alpha_2}^{k_2} Q_{\alpha_3}^{k_3}
\nn\\
~[1,1,0,0]_x s t^3 &\ra&
\left\{
\ba{l}
{A_{001}}^{i j k} =
\epsilon^{\alpha_1 \alpha_2 \alpha_3}
Q_{\alpha_1}^{i} Q_{\alpha_2}^{j} \phi_{\alpha_3}^{\beta} Q_{\beta}^{k}
\nn\\
\epsilon_{ijkmn} {A_{001}}^{ijk}=0
\ea
\right.
\nn\\
~[1,1,0,0]_x s^2 t^3 &\ra&
\left\{
\ba{l}
{A_{002}}^{ijk} =
\epsilon^{\alpha_1 \alpha_2 \alpha_3}
Q_{\alpha_1}^{i} Q_{\alpha_2}^{j} \phi_{\alpha_3}^{\beta_1}\phi_{\beta_1}^{\beta_2} Q_{\beta_2}^{k}
\nn\\
\epsilon_{ijkmn} {A_{002}}^{ijk} = \frac{1}{3} u_2 {B}_{mn}
\nn\\
{A_{011}}^{ijk} =
\epsilon^{\alpha_1 \alpha_2 \alpha_3}
Q_{\alpha_1}^{k} \phi_{\alpha_2}^{\beta_1} Q_{\beta_1}^{i} \phi_{\alpha_3}^{\beta_2} Q_{\beta_2}^{j}
\nn\\
\epsilon_{ijkmn} {A_{011}}^{ijk}= - \frac{1}{6} u_2 {B}_{mn}
\nn\\
\ra {A_{002}}^{ijk} = {A_{011}}^{ijk} + \frac{1}{24} u_2 \epsilon^{ijkmn} B_{mn}
\ea\right.
\nn\\
~[3,0,0,0]_x s^3 t^3 &\ra&
\left\{\ba{l}
{A_{012}}^{ijk}
= 
\epsilon^{\alpha_1 \alpha_2 \alpha_3} 
Q_{\alpha_1}^{i}
\phi_{\alpha_2}^{\beta_1} Q_{\beta_1}^{j}
\phi_{\alpha_3}^{\beta_2} \phi_{\beta_2}^{\beta_3} Q_{\beta_3}^{k}
\nn\\
{S_{012}}^{ijk} =
{A_{012}}^{ijk}
+ {A_{012}}^{jki}
+ {A_{012}}^{kij}
\ea\right.
~~.
\eea
\\

\paragraph{Quadratic relations:} The terms of the plethystic logarithm corresponding to quadratic relations between master space generators are as follows,
\beal{es1160}
&
- [1,0,0,0]_x t^6
&
\nn\\
&
- [0,1,0,1]_x s t^6
- [2,0,0,1]_x s t^6
- [1,0,0,0]_x s t^6
&
\nn\\
&
- [0,1,0,1]_x s^2 t^6
- 2 [2,0,0,1]_x s^2 t^6
- [0,0,2,0]_x s^2 t^6
- [1,1,1,0]_x s^2 t^6
- [1,0,0,0]_x s^2 t^6
&
\nn\\
&
- [0,1,0,1]_x s^3 t^6
- 2 [2,0,0,1]_x s^3 t^6
- [0,0,2,0]_x s^3 t^6
- 2 [1,1,1,0]_x s^3 t^6
- [3,0,1,0]_x s^3 t^6
&
\nn\\
&
- [2,0,0,1]_x s^4 t^6
- [0,0,2,0]_x s^4 t^6
- 2 [1,1,1,0]_x s^4 t^6
- [2,2,0,0]_x s^4 t^6
- [3,0,1,0]_x s^4 t^6
&
\nn\\
&
- [1,1,1,0]_x s^5 t^6 - [2,2,0,0]_x s^5 t^6 - [3,0,1,0]_x s^5 t^6 
&
\nn\\
&
- [2,2,0,0]_x s^6t^6
&
~.~
\nn\\
\eea
From \eref{es1154}, we remind ourselves about the generators of the master space which form the quadratic relations corresponding to the negative terms of the plethystic logarithm in \eref{es1160} above,
\beal{es1161}
u_2 ~,~ 
u_3 ~,~
{B}_{ij}~,~
{A_{001}}^{ijk} ~,~
{A_{002}}^{ijk} ~,~
{S_{012}}^{ijk} ~.~
\eea
Let us go through the quadratic relations as follows, by first considering the relations associated to the order $t^6$. At this order, we have to consider the following symmetric product of $SU(5)$ representations,
\beal{es1165}
\text{Sym}^{2}[0,0,1,0]_x = [0,0,2,0]_x + [1,0,0,0]_x ~,~
\eea
which leads to the following relations:
\begin{itemize}
\item \underline{$-[1,0,0,0]_x t^6$ relations:} The following generator product corresponds to this order,
\beal{es1170}
H^{i} = \epsilon^{i j k l m} {B}_{jk} {B}_{lm} = 0 ~.~
\eea
The product vanishes exactly and corresponds to the quadratic relation at this order.
\\

\end{itemize}

The next order of relations to consider is $s t^6$. For this order, we consider the following representation product,
\beal{es1170b}
[1,1,0,0]_x \times [0,0,1,0]_x = [1,1,1,0]_x + [2,0,0,1]_x + [0,1,0,1]_x + [1,0,0,0]_x ~.~
\eea
From the above product, we construct the following quadratic relations of generators for orders of $s t^6$:
\begin{itemize}
\item \underline{$-[0,1,0,1]_x s t^6$ relations:} The following product of generators transforms in $-[0,1,0,1]_x s t^6$,
\beal{es1171}
{R_{(I)}}^{ij}_{~~k} = {A_{001}}^{ijm} {B}_{mk} = 0~,~
\eea
which exactly vanishes. This is precisely the quadratic relations at this order of the plethystic logarithm.
\\

\item \underline{$-[2,0,0,1]_x s t^6$ relations:} For the quadratic relation at this order, we consider the following generator product,
\beal{es1172}
{R_{(II)}}^{ij}_{~~k} = {A_{001}}^{imj} B_{mk} = 0 ~,~
\eea
which exactly vanishes and accordingly corresponds to the quadratic relation at order $-[2,0,0,1]_x s t^6$.
\\

\item \underline{$-[1,0,0,0]_x s t^6$ relations:} At this order, we have the following generator product,
\beal{es1175}
{R_{(III)}}^{i} = {A_{001}}^{imn} {B}_{mn} =0
~,~
\eea
which vanishes exactly. This is the quadratic relation which transforms in $-[1,0,0,0]_x s t^6$.
\\

\end{itemize}

The next order to consider is $s^2 t^6$. For this order, we have to look at the following $SU(5)$ representation products,
\beal{es1175b}
\text{Sym}^{2} [1,1,0,0]_x &=& [2,2,0,0]_x + [1,1,1,0]_x + [2,0,0,1]_x + [0,0,2,0]_x ~,~
\nn\\
~[1,1,0,0]_x \times [0,0,1,0]_x &=& [1,1,0,0]_x + [2,0,0,1]_x + [0,1,0,1]_x + [1,0,0,0]_x ~,~
\nn\\
\text{Sym}^{2} [0,0,1,0]_x &=& [0,0,2,0]_x + [1,0,0,0]_x ~.~
\eea
The above products lead to the following quadratic relations:
\begin{itemize}

\item \underline{$-[0,1,0,1]_x s^2 t^6$ relations:} The following generator product corresponds to this order,
\beal{es1176}
{O_{(I)}}^{ij}_{~~k} = {A_{002}}^{ijm} {B}_{mk} ~,~
\eea
and vanishes exactly. This is the quadratic relation corresponding to $-[0,1,0,1]_x s^2 t^6$.
\\

\item \underline{$-2[2,0,0,1]_x s^2 t^6$ relations:} There are precisely two distinct generator products at this order which are
\beal{es1177}
{O_{(II)}}^{ij}_{~~k} &=& {A_{001}}^{mn i} {A_{001}}^{pq j} \epsilon_{mnpq k} =0 ~,~
\nn\\
{O_{(III)}}^{ij}_{~~k} &=& {A_{002}}^{im j} B_{mk} =0 ~,~
\eea
which individually exactly vanish. These are the two quadratic relations which transform as $-[2,0,0,1]_x s^2 t^6$.
\\

\item \underline{$-[0,0,2,0]_x s^2 t^6$ relations:} The following generator products correspond to the order $-[0,0,2,0]_x s^2 t^6$,
\beal{es1178}
{o_{(I)}}_{ijkl} &=& \epsilon_{pqsij} \epsilon_{mnrkl} {A_{001}}^{pqr} {A_{001}}^{mns} ~,~
\nn\\
{o_{(II)}}_{ijkl} &=& u_2 B_{ij} B_{kl} ~,~
\eea
which satisfy
\beal{es1179}
&{o_{(I)}}_{ijkl} = - {o_{(I)}}_{jikl} ~,~ {o_{(I)}}_{ijkl} = - {o_{(I)}}_{ijlk} ~,~ {o_{(I)}}_{ijkl} = {o_{(I)}}_{jilk} &  ~,~
\nn\\
&{o_{(II)}}_{ijkl} = - {o_{(II)}}_{jikl} ~,~ {o_{(II)}}_{ijkl} = - {o_{(II)}}_{ijlk} ~,~ {o_{(II)}}_{ijkl} = {o_{(II)}}_{jilk} &  ~.~
\eea
The above generator products satisfy the following quadratic relation,
\beal{es1180}
{O_{(IV)}}_{ijkl} = {o_{(I)}}_{ijkl} - \frac{1}{36} {o_{(II)}}_{ijkl} = 0 ~,~
\eea
which is the relations corresponding to $-[0,0,2,0]_x s^2 t^6$.
\\

\item \underline{$-[1,1,1,0]_x s^2 t^6$ relations:} At this order, we can consider the following generator products,
\beal{es1181}
{o_{(III)}}^{ijk}_{~~~mn} &=& {A_{001}}^{pqk} {A_{001}}^{ij r} \epsilon_{pqr mn} ~,~
\nn\\
{o_{(IV)}}^{ijk}_{~~~mn} &=& {A_{002}}^{ijk} B_{mn} ~,~
\eea
which satisfy the following quadratic relation,
\beal{es1182}
{O_{(V)}}^{ijk}_{~~~mn} = {o_{(III)}}^{ijk}_{~~~mn} - \frac{1}{3} {o_{(IV)}}^{ijk}_{~~~mn} = 0 ~.~
\eea
This is precisely the relations transforming in $-[1,1,1,0]_x s^2 t^6$.
\\

\item \underline{$-[1,0,0,0]_x s^2 t^6$ relations:} The following generator product vanishes exactly,
\beal{es1183}
{O_{(VI)}}^{i} = {A_{002}}^{mni} B_{mn} = 0 ~,~
\eea
which is the quadratic relation at the order $-[1,0,0,0]_x s^2 t^6$.
\\

\end{itemize}

For the orders containing $s^3 t^6$, we consider the following products of $SU(5)$ representations,
\beal{es1183b}
~[1,1,0,0]_x \times [1,1,0,0]_x &=& [2,2,0,0]_x + [3,0,1,0]_x + [0,3,0,0]_x + 2 [1,1,1,0]_x 
\nn\\
&& + [2,0,0,1] + [0,0,2,0]_x + [0,1,0,1]_x ~,~
\nn\\
~[3,0,0,0]_x \times [0,0,1,0]_x &=& [3,0,1,0]_x + [2,0,0,1]_x ~,~
\nn\\
~[1,1,0,0]_x \times [0,0,1,0]_x &=& [1,1,1,0]_x + [2,0,0,1]_x + [0,1,0,1]_x + [1,0,0,0]_x ~,~
\nn\\
\text{Sym}^{2} [0,0,1,0]_x &=& [0,0,2,0]_x + [1,0,0,0]_x ~.~
\eea
The above products are used to construct the following quadratic relations corresponding to order containing $s^3 t^6$:
\begin{itemize}

\item \underline{$-[0,1,0,1]_x s^3 t^6$ relations:} At this order, we have the following generator product
\beal{es1184}
{P_{(I)}}^{ij}_{~~k} = {A_{001}}^{ijm} {A_{002}}^{pqr} \epsilon_{mpqrk} =0 ~,~
\eea
which exactly vanishes. It corresponds to the quadratic relation for this order.
\\

\item \underline{$-2[2,0,0,1]_x s^3 t^6$ relations:} The following two generator product correspond to this order,
\beal{es1185}
{P_{(II)}}^{ij}_{~~k} &=& {A_{001}}^{mni} {A_{002}}^{pqj} \epsilon_{mnpqk} =0 ~,~
\\
{P_{(III)}}^{ij}_{~~k} &=& {S_{012}}^{ijm} B_{mk} = 0 ~,~
\eea
and exactly vanish. These are precisely the quadratic relations at this order.
\\

\item \underline{$-[0,0,2,0]_x s^3 t^6$ relations:} At this order, we consider the following generator products,
\beal{es1186}
{p_{(I)}}_{ijkl} &=& {A_{001}}^{pqm} {A_{002}}^{rsn} \epsilon_{pqnij} \epsilon_{rsmkl} ~,~
\nn\\
{p_{(II)}}_{ijkl} &=& u_3 B_{ij} B_{kl} ~,~
\eea
which satisfy the following quadratic relation,
\beal{es1187}
{P_{(IV)}}_{ijkl} = {p_{(I)}}_{ijkl} - \frac{1}{9} {p_{(II)}}_{ijkl} =0 ~.~
\eea
The above is precisely the quadratic relation at the order $-[0,0,2,0]_x s^3 t^6$.
\\

\item \underline{$-2[1,1,1,0]_x s^3 t^6$ relations:} The following generator products can be considered at this order,
\beal{es1188}
{p_{(III)}}^{ijk}_{~~~mn} &=&
{A_{001}}^{ijp} {A_{002}}^{qrk} \epsilon_{pqrmn} ~,~
\nn\\
{p_{(IV)}}^{ijk}_{~~~mn} &=&
{A_{002}}^{ijp} {A_{001}}^{qrk} \epsilon_{pqrmn} ~,~
\nn\\
{p_{(V)}}^{ijk}_{~~~mn} &=&
u_2 {A_{001}}^{ijk} B_{mn} ~,~
\nn\\
{p_{(VI)}}^{ijk}_{~~~mn} &=&
u_3 \epsilon^{ijkpq} B_{pq} B_{mn} ~,~
\eea
which satisfy the following quadratic relations,
\beal{es1189}
{P_{(V)}}^{ijk}_{~~~mn} &=& {p_{(III)}}^{ijk}_{~~~mn} - {p_{(IV)}}^{ijk}_{~~~mn} = 0 ~,~
\\
{P_{(VI)}}^{ijk}_{~~~mn} &=& {p_{(III)}}^{ijk}_{~~~mn} - \frac{1}{6} {p_{(V)}}^{ijk}_{~~~mn} - \frac{1}{108} {p_{(VI)}}^{ijk}_{~~~mn} = 0 ~.~
\eea
The above quadratic relations transform in the correct representation of this order, and hence are the relations we are looking for.
\\

\item \underline{$-[3,0,1,0]_x s^3 t^6$ relations:} For this order, we have to consider the following generator products,
\beal{es1189b1}
{p_{(VII)}}^{ijk}_{~~~mn} &=&
(
{A_{001}}^{ipj} {A_{002}}^{qrk} +
{A_{001}}^{ipj} {A_{002}}^{qrk} +
{A_{001}}^{ipj} {A_{002}}^{qrk} 
)
\epsilon_{pqrmn}
~,~
\nn\\
{p_{(VIII)}}^{ijk}_{~~~mn} &=&
{S_{012}}^{ijk} B_{nm} ~,~
\eea
where for the first product above we have symmetrized the product
\beal{es1189b2}
{A_{001}}^{ipj} {A_{002}}^{qrk}
\eea
in the indices $ijk$. The products above satisfy the following quadratic relation,
\beal{es1189b3}
{P_{(VII)}}^{ijk}_{~~~mn}
= {p_{(VII)}}^{ijk}_{~~~mn} + \frac{1}{3} {p_{(VIII)}}^{ijk}_{~~~mn} = 0 ~.~
\eea
The above is precisely the quadratic relation at this order.
\\

\end{itemize}

The next orders contain $s^4 t^6$ for which we have to consider the following representation products,
\beal{es1190}
\text{Sym}^{2} [1,1,0,0]_x &=& [2,2,0,0]_x + [1,1,1,0]_x + [2,0,0,1]_x + [0,0,2,0]_x ~,~
\nn\\
~[1,1,0,0]_x \times [3,0,0,0]_x &=& [4,1,0,0]_x + [2,2,0,0]_x + [3,0,1,0]_x + [1,1,1,0]_x ~,~
\nn\\
~[1,1,0,0]_x \times [0,0,1,0]_x &=& [1,1,1,0]_x + [2,0,0,1]_x + [0,1,0,1]_x + [1,0,0,0]_x ~,~
\nn\\
\text{Sym}^{2} [0,0,1,0]_x &=& [0,0,2,0]_x + [1,0,0,0]_x ~.~
\eea
The above representation products lead to the following quadratic relations of vortex moduli space generators:
\begin{itemize}
\item \underline{$-[2,0,0,1]_x s^4 t^6$ relations:} The following generator product corresponds to the order $-[2,0,0,1]_x s^4 t^6$,
\beal{es2000}
{U_{(I)}}^{ij}_{~~k} = {A_{002}}^{pqi} {A_{002}}^{mnj} \epsilon_{pqmsk} = 0 ~,~
\eea
which vanishes exactly. This is exactly the quadratic relation at this order.
\\

\item \underline{$-[0,0,2,0]_x s^4 t^6$ relations:} This order refers to the following generator products,
\beal{es2001}
{u_{(I)}}_{ijkl} &=&
{A_{002}}^{pqr} {A_{002}}^{mns} \epsilon_{pqsij} \epsilon_{mnrkl} ~,~
\nn\\
{u_{(II)}}_{ijkl} &=&
u_2 u_2 B_{ij} B_{kl} ~,~
\eea
which satisfy then following relation,
\beal{es2002}
{U_{(II)}}_{ijkl} = {u_{(I)}}_{ijkl} - \frac{1}{18} {u_{(II)}}_{ijkl} = 0
~.~
\eea 
This is precisely the quadratic relation corresponding to the order $-[0,0,2,0]_x s ^4 t^6$.
\\

\item \underline{$-2 [1,1,1,0]_x s^4 t^6$ relations:} At this order, we need to consider the following generator products,
\beal{es2005}
{u_{(III)}}^{ijk}_{~~~mn} &=&
{A_{002}}^{ijp} {A_{002}}^{qrk} \epsilon_{pqrmn} ~,~
\nn\\
{u_{(IV)}}^{ijk}_{~~~mn} &=&
{A_{001}}^{ijp} {S_{012}}^{qrk} \epsilon_{pqrmn} ~,~
\nn\\
{u_{(V)}}^{ijk}_{~~~mn} &=&
u_2 {A_{001}}^{ijp} {A_{001}}^{qrk} \epsilon_{pqrmn} ~,~
\nn\\
{u_{(VI)}}^{ijk}_{~~~mn} &=&
u_3 {A_{001}}^{ijk} B_{mn} ~.~
\eea
The above products satisfy the following quadratic relations,
\beal{es2006}
{U_{(III)}}^{ijk}_{~~~mn} &=&
{u_{(V)}}^{ijk}_{~~~mn} - 2 {u_{(III)}}^{ijk}_{~~~mn} + \frac{2}{9} {u_{(VI)}}^{ijk}_{~~~mn} = 0~,~
\nn\\
{U_{(IV)}}^{ijk}_{~~~mn} &=& {u_{(IV)}}^{ijk}_{~~~mn} + \frac{1}{6} {u_{(VI)}}^{ijk}_{~~~mn} = 0 ~,~
\eea
which are precisely the relations corresponding to the order $- 2 [1,1,1,0]_x s^4 t^6$.
\\

\item \underline{$-[2,2,0,0]_x s^4 t^6$ relations:} For the quadratic relation at this order we have to consider the following generator product and additional symmetrization and anti-symmetrizations of its indices,
\beal{es2007}
{u_{(V)}}^{ijklmn} = {A_{001}}^{ijk} {S_{012}}^{lmn} ~.~
\eea
We anti-symmetrize first the indices $[kl]$ and $[mn]$ as follows,
\beal{es2008}
{u_{(VI)}}^{ijklmn} = {u_{(V)}}^{ijklmn} - {u_{(V)}}^{ijlkmn} - {u_{(V)}}^{ijklnm} + {u_{(V)}}^{ijlknm} ~,~
\eea
and symmetrize the pairs of indices $[kl]$ and $[mn]$ to give
\beal{es2009}
{U_{(V)}}^{ijklmn} = {u_{(VI)}}^{ijklmn} + {u_{(VI)}}^{ijmnkl} = 0 ~.~
\eea
The above vanishes non-trivially, and given that the above quadratic relation transforms in the correct representation we have found the quadratic relation at this order.
\\

\item \underline{$-[3,0,1,0]_x s^4 t^6$ relations:} For this order, we consider the following generator product,
\beal{es2020}
{U_{(VI)}}^{ijk}_{~~~mn} =
(
{A_{001}}^{ipq} {S_{012}}^{rjk}
+ {A_{001}}^{jpq} {S_{012}}^{rki}
+ {A_{001}}^{kpq} {S_{012}}^{rij}
) \epsilon_{pqrmn}
= 0 ~,~
\nn\\
\eea
which exactly vanishes. This is precisely the quadratic relation at this order. We have in above symmetrized the product
\beal{es2021}
{A_{001}}^{ipq} {S_{012}}^{rjk}
\eea
in the indices $ijk$.
\\

\end{itemize}

The next set of quadratic relations refer to orders containing $s^5 t^6$. For these relations, we consider the following representation products,
\beal{es2200}
~[3,0,0,0]_x \times [1,1,0,0]_x &=& [4,1,0,0]_x + [2,2,0,0]_x + [3,0,1,0]_x + [1,1,1,0]_x ~,~
\nn\\
~[1,1,0,0]_x \times [1,1,0,0]_x &=& [2,2,0,0]_x + [3,0,1,0]_x + [0,3,0,0]_x + 2 [1,1,1,0]_x 
\nn\\
&&
+ [2,0,0,1]_x + [0,0,2,0]_x + [0,1,0,1]_x ~,~
\nn\\
~[3,0,0,0]_x \times [0,0,1,0]_x &=& [3,0,1,0]_x + [2,0,0,1]_x ~,~
\nn\\
~[1,1,0,0]_x \times [0,0,1,0]_x &=& [1,1,1,0]_x + [2,0,0,1]_x + [0,1,0,1]_x + [1,0,0,0]_x ~,~
\nn\\
~\text{Sym}^{2} [1,1,0,0]_x &=& [2,2,0,0]_x + [1,1,1,0]_x + [2,0,0,1]_x + [0,0,2,0]_x ~,~
\nn\\
~\text{Sym}^{2} [0,0,1,0]_x &=& [0,0,2,0]_x + [1,0,0,0]_x ~.~
\eea
The above products are used to identify the following quadratic relations between moduli space generators:
\begin{itemize}

\item \underline{$-[1,1,1,0]_x s^5 t^6$ relations:} For the order $-[1,1,1,0]_x s^5 t^6$, we consider the following generator products,
\beal{es2201}
{v_{(I)}}^{ijk}_{~~~mn} &=&
{A_{002}}^{ijp} {S_{012}}^{qrk} \epsilon_{pqrmn} ~,~
\nn\\
{v_{(II)}}^{ijk}_{~~~mn} &=&
u_3 {A_{001}}^{ijp} {A_{001}}^{qrk} \epsilon_{pqrmn} ~.~
\eea
Together the above products satisfy the following quadratic relations
\beal{es2202}
{V_{(I)}}^{ijk}_{~~~mn} = 
{v_{(I)}}^{ijk}_{~~~mn} + \frac{1}{2} {v_{(II)}}^{ijk}_{~~~mn} = 0 ~,~
\eea
which is precisely the relation we are looking for at order $-[1,1,1,0]_x s^5 t^6$.
\\

\item \underline{$-[2,2,0,0]_x s^5 t^6$ relations:} For the quadratic relation at this order, we have to consider the following generator products and its symmetrization and anti-symmetrization of indices. The first product to consider is the following,
\beal{es2202a}
{v_{(III)}}^{ijklmn} = ({A_{002}}^{ijk} - \frac{1}{2} u_2 B^{ijk}) {S_{012}}^{lmn} ~,~
\eea
which we anti-symmetrize in the indices $[kl]$ and $[mn]$ to give
\beal{es2202b}
{v_{(IV)}}^{ijklmn} = {v_{(III)}}^{ijklmn} - {v_{(III)}}^{ijlkmn} - {v_{(III)}}^{ijklnm} + {v_{(III)}}^{ijlknm} ~.~
\eea
We further anti-symmetrize the above in the pairs of indices $[ij]$ and $[mn]$ to give
\beal{es2202c}
{v_{(V)}}^{ijklmn} = {v_{(IV)}}^{ijklmn} - {v_{(IV)}}^{mnklij} ~.~
\eea
The second generator product to consider is the following
\beal{es2202d}
{v_{(VI)}}^{ijklmn} = u_3 {A_{001}}^{ijk} {A_{001}}^{lmn} ~,~
\eea
which is anti-symmetrized in the indices $[kl]$ and $[mn]$ as follows
\beal{es2202e}
{v_{(VII)}}^{ijklmn} = {v_{(VI)}}^{ijklmn} - {v_{(VI)}}^{ijlkmn} - {v_{(VI)}}^{ijklnm} + {v_{(VI)}}^{ijlknm}
~.~
\eea
In addition, we symmetrize the pairs of indices $[ij]$ and $[kl]$ to give
\beal{es2202f}
{v_{(VIII)}}^{ijklmn} = {v_{(VII)}}^{ijklmn} + {v_{(VII)}}^{klijmn} ~.~
\eea
The above generator products transform in the correct representation at this order of the plethystic logarithm, and satisfy the following quadratic relation
\beal{es2202g}
V_{(II)}^{ijklmn} = {v_{(V)}}^{ijklmn} + {v_{(VIII)}}^{ijmnkl} = 0.
\eea
The above is the quadratic relation we are searching for at this order.
\\

\item \underline{$-[3,0,1,0]_x s^5 t^6$ relations:} For this order, we have to consider,
\beal{es2203}
{v_{(IX)}}^{ijk}_{~~~mn} &=&
(
{A_{002}}^{ipq} {S_{012}}^{rjk} 
+ {A_{002}}^{jpq} {S_{012}}^{rki} 
+ {A_{002}}^{kpq} {S_{012}}^{rij} 
)
\epsilon_{pqrmn}
~,~
\nn\\
{v_{(X)}}^{ijk}_{~~~mn} &=&
u_2 {S_{012}}^{ijk} B_{mn}
~,~
\nn\\
\eea
where above the generator product
\beal{es2204}
{A_{002}}^{ipq} {S_{012}}^{rjk} \epsilon_{pqrmn}
~,~
\eea
have been symmetrised in the indices $ijk$. The above satisfy the following quadratic relation,
\beal{es2205}
{V_{(III)}}^{ijk}_{~~~mn} = 
{v_{(IX)}}^{ijk}_{~~~mn}
- \frac{1}{3} {v_{(X)}}^{ijk}_{~~~mn} = 0 ~,~
\eea
which is the relation at order $-[3,0,1,0]_x s^5 t^6$.
\\

\end{itemize}

The following order of $s^6 t^6$ leads us to use the following $SU(5)$ representation products,
\beal{es2250}
~\text{Sym}^2 [3,0,0,0]_x &=& [6,0,0,0]_x + [2,2,0,0]_x ~,~
\nn\\
~\text{Sym}^2 [1,1,0,0]_x &=& [2,2,0,0]_x + [1,1,0,0]_x + [2,0,0,1]_x + [0,0,2,0]_x ~,~
\nn\\
~[1,1,0,0]_x \times [3,0,0,0]_x &=& [4,1,0,0]_x + [2,2,0,0]_x + [3,0,1,0]_x + [1,1,1,0]_x ~,~
\nn\\
~[1,1,0,0]_x \times [1,1,0,0]_x &=& [2,2,0,0]_x + [3,0,1,0]_x + [0,3,0,0]_x + 2 [1,1,1,0]_x 
\nn\\
&& + [2,0,0,1]_x + [0,0,2,0]_x + [0,1,0,1]_x
~.~
\eea
From the above representation products we select the appropriate one to identify the following quadratic relations between generators:
\begin{itemize}
\item \underline{$- [2,2,0,0]_x s^6 t^6$ relations:} The quadratic relation at this order is constructed from the following generator product
\beal{es2251}
{z_{(I)}}^{ijklmn} = {S_{012}}^{ijk} {S_{012}}^{lmn} ~,~
\eea
which we first anti-symmetrize in the indices $[kl]$ and $[mn]$ as follows,
\beal{es2252}
{z_{(II)}}^{ijklmn} = {z_{(I)}}^{ijklmn} - {z_{(I)}}^{ijlkmn} - {z_{(I)}}^{ijklnm} + {z_{(I)}}^{ijlknm} ~.~
\eea
Then, we symmetrize the pairs of indices $[kl]$ and $[mn]$ such that
\beal{es2253}
Z^{ijklmn} = {z_{(II)}}^{ijklmn} + {z_{(II)}}^{ijmnkl} = 0 ~,~
\eea
vanishes exactly and hence is the quadratic relation we are looking for at this order.
\\

\end{itemize}

\paragraph{Vortex moduli space:} The $\mathbb{C}^{*}$ projection of the vortex master space $\master_{3,5}$ gives the full moduli space $\wV_{3,5}$ of the $3$ $U(5)$ vortex theory. The vortex moduli space is expressed as the following $\mathbb{C}^{*}$ projection,
\beal{es2251xx1}
&&
\wV_{3,5} = \wmaster_{3,5}/\{
B_{ij} \simeq \lambda^{3} B_{ij} ,
{A_{001}}^{ijk} \simeq \lambda^{3} {A_{001}}^{ijk} ,
\nn\\
&&
\hspace{3cm}
{A_{002}}^{ijk} \simeq \lambda^{3} {A_{002}}^{ijk} ,
{S_{012}}^{i j k} \simeq \lambda^{3} {S_{012}}^{i j k} 
\}
~,~
\eea
where $\lambda$ is the $\mathbb{C}^{*}$ parameter. The master space is expressed as follows,
\beal{es2251xx2}
&&
\wmaster_{3,5} = \mathbb{C}[
u_2, u_3, B_{ij}, {A_{001}}^{ijk}, {A_{002}}^{ijk}, {S_{012}}^{ijk}
]/\{
\nn\\
&&
\hspace{2cm}
H^{i} = 0 ~,~
\nn\\
&&
\hspace{2cm}
{R_{(I)}}^{ij}_{~~k} = 0 ~,~
{R_{(II)}}^{ij}_{~~k} = 0 ~,~
{R_{(III)}}^{i} = 0 ~,~
\nn\\
&&
\hspace{2cm}
{O_{(I)}}^{ij}_{~~k} = 0 ~,~
{O_{(II)}}^{ij}_{~~k} = 0 ~,~
{O_{(III)}}^{ij}_{~~k} = 0 ~,~
{O_{(IV)}}_{ijkl} = 0 ~,~
\nn\\
&&
\hspace{2cm}
{O_{(V)}}^{ijk}_{~~~mn} = 0 ~,~
{O_{(VI)}}^{i} = 0 ~,~
\nn\\
&&
\hspace{2cm}
{P_{(I)}}^{ij}_{~~k} = 0 ~,~
{P_{(II)}}^{ij}_{~~k} = 0 ~,~
{P_{(III)}}^{ij}_{~~k} = 0 ~,~
{P_{(IV)}}_{ijkl} = 0 ~,~
\nn\\
&&
\hspace{2cm}
{P_{(V)}}^{ijk}_{~~~mn} = 0 ~,~
{P_{(VI)}}^{ijk}_{~~~mn} = 0 ~,~
{P_{(VII)}}^{ijk}_{~~~mn} = 0 ~,~
\nn\\
&&
\hspace{2cm}
{U_{(I)}}^{ij}_{~~k} = 0 ~,~
{U_{(II)}}_{ijkl} = 0 ~,~
{U_{(III)}}^{ijk}_{~~~mn} = 0 ~,~
{U_{(IV)}}^{ijk}_{~~~mn} = 0 ~,~
\nn\\
&&
\hspace{2cm}
{U_{(V)}}^{ijklmn}= 0 ~,~
{U_{(VI)}}^{ijk}_{~~~mn}= 0 ~,~
\nn\\
&&
\hspace{2cm}
{V_{(I)}}^{ijk}_{~~~mn} = 0 ~,~
{V_{(II)}}^{ijklmn} = 0 ~,~
{V_{(III)}}^{ijk}_{~~~mn} = 0 ~,~
\nn\\
&&
\hspace{2cm}
Z^{ijklmn} = 0
~~~
\}
~.~
\eea
\\

\subsection{$3$ $U(6)$ vortices on $\mathbb{C}$ \label{s3b6}}

The Hilbert series for the $3$ $U(6)$ vortex master space is given by the Molien integral,
\beal{es2500}
g(t,s,x;\wmaster_{3,6}) = \oint \ud\mu_{SU(3)} \PE\Big[
[0,1]_w [1,0,0,0,0]_x + [1,1]_w s
\Big]~~,
\eea
where $[1,0,0,0,0]_w$ is the fundamental representation of the global $SU(6)$. The integral leads to the following character expansion of the Hilbert series
\beal{es2501}
&&
g(t,s,x;\wmaster_{3,6}) = 
\frac{1}{(1-s^2)(1-s^3)} \times
\nn\\
&&
\sum_{n_0=0}^{\infty}
\sum_{n_1=0}^{\infty}
\sum_{n_2=0}^{\infty}
\sum_{n_3=0}^{\infty}
\Big[
[n_1+n_2+3n_3,n_1+n_2,n_0,0,0]_x s^{n_1+2n_2+3n_3} t^{3n_0+3n_1+3n_2 +3n_3}
\nn\\
&&
\hspace{1cm}
+ [n_1+n_2,n_1+n_2+3n_3+3,n_0,0,0]_x s^{n_1+2n_2+3n_3+3} t^{3n_0+3n_1+3n_2+6n_3+6}
\Big]
~,~
\nn\\
\eea
where $[m_1,m_2,m_3,m_4]_x$ is a character of a $SU(5)$ irreducible representation.

The plethystic logarithm of the Hilbert series is
\beal{es2510}
\PL\Big[
g(t,s,x;\wmaster_{3,6})
\Big] 
&=&
s^2 + s^3 + [0,0,1,0,0]_x t^3
+ [1,1,0,0,0]_x s t^3
+ [1,1,0,0,0]_x s^2 t^3
\nn\\
&&
+ [3,0,0,0,0]_x s^3 t^3
- [1,0,0,0,1]_x t^6
- (
[2,0,0,1,0]_x
\nn\\
&&
+ [0,1,0,1,0]_x
+ [1,0,0,0,1]_x
) s t^6
- (
[1,1,1,0,0]_x 
+ [0,0,2,0,0]_x
\nn\\
&&
+ [2,0,0,1,0]_x
+ [0,1,0,1,0]_x
+ [1,0,0,0,1]_x
) s^2 t^6
\nn\\
&&
+ (
[1,1,0,0,0]_x
+ [0,0,0,1,1]_x
) t^9
- (
[3,0,1,0,0]_x
\nn\\
&&
+2 [1,1,1,0,0]_x
+[0,0,2,0,0]_x
+2 [2,0,0,1,0]_x
+ [0,1,0,1,0]_x
\nn\\
&&
+ [1,0,0,0,1]_x
) s^3 t^6
+\dots
- (
[0,0,2,0,0]_x
+ [1,1,1,0,0]_x
\nn\\
&&
+ [2,0,0,1,0]_x
+ [2,2,0,0,0]_x
+ [3,0,1,0,0]_x
) s^4 t^6 
+ \dots 
\nn\\
&&
- (
[1,1,1,0,0]_x
+ [2,2,0,0,0]_x
+ [3,0,1,0,0]_x
) s^5 t^6
+ \dots
\nn\\
&&
- [2,2,0,0,0]_x s^6 t^6
+ \dots ~~.
\eea
The generators of the vortex moduli space are indicated by the above plethystic logarithm. They are as follows,
\beal{es2520}
s^2 &\ra& u_2=\Tr(\phi^2)
\nn\\
s^3 &\ra& u_3=\Tr(\phi^3)
\nn\\
~[0,0,1,0,0]_x t^3 &\ra&
{B}^{ijk} = \epsilon^{\alpha_1 \alpha_2 \alpha_3} Q_{\alpha_1}^{i} Q_{\alpha_2}^{j} Q_{\alpha_3}^{k}
\nn\\
~[1,1,0,0,0]_x s t^3 &\ra&
\left\{
\ba{l}
{A_{001}}^{i j k} =
\epsilon^{\alpha_1 \alpha_2 \alpha_3}
Q_{\alpha_1}^{i} Q_{\alpha_2}^{j} \phi_{\alpha_3}^{\beta} Q_{\beta}^{k}
\nn\\
\epsilon_{ijkmnp} {A_{001}}^{ijk}=0 
\ea
\right.
\nn\\
~[1,1,0,0,0]_x s^2 t^3 &\ra&
\left\{
\ba{l}
{A_{002}}^{ijk} =
\epsilon^{\alpha_1 \alpha_2 \alpha_3}
Q_{\alpha_1}^{i} Q_{\alpha_2}^{j} \phi_{\alpha_3}^{\beta_1}\phi_{\beta_1}^{\beta_2} Q_{\beta_2}^{k}
\nn\\
\epsilon_{ijkmnp} {A_{002}}^{ijk} = - \frac{1}{3} u_2 \epsilon_{mnprsu} {B}^{rsu}
\nn\\
{A_{011}}^{ijk} =
\epsilon^{\alpha_1 \alpha_2 \alpha_3}
Q_{\alpha_1}^{k} \phi_{\alpha_2}^{\beta_1} Q_{\beta_1}^{i} \phi_{\alpha_3}^{\beta_2} Q_{\beta_2}^{j}
\nn\\
\epsilon_{ijkmnp} {A_{011}}^{ijk}= \frac{1}{6} u_2 \epsilon_{mnprsu} {B}^{rsu}
\nn\\
\ra {A_{002}}^{ijk} = {A_{011}}^{ijk} + \frac{1}{2} u_2 B^{ijk}
\ea\right.
\nn\\
~[3,0,0,0,0]_x s^3 t^3 &\ra&
\left\{\ba{l}
{A_{012}}^{ijk}
= 
\epsilon^{\alpha_1 \alpha_2 \alpha_3} 
Q_{\alpha_1}^{i}
\phi_{\alpha_2}^{\beta_1} Q_{\beta_1}^{j}
\phi_{\alpha_3}^{\beta_2} \phi_{\beta_2}^{\beta_3} Q_{\beta_3}^{k}
\nn\\
{S_{012}}^{ijk} =
{A_{012}}^{ijk}
+ {A_{012}}^{jki}
+ {A_{012}}^{kij}
\ea\right.
~~.
\eea
\\

\paragraph{Quadratic relations:} The terms of the plethystic logarithm corresponding to quadratic relations between master space generators are as follows,
\beal{es2550}
&
- [1,0,0,0,1]_x t^6
&
\nn\\
&
- [0,1,0,1,0]_x s t^6
- [2,0,0,1,0]_x s t^6
- [1,0,0,0,1]_x s t^6
&
\nn\\
&
- [0,1,0,1,0]_x s^2 t^6
- 2 [2,0,0,1,0]_x s^2 t^6
- [0,0,2,0,0]_x s^2 t^6
&
\nn\\
&
- [1,1,1,0,0]_x s^2 t^6
- [1,0,0,0,1]_x s^2 t^6
&
\nn\\
&
- [0,1,0,1,0]_x s^3 t^6
- 2 [2,0,0,1,0]_x s^3 t^6
- [0,0,2,0,0]_x s^3 t^6
&
\nn\\
&
- 2 [1,1,1,0,0]_x s^3 t^6
- [3,0,1,0,0]_x s^3 t^6
&
\nn\\
&
- [2,0,0,1,0]_x s^4 t^6
- [0,0,2,0,0]_x s^4 t^6
- 2 [1,1,1,0,0]_x s^4 t^6
&
\nn\\
&
- [2,2,0,0,0]_x s^4 t^6
- [3,0,1,0,0]_x s^4 t^6
&
\nn\\
&
- [1,1,1,0,0]_x s^5 t^6 - [2,2,0,0,0]_x s^5 t^6 - [3,0,1,0,0]_x s^5 t^6 
&
\nn\\
&
- [2,2,0,0,0]_x s^6t^6
&
~.~
\nn\\
\eea
The quadratic relations are formed by the generators of the vortex master space which are as discussed above as follows,
\beal{es2600}
u_2~,~
u_3~,~
B^{ijk}~,~
{A_{001}}^{ijk}~,~
{A_{002}}^{ijk}~,~
{S_{012}}^{ijk}~.~
\eea
Let us consider the first quadratic relation at order $t^6$ which requires consideration of the following $SU(6)$ representation product,
\beal{es2601}
\text{Sym}^{2} [0,0,1,0,0]_x = [0,0,2,0,0]_x + [1,0,0,0,1]_x ~.~
\eea
The above symmetric product allows us to construct the following quadratic relation:
\begin{itemize}
\item \underline{$-[1,0,0,0,1]_x t^6$ relations:} The following product of generators exactly vanishes,
\beal{es2602}
H_{i}^{~j} = \epsilon_{ipqrsu} B^{pqr} B^{suj} ~,~
\eea
and given that it transforms in the adjoint of $SU(6)$ is exactly the quadratic relation at this order.
\\

\end{itemize}

For the next set of quadratic relations containing the order $s t^6$, we consider the following representation products,
\beal{es2603}
[1,1,0,0,0]_x \times [0,0,1,0,0]_x &= &
[1,1,1,0,0]_x + [2,0,0,1,0]_x + [0,1,0,1,0]_x 
\nn\\
&&
+ [1,0,0,0,1]_x
~.~
\eea
The above tensor product allows us to construct the following quadratic relations:
\begin{itemize}
\item \underline{$-[0,1,0,1,0]_x s t^6$ relations:} At this order, we consider the following generator product,
\beal{es2610}
{R_{(I)}}^{ij}_{~~kl} = {A_{001}}^{ijm} B^{pqr} \epsilon_{mpqrkl} =0 ~,~
\eea
which vanishes exactly. This transforms in the correct representation of $SU(6)$ and hence is the quadratic relation at this order.
\\

\item \underline{$-[2,0,0,1,0]_x s t^6$ relations:} The following generator product vanishes exactly,
\beal{es2611}
{R_{(II)}}^{ij}_{~~kl} = {A_{001}}^{imj} B^{pqr} \epsilon_{mpqrkl} = 0 ~.~
\eea
It transforms in the correct representation and hence is the quadratic relation at this order.
\\

\item \underline{$-[1,0,0,0,1]_x s t^6$ relations:} For this order, the following generator product is considered,
\beal{es2620}
{R_{(III)}}^{i}_{~j} = {A_{001}}^{imn} B^{klp} \epsilon_{mnklpj} =0 ~.~
\eea
The above product exactly vanishes and since it is in the adjoint of $SU(6)$ it is the quadratic relation at this order.
\\

\end{itemize}

The next set of quadratic relations contains the order $s^2 t^6$. For these relations, we consider the following representation products of $SU(6)$,
\beal{es2621}
\text{Sym}^{2} [1,1,0,0,0]_x &=& [2,2,0,0,0]_x + [1,1,1,0,0]_x + [2,0,0,1,0]_x + [0,0,2,0,0]_x ~,~
\nn\\
~[1,1,0,0,0]_x \times [0,0,1,0,0]_x &= &
[1,1,1,0,0]_x + [2,0,0,1,0]_x + [0,1,0,1,0]_x 
\nn\\
&&
+ [1,0,0,0,1]_x ~,~
\nn\\
\text{Sym}^{2} [0,0,1,0,0]_x &=& [0,0,2,0,0]_x + [1,0,0,0,1]_x ~.~
\eea
The following quadratic relations can be identified from the above representation products as follows:
\begin{itemize}

\item \underline{$-[0,1,0,1,0]_x s^2 t^6$ relations:} For this order, we consider the following generator product,
\beal{es2625}
{O_{(I)}}^{ij}_{~~kl} = {A_{002}}^{ijp} B^{qrs} \epsilon_{pqrskl} = 0 ~,~
\eea
which exactly vanishes. The above transforms in the correct representation and therefore is precisely the quadratic relation for this order.
\\

\item \underline{$- 2[2,0,0,1,0]_x s^2 t^6$ relations:} For the quadratic relations at this order, we consider the following generator products,
\beal{es2626}
{O_{(II)}}^{ij}_{~~kl} &=& {A_{001}}^{mni} {A_{001}}^{pqj} \epsilon_{mnpqkl} = 0 ~,~
\nn\\
{O_{(III)}}^{ij}_{~~kl} &=& {A_{002}}^{imj} B^{npq} \epsilon_{mnpqkl} = 0 ~,~
\eea
where the products above both vanish exactly. Given that they transform in the correct representation, we identify them as the 2 quadratic relations at this order.
\\

\item \underline{$- [0,0,2,0,0]_x s^2 t^6$ relations:} The following generator products are helpful in constructing the quadratic relations at this order,
\beal{es2627}
{o_{(I)}}_{ijklmn} &=& \epsilon_{pqsijk} \epsilon_{uvrlmn} {A_{001}}^{pqr} A_{001}^{uvs} ~,~
\nn\\
{o_{(II)}}_{ijklmn} &=& u_2 B^{pqs} B^{uvr} \epsilon_{pqsijk} \epsilon_{uvrlmn} ~.~
\eea
The above products transform in the representation of this order. They satisfy the following quadratic relation,
\beal{es2628} 
{O_{(IV)}}_{ijklmn} = {o_{(I)}}_{ijklmn} - \frac{1}{9} {o_{(II)}}_{ijklmn} ~,~
\eea
which is precisely the relation at this order.
\\

\item \underline{$- [1,1,1,0,0]_x s^2 t^6$ relations:} For this order, we consider the following generator products,
\beal{es2628}
{o_{(III)}}^{ijk}_{~~~lmn} &=& {A_{001}}^{pqk} {A_{001}}^{ijr} \epsilon_{pqrlmn} 
~,~
\nn\\
{o_{(IV)}}^{ijk}_{~~~lmn} &=& {A_{002}}^{ijk} B^{pqr} \epsilon_{pqrlmn} 
~,~
\eea
which transform in the correct representation at this order. The above products satisfy the following quadratic relation,
\beal{es2629}
{O_{(V)}}^{ijk}_{~~~lmn} = {o_{(III)}}^{ijk}_{~~~lmn} - \frac{1}{3} {o_{(IV)}}^{ijk}_{~~~lmn} = 0 ~,~
\eea
which is precisely the relation at this order.
\\

\item \underline{$- [1,0,0,0,1]_x s^2 t^6$ relations:} The following generator product vanishes exactly,
\beal{es2630}
{O_{(VI)}}^{i}_{~j} = {A_{002}}^{pqi} B^{lmn} \epsilon_{pqlmnj} = 0~.~
\eea
The above quadratic relations transform in the adjoint of $SU(6)$ and hence are the quadratic relations for this order.
\\

\end{itemize}

For the next set of quadratic relations at orders containing $s^3 t^6$, we first consider the following $SU(6)$ representation products,
\beal{es2640}
~[1,1,0,0,0]_x \times  [1,1,0,0,0]_x &=& [2,2,0,0,0]_x + [3,0,1,0,0]_x + [0,3,0,0,0]_x 
\nn\\
&&
+ 2 [1,1,1,0,0]_x + [2,0,0,1,0]_x + [0,0,2,0,0]_x 
\nn\\
&&
+ [0,1,0,1,0]_x ~,~
\nn\\
~[3,0,0,0,0]_x \times [0,0,1,0,0]_x &=& [3,0,1,0,0]_x + [2,0,0,1,0]_x ~,~
\nn\\
~[1,1,0,0,0]_x \times [0,0,1,0,0]_x &=& [1,1,1,0,0]_x + [2,0,0,1,0]_x + [0,1,0,1,0]_x 
\nn\\
&&
+ [1,0,0,0,1]_x~,~
\nn\\
\text{Sym}^2 [0,0,1,0,0]_x &=& [0,0,2,0,0]_x + [1,0,0,0,1]_x ~.~
\eea
The above representation products help us in constructing the following quadratic relations between master space generators at order $s^3 t^6$:
\begin{itemize}

\item \underline{$-[0,1,0,1,0]_x s^3 t^6$ relations:} The following generator product vanishes exactly,
\beal{es2650}
{P_{(I)}}^{ij}_{~~kl} = {A_{001}}^{ijm} {A_{002}}^{pqr} \epsilon_{mpqrkl} = 0 ~,~
\eea
and transforms in the representation of this order. Accordingly, it is exactly the quadratic relation we are looking for this order.
\\

\item \underline{$-2 [2,0,0,1,0]_x s^3 t^6$ relations:} The following generator products are relevant for the quadratic relations at this order,
\beal{es2651}
{P_{(II)}}^{ij}_{~~kl} &=&
{A_{001}}^{mni} {A_{002}}^{pqj} \epsilon_{mnpqkl} = 0 ~,~
\nn\\
{P_{(III)}}^{ij}_{~~kl} &=&
{S_{012}}^{ijm} B^{npq} \epsilon_{mnpqkl} = 0 ~.~
\eea
Both above vanish and satisfy the correct transformation property for this order. They are precisely the two quadratic relations at this order.
\\

\item \underline{$-[0,0,2,0,0]_x s^3 t^6$ relations:} For this order, we consider the following generator products,
\beal{es2652}
{p_{(I)}}_{ijklmn} &=&
{A_{001}}^{pqu} {A_{002}}^{rsv} \epsilon_{pqvijk} \epsilon_{rsulmn} ~,~
\nn\\
{p_{(II)}}_{ijklmn} &=& 
u_3 B^{pqu} B^{rsv} \epsilon_{pquijk} \epsilon_{rsvlmn} ~,~
\eea
which transform in the correct representation corresponding to this order. The above products satisfy the following quadratic relation 
\beal{es2653}
{P_{(IV)}}_{ijklmn} = {p_{(I)}}_{ijklmn} - \frac{1}{9} {p_{(II)}}_{ijklmn} ~,~
\eea
which is the relation for this order.
\\

\item \underline{$- 2 [1,1,1,0,0]_x s^3 t^6$ relations:}  The following generator products are useful in constructing the quadratic relations for this order,
\beal{es2654}
{p_{(III)}}^{ijk}_{~~~lmn} &=&
{A_{001}}^{ijp} {A_{002}}^{qrk} \epsilon_{pqrlmn} ~,~
\nn\\
{p_{(IV)}}^{ijk}_{~~~lmn} &=&
{A_{002}}^{ijp} {A_{001}}^{qrk} \epsilon_{pqrlmn} ~,~
\nn\\
{p_{(V)}}^{ijk}_{~~~lmn} &=&
u_2 {A_{001}}^{ijp} B^{qrk} \epsilon_{pqrlmn} ~.~
\eea
The above products transform in the correct representation for this order. The quadratic relations formed by the above are
\beal{es2655}
{P_{(V)}}^{ijk}_{~~~lmn} = {p_{(III)}}^{ijk}_{~~~lmn} - {p_{(IV)}}^{ijk}_{~~~lmn} = 0 ~,~
\nn\\
{P_{(VI)}}^{ijk}_{~~~lmn} = {p_{(III)}}^{ijk}_{~~~lmn} - \frac{1}{2} {p_{(V)}}^{ijk}_{~~~lmn} = 0 ~,~
\eea
exactly corresponding to the two expected quadratic relations at this order.
\\

\item \underline{$-[3,0,1,0,0]_x s^3 t^6$ relations:} For this order, we consider the following generator products,
\beal{es2660}
{p_{(VI)}}^{ijk}_{~~~lmn}
&=&
 (
{A_{001}}^{ipj} {A_{002}}^{qrk} 
+  {A_{001}}^{jpk} {A_{002}}^{qri} 
+ {A_{001}}^{kpi} {A_{002}}^{qrj} 
)\epsilon_{pqrlmn}
~,~
\nn\\
{p_{(VII)}}^{ijk}_{~~~lmn}
&=&
{S_{012}}^{ijk} B^{pqr} \epsilon_{pqrlmn} ~,~
\eea
which transform in the correct representation for this order. The above products satisfy the following quadratic relation 
\beal{es2661}
{P_{(VII)}}^{ijk}_{~~~lmn} = {p_{(VI)}}^{ijk}_{~~~lmn} + \frac{1}{3} {p_{(VII)}}^{ijk}_{~~~lmn} ~,~
\eea
which is precisely the relation for this order.
\\

\end{itemize}

We can now consider the next set of quadratic generator relations which are at orders containing $s^4 t^6$. In order to construct the relations, we consider the following $SU(6)$ representation products,
\beal{es2662}
\text{Sym}^{2} [1,1,0,0,0]_x &=& [2,2,0,0,0]_x + [1,1,1,0,0]_x + [2,0,0,1,0]_x + [0,0,2,0,0]_x ~,~
\nn\\
~[1,1,0,0,0]_x \times [3,0,0,0,0]_x &=& [4,1,0,0,0]_x + [2,2,0,0,0]_x + [3,0,1,0,0]_x 
\nn\\
&&
+ [1,1,1,0,0]_x ~,~
\nn\\
~[1,1,0,0,0]_x \times [0,0,1,0,0]_x &=& [1,1,1,0,0]_x + [2,0,0,1,0]_x + [0,1,0,1,0]_x 
\nn\\
&&
+ [1,0,0,0,1]_x ~,~
\nn\\
~\text{Sym}^{2} [0,0,1,0,0]_x &=& [0,0,2,0,0]_x + [1,0,0,0,1]_x ~.~
\eea
With the above representation products in mind, we construct the following quadratic relations:
\begin{itemize}

\item \underline{$-[2,0,0,1,0]_x s^4 t^6$ relations:} For this order, we consider the following generator product,
\beal{es2665}
{U_{(I)}}^{ij}_{~~kl} = {A_{002}}^{pqi} {A_{002}}^{mnj} \epsilon_{pqmnkl} = 0~,~
\eea	
which vanishes exactly. This is precisely the quadratic relation at this order.
\\

\item \underline{$-[0,0,2,0,0]_x s^4 t^6$ relations:} In order to construct the quadratic relation for this order, we consider the following products,
\beal{es2666}
{u_{(I)}}_{ijklmn} 
&=& 
{A_{002}}^{pqr} {A_{002}}^{uvs} \epsilon_{pqsijk} \epsilon_{uvrlmn} 
~,~
\nn\\
{u_{(II)}}_{ijklmn}
 &=&
u_2 u_2 B^{pqr} B^{uvw} \epsilon_{pqrijk} \epsilon_{uvwlmn}
~.~
\eea
The above products transform in the correct $SU(6)$ representation of this order. They satisfy the following quadratic relation,
\beal{es2667}
{U_{(II)}}_{ijklmn} &=& {u_{(I)}}_{ijklmn}  -  \frac{1}{3} {u_{(II)}}_{ijklmn} = 0 ~,~
\eea
which is precisely the relation we are looking for here.
\\

\item \underline{$-2 [1,1,1,0,0]_x s^4 t^6$ relations:} Corresponding to this order, there are two distinct quadratic relations. In order to construct them, we consider the following generator products,
\beal{es2670}
{u_{(III)}}^{ijk}_{~~~lmn} &=&
{A_{002}}^{ijp} {A_{002}}^{qrk} \epsilon_{pqrlmn} 
~,~
\nn\\
{u_{(IV)}}^{ijk}_{~~~lmn} &=& 
{A_{001}}^{ijp} {S_{012}}^{qrk} \epsilon_{pqrlmn} 
~,~
\nn\\
{u_{(V)}}^{ijk}_{~~~lmn} &=&
u_2 {A_{001}}^{ijk} {A_{001}}^{pqr} \epsilon_{pqrlmn} 
~,~
\nn\\
{u_{(VI)}}^{ijk}_{~~~lmn} &=&
u_3 {A_{001}}^{ijk} B^{pqr} \epsilon_{pqrlmn}
~,~
\eea
which transform in the representation of this order. The above products form the following two quadratic relations,
\beal{es2671}
{U_{(III)}}^{ijk}_{~~~lmn} &=&
{u_{(V)}}^{ijk}_{~~~lmn} - 2 {u_{(III)}}^{ijk}_{~~~lmn} + \frac{2}{9} {u_{(VI)}}^{ijk}_{~~~lmn} = 0
~,~
\nn\\
{U_{(IV)}}^{ijk}_{~~~lmn} &=&
{u_{(IV)}}^{ijk}_{~~~lmn} + \frac{1}{6} {u_{(VI)}}^{ijk}_{~~~lmn} = 0 ~.~
\eea
The two quadratic relations above are precisely corresponding to this order.
\\

\item \underline{$-[2,2,0,0,0]_x s^4 t^6$ relations:} The relation at this order requires us to have a look at the following product of generators,
\beal{es2672}
{u_{(VII)}}^{ijklmn} = {A_{001}}^{ijk} {S_{012}}^{lmn} ~,~
\eea
where we antisymmetrize on the indices $[kl]$ and $[mn]$ as follows,
\beal{es2673}
{u_{(VIII)}}^{ijklmn} = {u_{(VII)}}^{ijklmn} - {u_{(VII)}}^{ijlkmn} -  {u_{(VII)}}^{ijklnm} + {u_{(VII)}}^{ijlknm} ~.~
\nn\\
\eea
A further symmetrization on the two paired indices $[kl]$ and $[mn]$ leads to the following
\beal{es2674}
{U_{(V)}}^{ijklmn} = {u_{(VIII)}}^{ijklmn} + {u_{(VIII)}}^{ijmnkl} = 0
\eea
which exactly vanishes. Given that the above expression is precisely of order $-[2,2,0,0,0]_x s^4 t^6$, this is the quadratic relation of generators we are looking for.
\\

\item \underline{$-[3,0,1,0,0]_x s^4 t^6$ relations:} The quadratic relation at this order is formed by
\beal{es2680}
{U_{(VI)}}^{ijk}_{~~~lmn} = 
(
{A_{001}}^{ipq} {S_{012}}^{rjk}
+ {A_{001}}^{jpq} {S_{012}}^{rki}
+ {A_{001}}^{jpq} {S_{012}}^{rij} 
) \epsilon_{pqrlmn}
= 0
~,~
\nn\\
\eea
where the above contains the symmetrization of the generator product
\beal{es2681}
{A_{001}}^{ipq} {S_{012}}^{rjk} \epsilon_{pqrlmn}
\eea
in the indices $ijk$. The above quadratic relation satisfies precisely the transformation properties for this order and is the relation we are looking for.
\\
\end{itemize}

The next set of quadratic relation are at orders which contain $s^5 t^6$. We consider the following representation products in order to construct the relations,
\beal{es2690}
~[3,0,0,0,0]_x \times [1,1,0,0,0]_x  &=&
[4,1,0,0,0]_x + [2,2,0,0,0]_x + [3,0,1,0,0]_x 
\nn\\
&&
+ [1,1,1,0,0]_x
~,~
\nn\\
~[1,1,0,0,0]_x \times [1,1,0,0,0]_x  &=&
[2,2,0,0,0]_x + [3,0,1,0,0]_x + [0,3,0,0,0]_x 
\nn\\
&&
+ 2 [1,1,1,0,0]_x 
+ [2,0,0,1,0]_x 
+ [0,0,2,0,0]_x 
\nn\\
&&
+ [0,1,0,1,0]_x
~,~
\nn\\
~[3,0,0,0,0]_x \times [0,0,1,0,0]_x  &=&
[3,0,1,0,0]_x + [2,0,0,1,0]_x 
~,~
\nn\\
~[1,1,0,0,0]_x \times [0,0,1,0,0]_x  &=&
[1,1,1,0,0]_x + [2,0,0,1,0]_x + [0,1,0,1,0]_x 
\nn\\
&&
+ [1,0,0,0,1]_x
~,~
\nn\\
\text{Sym}^{2} [1,1,0,0,0]_x &=&
[2,2,0,0,0]_x + [1,1,1,0,0]_x + [2,0,0,1,0]_x 
\nn\\
&&
+ [0,0,2,0,0]_x
~,~
\nn\\
\text{Sym}^{2} [0,0,1,0,0]_x &=&
[0,0,2,0,0]_x + [1,0,0,0,1]_x
~~.
\\
\nn
\eea
The above products are used to construct the following quadratic relations of generators for vortex master spaces:
\begin{itemize}
\item \underline{$-[1,1,1,0,0]_x s^5 t^6$ relations:} For the quadratic relation at this order, we consider the following products of generators,
\beal{es2805}
{v_{(I)}}^{ijk}_{~~~lmn} &=& 
{A_{002}}^{ijp} {S_{012}}^{qrk}
\epsilon_{pqrlmn}
~,~
\nn\\
{v_{(II)}}^{ijk}_{~~~lmn} &=&
u_3 {A_{001}}^{ijp} {A_{001}}^{qrk}
\epsilon_{pqrlmn}
~,~
\eea
which transform in the correct irreducible representation of this order. The above products satisfy the following quadratic relation,
\beal{es2806}
{V_{(I)}}^{ijk}_{~~~lmn} = {v_{(I)}}^{ijk}_{~~~lmn} + \frac{1}{2} {v_{(II)}}^{ijk}_{~~~lmn} = 0 ~.~
\eea
The above precisely is the quadratic relation we are looking for at this order.
\\

\item \underline{$-[2,2,0,0,0]_x s^5 t^6$ relations:} For the quadratic relation at this order, we need to consider several generator products with various symmetrizations and anti-symmetrizations of indices. The first product to consider is the following,
\beal{es2810}
{v_{(III)}}^{ijklmn} = ( {A_{002}}^{ijk} - \frac{1}{2} u_2 B^{ijk}) {S_{012}}^{lmn} ~,~
\eea
where we recall that ${A_{011}}^{ijk} = {A_{002}}^{ijk} - \frac{1}{2} u_2 B^{ijk}$.
Above, we anti-symmetrize in the indices $[kl]$ and $[mn]$ to give
\beal{es2811}
{v_{(IV)}}^{ijklmn} = {v_{(III)}}^{ijklmn} - {v_{(III)}}^{ijlkmn} - {v_{(III)}}^{ijklnm} + {v_{(III)}}^{ijlknm}  ~,~
\eea
and further anti-symmetrize in the pairs of indices $[ij]$ and $[mn]$ to obtain,
\beal{es2812}
{v_{(V)}}^{ijklmn} = {v_{(IV)}}^{ijklmn} - {v_{(IV)}}^{mnklij} ~.~
\eea
The second generator product to consider is the following,
\beal{es2813}
{v_{(VI)}}^{ijklmn} = u_3 {A_{001}}^{ijk} {A_{001}}^{lmn} ~,~
\eea
which we anti-symmetrize in the indices $[kl]$ and $[mn]$ giving,
\beal{es2814}
{v_{(VII)}}^{ijklmn} = {v_{(VI)}}^{ijklmn} - {v_{(VI)}}^{ijlkmn} - {v_{(VI)}}^{ijklnm} + {v_{(VI)}}^{ijlknm} ~.~
\eea
A further symmetrization of the pair of indices $[ij]$ and $[kl]$ gives
\beal{es2815}
{v_{(VIII)}}^{ijklmn} = {v_{(VII)}}^{ijklmn} + {v_{(VII)}}^{klijmn} ~.`
\eea
From the above generator products, which correctly transform at this order, we can construct the following quadratic relation,
\beal{es2816}
{V_{(II)}}^{ijklmn} = {v_{(V)}}^{ijklmn} + {v_{(VIII)}}^{ijmnkl} = 0 ~.~
\eea
This is precisely we are looking for at this order.
\\

\item \underline{$-[3,0,1,0,0]_x s^5 t^6$ relations:} The following generator products are of importance in order to construct the quadratic relation at this order. The first product is as follows
\beal{es2817}
{v_{(IX)}}^{ijk}_{~~~lmn} = {A_{002}}^{ipq} {S_{012}}^{rjk} \epsilon_{pqrlmn} ~,~
\eea    
which we symmetrize in the indices $ijk$ as follows,
\beal{es2818}
{v_{(X)}}^{ijk}_{~~~lmn} = {v_{(IX)}}^{ijk}_{~~~lmn} + {v_{(IX)}}^{jki}_{~~~lmn} + {v_{(IX)}}^{kij}_{~~~lmn} ~.~
\eea
The second generator product which we need to consider is the following,
\beal{es2819}
{v_{(XI)}}^{ijk}_{~~~lmn} = u_2 {S_{012}}^{ijk} B^{pqr} \epsilon_{pqrlmn} ~.~
\eea
From the above products, we construct the following quadratic relation,
\beal{es2921}
{V_{(III)}}^{ijk}_{~~~lmn} = {v_{(X)}}^{ijk}_{~~~lmn}  - \frac{1}{3} {v_{(XI)}}^{ijk}_{~~~lmn} = 0 ~.~
\eea
This is precisely the relation we are looking for at this order of the plethystic logarithm. 
\\

\end{itemize}

The next order of $s^6 t^6$ leads us to the following products of $SU(6)$ representations,
\beal{es2820}
\text{Sym}^{2} [3,0,0,0,0]_x &=& [6,0,0,0,0]_x + [2,2,0,0,0]_x ~,~
\nn\\
\text{Sym}^{2} [1,1,0,0,0]_x &=& [2,2,0,0,0]_x + [1,1,1,0,0]_x + [2,0,0,1,0]_x 
\nn\\
&&
+ [0,0,2,0,0]_x ~,~
\nn\\
~[1,1,0,0,0]_x \times [3,0,0,0,0]_x &=&
[4,1,0,0,0]_x + [2,2,0,0,0]_x + [3,0,1,0,0]_x 
\nn\\
&&
+ [1,1,1,0,0]_x ~,~
\nn\\
~[1,1,0,0,0]_x \times [1,1,0,0,0]_x &=& 
[2,2,0,0,0]_x + [3,0,1,0,0]_x + [0,3,0,0,0]_x 
\nn\\
&&
+ 2 [1,1,1,0,0]_x + [2,0,0,1,0]_x + [0,0,2,0,0]_x 
\nn\\
&&
+ [0,1,0,1,0]_x 
~.~
\eea
From the above, we can construct candidate products of generators of the vortex master space, in order to identify quadratic relations amongst them: 
\begin{itemize}
\item \underline{$-[2,2,0,0,0]_x s^6 t^6$ relations:} For the quadratic relations at this order, we consider the following product of generators of the vortex master space,
\beal{es2821}
{z_{(I)}}^{ijklmn} = {S_{012}}^{ijk} {S_{012}}^{lmn} ~.~
\eea
We anti-symmetrize the above on the indices $[kl]$ and $[mn]$ 
\beal{es2822}
{z_{(II)}}^{ijklmn} = {z_{(I)}}^{ijklmn} - {z_{(I)}}^{ijlkmn} - {z_{(I)}}^{ijklnm} + {z_{(I)}}^{ijlknm} ~,~
\eea
and further symmetrize on the pairs of indices $[kl]$ and $[mn]$ to give
\beal{es2823}
Z^{ijklmn} = {z_{(II)}}^{ijklmn} + {z_{(II)}}^{ijmnkl} = 0 ~.~
\eea
The above quadratic relation vanishes non-trivially and by construction transforms in the representation of this order of the plethystic logarithm. The above is the quadratic relation corresponding to this order.
\\
\end{itemize}

\paragraph{Vortex moduli space:} Given the above explicit computation of the quadratic relations between generators, we can proceed in identifying the vortex moduli space for the $3$ $U(6)$ vortex theory. The vortex moduli space can be expressed as a $\mathbb{C}^{*}$ projection as follows,
\beal{es2823xx1}
&&
\wV_{3,6} = \wmaster_{3,6}/\{
B^{ijk} \simeq \lambda^{3} B^{ijk} ,
{A_{001}}^{ijk} \simeq \lambda^{3} {A_{001}}^{ijk} ,
\nn\\
&&
\hspace{3cm}
{A_{002}}^{ijk} \simeq \lambda^{3} {A_{002}}^{ijk} ,
{S_{012}}^{i j k} \simeq \lambda^{3} {S_{012}}^{i j k} 
\}
~.~
\eea
The master space of the vortex theory can be identified with the help of the quadratic relations as follows,
\beal{es2823xx2}
&&
\wmaster_{3,6} = \mathbb{C}[
u_2, u_3, B^{ijk}, {A_{001}}^{ijk}, {A_{002}}^{ijk}, {S_{012}}^{ijk}
]/\{
\nn\\
&&
\hspace{2cm}
H_{i}^{~j} = 0 ~,~
\nn\\
&&
\hspace{2cm}
{R_{(I)}}^{ij}_{~~kl} = 0 ~,~
{R_{(II)}}^{ij}_{~~kl} = 0 ~,~
{R_{(III)}}^{i}_{~j} = 0 ~,~
\nn\\
&&
\hspace{2cm}
{O_{(I)}}^{ij}_{~~kl} = 0 ~,~
{O_{(II)}}^{ij}_{~~kl} = 0 ~,~
{O_{(III)}}^{ij}_{~~kl} = 0 ~,~
{O_{(IV)}}_{ijklmn} = 0 ~,~
\nn\\
&&
\hspace{2cm}
{O_{(V)}}^{ijk}_{~~~lmn} = 0 ~,~
{O_{(VI)}}^{i}_{~j} = 0 ~,~
\nn\\
&&
\hspace{2cm}
{P_{(I)}}^{ij}_{~~kl} = 0 ~,~
{P_{(II)}}^{ij}_{~~kl} = 0 ~,~
{P_{(III)}}^{ij}_{~~kl} = 0 ~,~
{P_{(IV)}}_{ijklmn} = 0 ~,~
\nn\\
&&
\hspace{2cm}
{P_{(V)}}^{ijk}_{~~~lmn} = 0 ~,~
{P_{(VI)}}^{ijk}_{~~~lmn} = 0 ~,~
{P_{(VII)}}^{ijk}_{~~~lmn} = 0 ~,~
\nn\\
&&
\hspace{2cm}
{U_{(I)}}^{ij}_{~~kl} = 0 ~,~
{U_{(II)}}_{ijklmn} = 0 ~,~
{U_{(III)}}^{ijk}_{~~~lmn} = 0 ~,~
{U_{(IV)}}^{ijk}_{~~~lmn} = 0 ~,~
\nn\\
&&
\hspace{2cm}
{U_{(V)}}^{ijklmn}= 0 ~,~
{U_{(VI)}}^{ijk}_{~~~lmn}= 0 ~,~
\nn\\
&&
\hspace{2cm}
{V_{(I)}}^{ijk}_{~~~lmn} = 0 ~,~
{V_{(II)}}^{ijklmn} = 0 ~,~
{V_{(III)}}^{ijk}_{~~~lmn} = 0 ~,~
\nn\\
&&
\hspace{2cm}
Z^{ijklmn} = 0
~~~
\}
~.~
\eea
\\

\subsection{$3$ $U(7)$ vortices on $\mathbb{C}$ \label{s3b7}}

The Hilbert series for the $3$ $U(7)$ vortex master space can be obtained by solving the following Molien integral,
\beal{es72500}
g(t,s,x;\wmaster_{3,7}) = \oint \ud\mu_{SU(3)} \PE\Big[
[0,1]_w [1,0,0,0,0,0]_x + [1,1]_w s
\Big]~~,
\eea
where $[1,0,0,0,0,0]_w$ is the fundamental representation of the global $SU(7)$. The integral leads to the following character expansion of the Hilbert series
\beal{es72501}
&&
g(t,s,x;\wmaster_{3,7}) = 
\frac{1}{(1-s^2)(1-s^3)} \times
\nn\\
&&
\sum_{n_0=0}^{\infty}
\sum_{n_1=0}^{\infty}
\sum_{n_2=0}^{\infty}
\sum_{n_3=0}^{\infty}
\Big[
[n_1+n_2+3n_3,n_1+n_2,n_0,0,0,0]_x s^{n_1+2n_2+3n_3} t^{3n_0+3n_1+3n_2 +3n_3}
\nn\\
&&
\hspace{1cm}
+ [n_1+n_2,n_1+n_2+3n_3+3,n_0,0,0,0]_x s^{n_1+2n_2+3n_3+3} t^{3n_0+3n_1+3n_2+6n_3+6}
\Big]
~,~
\nn\\
\eea
where $[m_1,m_2,m_3,m_4,m_5,m_6]_x$ is a character of a $SU(7)$ irreducible representation with highest weights $m_1,\dots,m_6$.

The plethystic logarithm of the Hilbert series is
\beal{es72510}
\PL\Big[
g(t,s,x;\wmaster_{3,7})
\Big] 
&=&
s^2 + s^3 + [0,0,1,0,0,0]_x t^3
+ [1,1,0,0,0,0]_x s t^3
\nn\\
&&
+ [1,1,0,0,0,0]_x s^2 t^3
+ [3,0,0,0,0,0]_x s^3 t^3
\nn\\
&&
- [1,0,0,0,1,0]_x t^6
- (
[2,0,0,1,0,0]_x
+ [0,1,0,1,0,0]_x
\nn\\
&&
+ [1,0,0,0,1,0]_x
) s t^6
- (
[1,1,1,0,0,0]_x 
+ [0,0,2,0,0,0]_x
\nn\\
&&
+ [2,0,0,1,0,0]_x
+ [0,1,0,1,0,0]_x
+ [1,0,0,0,1,0]_x
) s^2 t^6
\nn\\
&&
+ (
[1,1,0,0,0,0]_x
+ [0,0,0,1,1,0]_x
) t^9
- (
[3,0,1,0,0,0]_x
\nn\\
&&
+2 [1,1,1,0,0,0]_x
+[0,0,2,0,0,0]_x
+2 [2,0,0,1,0,0]_x
\nn\\
&&
+ [0,1,0,1,0,0,0]_x
+ [1,0,0,0,1,0,0]_x
) s^3 t^6
+\dots
\nn\\
&&
- (
[0,0,2,0,0,0]_x
+ [1,1,1,0,0,0]_x
+ [2,0,0,1,0,0]_x
\nn\\
&&
+ [2,2,0,0,0,0]_x
+ [3,0,1,0,0,0]_x
) s^4 t^6 
+ \dots 
\nn\\
&&
- (
[1,1,1,0,0]_x
+ [2,2,0,0,0]_x
+ [3,0,1,0,0]_x
) s^5 t^6
+ \dots
\nn\\
&&
- [2,2,0,0,0]_x s^6 t^6
+ \dots ~~.
\eea
The generators of the vortex master space are indicated by the above plethystic logarithm. They are as follows,
\beal{es72520}
s^2 &\ra& u_2=\Tr(\phi^2)
\nn\\
s^3 &\ra& u_3=\Tr(\phi^3)
\nn\\
~[0,0,1,0,0,0]_x t^3 &\ra&
{B}^{ijk} = \epsilon^{\alpha_1 \alpha_2 \alpha_3} Q_{\alpha_1}^{i} Q_{\alpha_2}^{j} Q_{\alpha_3}^{k}
\nn\\
~[1,1,0,0,0,0]_x s t^3 &\ra&
\left\{
\ba{l}
{A_{001}}^{i j k} =
\epsilon^{\alpha_1 \alpha_2 \alpha_3}
Q_{\alpha_1}^{i} Q_{\alpha_2}^{j} \phi_{\alpha_3}^{\beta} Q_{\beta}^{k}
\nn\\
\epsilon_{ijkmnpo} {A_{001}}^{ijk}=0 
\ea
\right.
\nn\\
~[1,1,0,0,0,0]_x s^2 t^3 &\ra&
\left\{
\ba{l}
{A_{002}}^{ijk} =
\epsilon^{\alpha_1 \alpha_2 \alpha_3}
Q_{\alpha_1}^{i} Q_{\alpha_2}^{j} \phi_{\alpha_3}^{\beta_1}\phi_{\beta_1}^{\beta_2} Q_{\beta_2}^{k}
\nn\\
\epsilon_{ijkmnpo} {A_{002}}^{ijk} = - \frac{1}{3} u_2 \epsilon_{mnporsu} {B}^{rsu}
\nn\\
{A_{011}}^{ijk} =
\epsilon^{\alpha_1 \alpha_2 \alpha_3}
Q_{\alpha_1}^{k} \phi_{\alpha_2}^{\beta_1} Q_{\beta_1}^{i} \phi_{\alpha_3}^{\beta_2} Q_{\beta_2}^{j}
\nn\\
\epsilon_{ijkmnpo} {A_{011}}^{ijk}= \frac{1}{6} u_2 \epsilon_{mnporsu} {B}^{rsu}
\nn\\
\ra {A_{002}}^{ijk} = {A_{011}}^{ijk} + \frac{1}{2} u_2 B^{ijk}
\ea\right.
\nn\\
~[3,0,0,0,0,0]_x s^3 t^3 &\ra&
\left\{\ba{l}
{A_{012}}^{ijk}
= 
\epsilon^{\alpha_1 \alpha_2 \alpha_3} 
Q_{\alpha_1}^{i}
\phi_{\alpha_2}^{\beta_1} Q_{\beta_1}^{j}
\phi_{\alpha_3}^{\beta_2} \phi_{\beta_2}^{\beta_3} Q_{\beta_3}^{k}
\nn\\
{S_{012}}^{ijk} =
{A_{012}}^{ijk}
+ {A_{012}}^{jki}
+ {A_{012}}^{kij}
\ea\right.
~~.
\eea
\\

\paragraph{Quadratic relations:} The plethystic logarithm indicates the quadratic relations formed amongst the generators,
\beal{es72550}
&
- [1,0,0,0,1,0]_x t^6
&
\nn\\
&
- [0,1,0,1,0,0]_x s t^6
- [2,0,0,1,0,0]_x s t^6
- [1,0,0,0,1,0]_x s t^6
&
\nn\\
&
- [0,1,0,1,0,0]_x s^2 t^6
- 2 [2,0,0,1,0,0]_x s^2 t^6
- [0,0,2,0,0,0]_x s^2 t^6
&
\nn\\
&
- [1,1,1,0,0,0]_x s^2 t^6
- [1,0,0,0,1,0]_x s^2 t^6
&
\nn\\
&
- [0,1,0,1,0,0]_x s^3 t^6
- 2 [2,0,0,1,0,0]_x s^3 t^6
- [0,0,2,0,0,0]_x s^3 t^6
&
\nn\\
&
- 2 [1,1,1,0,0,0]_x s^3 t^6
- [3,0,1,0,0,0]_x s^3 t^6
&
\nn\\
&
- [2,0,0,1,0,0]_x s^4 t^6
- [0,0,2,0,0,0]_x s^4 t^6
- 2 [1,1,1,0,0,0]_x s^4 t^6
&
\nn\\
&
- [2,2,0,0,0,0]_x s^4 t^6
- [3,0,1,0,0,0]_x s^4 t^6
&
\nn\\
&
- [1,1,1,0,0,0]_x s^5 t^6 - [2,2,0,0,0,0]_x s^5 t^6 - [3,0,1,0,0,0]_x s^5 t^6 
&
\nn\\
&
- [2,2,0,0,0,0]_x s^6t^6
&
~.~
\nn\\
\eea
The quadratic relations are formed by the generators of the vortex master space which are:
\beal{es72600}
u_2~,~
u_3~,~
B^{ijk}~,~
{A_{001}}^{ijk}~,~
{A_{002}}^{ijk}~,~
{S_{012}}^{ijk}~.~
\eea
For the first quadratic relation at order $t^6$ we consider the following $SU(7)$ representation product,
\beal{es72601}
\text{Sym}^{2} [0,0,1,0,0,0]_x = [0,0,2,0,0,0]_x + [1,0,0,0,1,0]_x ~.~
\eea
The above symmetric product allows us to construct the following quadratic relation:
\begin{itemize}
\item \underline{$-[1,0,0,0,1,0]_x t^6$ relations:} We consider the following generator product,
\beal{es72602}
H_{ij}^{~~k} = \epsilon_{ijpqrsu} B^{pqr} B^{suk} = 0 ~,~
\eea
which vanishes exactly. It is exactly the quadratic relation at this order.
\\

\end{itemize}

For the second set of quadratic relations containing the order $s t^6$, we consider the following representation products,
\beal{es72603}
[1,1,0,0,0,0]_x \times [0,0,1,0,0,0]_x &= &
[1,1,1,0,0,0]_x + [2,0,0,1,0,0]_x 
\nn\\
&&
+ [0,1,0,1,0,0]_x  + [1,0,0,0,1,0]_x
~.~
\eea
The above tensor product guides us in constructing the following quadratic relations:
\begin{itemize}
\item \underline{$-[0,1,0,1,0,0]_x s t^6$ relations:} We consider the following generator product for this order,
\beal{es72610}
{R_{(I)}}^{ij}_{~~klo} = {A_{001}}^{ijm} B^{pqr} \epsilon_{mpqrklo} =0 ~,~
\eea
which vanishes exactly. This transforms in the correct representation of $SU(7)$ and is the quadratic relation at this order.
\\

\item \underline{$-[2,0,0,1,0,0]_x s t^6$ relations:} The following generator product vanishes exactly,
\beal{es72611}
{R_{(II)}}^{ij}_{~~klo} = {A_{001}}^{imj} B^{pqr} \epsilon_{mpqrklo} = 0 ~.~
\eea
It is in the correct representation and hence is the quadratic relation at this order.
\\

\item \underline{$-[1,0,0,0,1,0]_x s t^6$ relations:} For this order, the following generator product is considered,
\beal{es72620}
{R_{(III)}}^{i}_{~jo} = {A_{001}}^{imn} B^{klp} \epsilon_{mnklpjo} =0 ~.~
\eea
The above product exactly vanishes. It is the quadratic relation at this order.
\\

\end{itemize}

The next set of quadratic relations contains the order $s^2 t^6$. We consider the following $SU(7)$ representation products in order to construct the relations,
\beal{es72621}
\text{Sym}^{2} [1,1,0,0,0,0]_x &=& [2,2,0,0,0,0]_x + [1,1,1,0,0,0]_x + [2,0,0,1,0,0]_x 
\nn\\
&&
+ [0,0,2,0,0,0]_x ~,~
\nn\\
~[1,1,0,0,0,0]_x \times [0,0,1,0,0,0]_x &= &
[1,1,1,0,0,0]_x + [2,0,0,1,0,0]_x + [0,1,0,1,0,0]_x 
\nn\\
&&
+ [1,0,0,0,1,0]_x ~,~
\nn\\
\text{Sym}^{2} [0,0,1,0,0,0]_x &=& [0,0,2,0,0,0]_x + [1,0,0,0,1,0]_x ~.~
\eea
The above representation products guide us in constructing the following quadratic relations:
\begin{itemize}

\item \underline{$-[0,1,0,1,0,0]_x s^2 t^6$ relations:} We consider the following generator product for this order,
\beal{es72625}
{O_{(I)}}^{ij}_{~~klo} = {A_{002}}^{ijp} B^{qrs} \epsilon_{pqrsklo} = 0 ~,~
\eea
which exactly vanishes. The above is precisely the quadratic relation for this order.
\\

\item \underline{$- 2[2,0,0,1,0,0]_x s^2 t^6$ relations:} We consider the following generator products for the quadratic relation at this order,
\beal{es72626}
{O_{(II)}}^{ij}_{~~klo} &=& {A_{001}}^{mni} {A_{001}}^{pqj} \epsilon_{mnpqklo} = 0 ~,~
\nn\\
{O_{(III)}}^{ij}_{~~klo} &=& {A_{002}}^{imj} B^{npq} \epsilon_{mnpqklo} = 0 ~,~
\eea
where the products above both vanish exactly. The above are the two distinct quadratic relations at this order.
\\

\item \underline{$- [0,0,2,0,0,0]_x s^2 t^6$ relations:} The following generator products are helpful in constructing the quadratic relations at this order,
\beal{es72627}
{o_{(I)}}_{ijklmn o_1 o_2} &=& \epsilon_{pqsijk o_1} \epsilon_{uvrlmn o_2} {A_{001}}^{pqr} A_{001}^{uvs} ~,~
\nn\\
{o_{(II)}}_{ijklmn o_1 o_2} &=& u_2 B^{pqs} B^{uvr} \epsilon_{pqsijk o_1} \epsilon_{uvrlmn o_2} ~.~
\eea
The above products satisfy the following quadratic relation,
\beal{es72628} 
{O_{(IV)}}_{ijklmn o_1 o_2} = {o_{(I)}}_{ijklmn o_1 o_2} - \frac{1}{9} {o_{(II)}}_{ijklmn o_1 o_2} ~,~
\eea
which is precisely the relation at this order.
\\

\item \underline{$- [1,1,1,0,0,0]_x s^2 t^6$ relations:} We consider the following generator products for the relation at this order,
\beal{es72628}
{o_{(III)}}^{ijk}_{~~~lmn o} &=& {A_{001}}^{pqk} {A_{001}}^{ijr} \epsilon_{pqrlmn o} 
~,~
\nn\\
{o_{(IV)}}^{ijk}_{~~~lmn o} &=& {A_{002}}^{ijk} B^{pqr} \epsilon_{pqrlmn o} 
~,~
\eea
which transform in the correct representation at this order. The above products satisfy the following quadratic relation,
\beal{es72629}
{O_{(V)}}^{ijk}_{~~~lmn o} = {o_{(III)}}^{ijk}_{~~~lmn o} - \frac{1}{3} {o_{(IV)}}^{ijk}_{~~~lmn o} = 0 ~,~
\eea
which is precisely the relation at this order.
\\

\item \underline{$- [1,0,0,0,1,0]_x s^2 t^6$ relations:} The following generator product vanishes exactly,
\beal{es72630}
{O_{(VI)}}^{i}_{~jo} = {A_{002}}^{pqi} B^{lmn} \epsilon_{pqlmnjo} = 0~.~
\eea
The above is the quadratic relations for this order.
\\

\end{itemize}

The next set of quadratic relations are at orders of $s^3 t^6$. We first consider the following $SU(7)$ representation products,
\beal{es72640}
~[1,1,0,0,0,0]_x \times  [1,1,0,0,0,0]_x &=& [2,2,0,0,0,0]_x + [3,0,1,0,0,0]_x 
\nn\\
&&
+ [0,3,0,0,0,0]_x  + 2 [1,1,1,0,0,0]_x 
\nn\\
&&
+ [2,0,0,1,0,0]_x  + [0,0,2,0,0,0]_x  
\nn\\
&&
+ [0,1,0,1,0,0]_x ~,~
\nn\\
~[3,0,0,0,0,0]_x \times [0,0,1,0,0,0]_x &=& [3,0,1,0,0,0]_x + [2,0,0,1,0,0]_x ~,~
\nn\\
~[1,1,0,0,0,0]_x \times [0,0,1,0,0,0]_x &=& [1,1,1,0,0,0]_x + [2,0,0,1,0,0]_x 
\nn\\
&&
+ [0,1,0,1,0,0]_x  + [1,0,0,0,1,0]_x~,~
\nn\\
\text{Sym}^2 [0,0,1,0,0,0]_x &=& [0,0,2,0,0,0]_x + [1,0,0,0,1,0]_x ~.~
\eea
The above representation products lead us to the following quadratic relations:
\begin{itemize}

\item \underline{$-[0,1,0,1,0,0]_x s^3 t^6$ relations:} The following generator product vanishes exactly,
\beal{es72650}
{P_{(I)}}^{ij}_{~~klo} = {A_{001}}^{ijm} {A_{002}}^{pqr} \epsilon_{mpqrklo} = 0 ~,~
\eea
and transforms in the representation of this order. Accordingly, it is exactly the quadratic relation we are looking for this order.
\\

\item \underline{$-2 [2,0,0,1,0,0]_x s^3 t^6$ relations:} For this order, we consider the following generator products,
\beal{es72651}
{P_{(II)}}^{ij}_{~~klo} &=&
{A_{001}}^{mni} {A_{002}}^{pqj} \epsilon_{mnpqklo} = 0 ~,~
\nn\\
{P_{(III)}}^{ij}_{~~klo} &=&
{S_{012}}^{ijm} B^{npq} \epsilon_{mnpqklo} = 0 ~.~
\eea
Both above vanish and satisfy the correct transformation property for this order. They are precisely the two quadratic relations at this order.
\\

\item \underline{$-[0,0,2,0,0,0]_x s^3 t^6$ relations:} We first consider the following generator products,
\beal{es72652}
{p_{(I)}}_{ijklmn o_1 o_2} &=&
{A_{001}}^{pqu} {A_{002}}^{rsv} \epsilon_{pqvijk o_1} \epsilon_{rsulmn o_2} ~,~
\nn\\
{p_{(II)}}_{ijklmn o_1 o_2} &=& 
u_3 B^{pqu} B^{rsv} \epsilon_{pquijk o_1} \epsilon_{rsvlmn o_2} ~,~
\eea
which transform in the correct representation corresponding to this order. The products satisfy the following quadratic relation 
\beal{es72653}
{P_{(IV)}}_{ijklmn o_1 o_2} = {p_{(I)}}_{ijklmn o_1 o_2} - \frac{1}{9} {p_{(II)}}_{ijklmn o_1 o_2} ~,~
\eea
which is the relation for this order.
\\

\item \underline{$- 2 [1,1,1,0,0,0]_x s^3 t^6$ relations:}  For the quadratic relation at this order, we need to consider the following generator products,
\beal{es72654}
{p_{(III)}}^{ijk}_{~~~lmno} &=&
{A_{001}}^{ijp} {A_{002}}^{qrk} \epsilon_{pqrlmno} ~,~
\nn\\
{p_{(IV)}}^{ijk}_{~~~lmno} &=&
{A_{002}}^{ijp} {A_{001}}^{qrk} \epsilon_{pqrlmno} ~,~
\nn\\
{p_{(V)}}^{ijk}_{~~~lmno} &=&
u_2 {A_{001}}^{ijp} B^{qrk} \epsilon_{pqrlmno} ~.~
\eea
The above products transform in the correct representation for this order. The quadratic relations formed by the above are
\beal{es72655}
{P_{(V)}}^{ijk}_{~~~lmno} = {p_{(III)}}^{ijk}_{~~~lmno} - {p_{(IV)}}^{ijk}_{~~~lmno} = 0 ~,~
\nn\\
{P_{(VI)}}^{ijk}_{~~~lmno} = {p_{(III)}}^{ijk}_{~~~lmno} - \frac{1}{2} {p_{(V)}}^{ijk}_{~~~lmno} = 0 ~,~
\eea
exactly corresponding to the two expected quadratic relations at this order.
\\

\item \underline{$-[3,0,1,0,0,0]_x s^3 t^6$ relations:} We consider the following generator products for the quadratic relation at this order,
\beal{es72660}
{p_{(VI)}}^{ijk}_{~~~lmno}
&=&
 (
{A_{001}}^{ipj} {A_{002}}^{qrk} 
+  {A_{001}}^{jpk} {A_{002}}^{qri} 
+ {A_{001}}^{kpi} {A_{002}}^{qrj} 
)\epsilon_{pqrlmno}
~,~
\nn\\
{p_{(VII)}}^{ijk}_{~~~lmno}
&=&
{S_{012}}^{ijk} B^{pqr} \epsilon_{pqrlmno} ~,~
\eea
which transform in the correct representation for this order. The above products satisfy the following quadratic relation 
\beal{es2661}
{P_{(VII)}}^{ijk}_{~~~lmno} = {p_{(VI)}}^{ijk}_{~~~lmno} + \frac{1}{3} {p_{(VII)}}^{ijk}_{~~~lmno} ~,~
\eea
which is precisely the relation for this order.
\\

\end{itemize}

The next set of quadratic relations are of orders $s^4 t^6$. In order to construct the relations, we consider the following $SU(7)$ representation products,
\beal{es72662}
\text{Sym}^{2} [1,1,0,0,0,0]_x &=& [2,2,0,0,0,0]_x + [1,1,1,0,0,0]_x 
\nn\\
&&
+ [2,0,0,1,0,0]_x + [0,0,2,0,0,0]_x ~,~
\nn\\
~[1,1,0,0,0,0]_x \times [3,0,0,0,0,0]_x &=& [4,1,0,0,0,0]_x + [2,2,0,0,0,0]_x 
\nn\\
&&
+ [3,0,1,0,0,0]_x + [1,1,1,0,0,0]_x ~,~
\nn\\
~[1,1,0,0,0,0]_x \times [0,0,1,0,0,0]_x &=& [1,1,1,0,0,0]_x + [2,0,0,1,0,0]_x 
\nn\\
&&
+ [0,1,0,1,0,0]_x + [1,0,0,0,1,0]_x ~,~
\nn\\
~\text{Sym}^{2} [0,0,1,0,0,0]_x &=& [0,0,2,0,0,0]_x + [1,0,0,0,1,0]_x ~.~
\eea
The above representation products help us in constructing the quadratic relations as follows:
\begin{itemize}

\item \underline{$-[2,0,0,1,0,0]_x s^4 t^6$ relations:} We consider the following generator product for the quadratic relation at this order,
\beal{es72665}
{U_{(I)}}^{ij}_{~~klo} = {A_{002}}^{pqi} {A_{002}}^{mnj} \epsilon_{pqmnklo} = 0~,~
\eea	
which vanishes exactly. This is precisely the quadratic relation at this order.
\\

\item \underline{$-[0,0,2,0,0,0]_x s^4 t^6$ relations:} For this order, we consider the following generator products for the quadratic relation,
\beal{es72666}
{u_{(I)}}_{ijklmn o_1 o_2} 
&=& 
{A_{002}}^{pqr} {A_{002}}^{uvs} \epsilon_{pqsijk o_1} \epsilon_{uvrlmn o_2} 
~,~
\nn\\
{u_{(II)}}_{ijklmn o_1 o_2}
 &=&
u_2 u_2 B^{pqr} B^{uvw} \epsilon_{pqrijk o_1} \epsilon_{uvwlmn o_2}
~.~
\eea
The above products transform in the correct $SU(7)$ representation of this order. They satisfy the following quadratic relation,
\beal{es72667}
{U_{(II)}}_{ijklmn o_1 o_2} &=& {u_{(I)}}_{ijklmn o_1 o_2}  -  \frac{1}{3} {u_{(II)}}_{ijklmn o_1 o_2} = 0 ~,~
\eea
which is precisely the relation we are looking for here.
\\

\item \underline{$-2 [1,1,1,0,0,0]_x s^4 t^6$ relations:} There are two distinct quadratic relations at this order. In order to construct them, we consider the following generator products,
\beal{es72670}
{u_{(III)}}^{ijk}_{~~~lmno} &=&
{A_{002}}^{ijp} {A_{002}}^{qrk} \epsilon_{pqrlmno} 
~,~
\nn\\
{u_{(IV)}}^{ijk}_{~~~lmno} &=& 
{A_{001}}^{ijp} {S_{012}}^{qrk} \epsilon_{pqrlmno} 
~,~
\nn\\
{u_{(V)}}^{ijk}_{~~~lmno} &=&
u_2 {A_{001}}^{ijk} {A_{001}}^{pqr} \epsilon_{pqrlmno} 
~,~
\nn\\
{u_{(VI)}}^{ijk}_{~~~lmno} &=&
u_3 {A_{001}}^{ijk} B^{pqr} \epsilon_{pqrlmno}
~,~
\eea
which transform in the representation of this order. The above products form the following two quadratic relations,
\beal{es72671}
{U_{(III)}}^{ijk}_{~~~lmno} &=&
{u_{(V)}}^{ijk}_{~~~lmno} - 2 {u_{(III)}}^{ijk}_{~~~lmno} + \frac{2}{9} {u_{(VI)}}^{ijk}_{~~~lmno} = 0
~,~
\nn\\
{U_{(IV)}}^{ijk}_{~~~lmno} &=&
{u_{(IV)}}^{ijk}_{~~~lmno} + \frac{1}{6} {u_{(VI)}}^{ijk}_{~~~lmno} = 0 ~.~
\eea
The above are the two quadratic relations at this order.
\\

\item \underline{$-[2,2,0,0,0,0]_x s^4 t^6$ relations:} The following generator product with its symmetrization and anti-symmetrization of indices is required for the construction of the quadratic relation at this order,
\beal{es72672}
{u_{(VII)}}^{ijklmn} = {A_{001}}^{ijk} {S_{012}}^{lmn} ~,~
\eea
where we antisymmetrize on the indices $[kl]$ and $[mn]$ as follows,
\beal{es2673}
{u_{(VIII)}}^{ijklmn} = {u_{(VII)}}^{ijklmn} - {u_{(VII)}}^{ijlkmn} -  {u_{(VII)}}^{ijklnm} + {u_{(VII)}}^{ijlknm} ~.~
\nn\\
\eea
A further symmetrization on the two paired indices $[kl]$ and $[mn]$ leads to the following
\beal{es72674}
{U_{(V)}}^{ijklmn} = {u_{(VIII)}}^{ijklmn} + {u_{(VIII)}}^{ijmnkl} = 0 ~,~
\eea
which exactly vanishes. This is precisely the quadratic relation at this order we are looking for.
\\

\item \underline{$-[3,0,1,0,0,0]_x s^4 t^6$ relations:} The quadratic relation at this order is formed by
\beal{es72680}
{U_{(VI)}}^{ijk}_{~~~lmno} = 
(
{A_{001}}^{ipq} {S_{012}}^{rjk}
+ {A_{001}}^{jpq} {S_{012}}^{rki}
+ {A_{001}}^{jpq} {S_{012}}^{rij} 
) \epsilon_{pqrlmno}
= 0
~,~
\nn\\
\eea
where the above contains the symmetrization of the generator product
\beal{es72681}
{A_{001}}^{ipq} {S_{012}}^{rjk} \epsilon_{pqrlmno}
\eea
in the indices $ijk$. The above quadratic relation satisfies precisely the transformation properties for this order and is the relation we are looking for.
\\
\end{itemize}

For the next set of quadratic relations at orders of $s^5 t^6$, we consider the following representation products in order to construct the relations,
\beal{es72690}
~[3,0,0,0,0,0]_x \times [1,1,0,0,0,0]_x  &=&
[4,1,0,0,0,0]_x + [2,2,0,0,0,0]_x 
\nn\\
&&
+ [3,0,1,0,0,0]_x + [1,1,1,0,0,0]_x
~,~
\nn\\
~[1,1,0,0,0,0]_x \times [1,1,0,0,0,0]_x  &=&
[2,2,0,0,0,0]_x + [3,0,1,0,0,0]_x 
\nn\\
&&
+ [0,3,0,0,0,0]_x 
+ 2 [1,1,1,0,0,0]_x 
\nn\\
&&
+ [2,0,0,1,0,0]_x 
+ [0,0,2,0,0,0]_x 
\nn\\
&&
+ [0,1,0,1,0,0]_x
~,~
\nn\\
~[3,0,0,0,0,0]_x \times [0,0,1,0,0,0]_x  &=&
[3,0,1,0,0,0]_x + [2,0,0,1,0,0]_x 
~,~
\nn\\
~[1,1,0,0,0,0]_x \times [0,0,1,0,0,0]_x  &=&
[1,1,1,0,0,0]_x + [2,0,0,1,0,0]_x 
\nn\\
&&
+ [0,1,0,1,0,0]_x 
+ [1,0,0,0,1,0]_x
~,~
\nn\\
\text{Sym}^{2} [1,1,0,0,0,0]_x &=&
[2,2,0,0,0,0]_x + [1,1,1,0,0,0]_x 
\nn\\
&&
+ [2,0,0,1,0,0]_x 
+ [0,0,2,0,0,0]_x
~,~
\nn\\
\text{Sym}^{2} [0,0,1,0,0,0]_x &=&
[0,0,2,0,0,0]_x + [1,0,0,0,1,0]_x
~~.
\\
\nn
\eea
The above products are used to construct the following quadratic relations of generators:
\begin{itemize}
\item \underline{$-[1,1,1,0,0,0]_x s^5 t^6$ relations:} We consider the following products of generators for the quadratic relation at this order,
\beal{es72805}
{v_{(I)}}^{ijk}_{~~~lmno} &=& 
{A_{002}}^{ijp} {S_{012}}^{qrk}
\epsilon_{pqrlmno}
~,~
\nn\\
{v_{(II)}}^{ijk}_{~~~lmno} &=&
u_3 {A_{001}}^{ijp} {A_{001}}^{qrk}
\epsilon_{pqrlmno}
~,~
\eea
which transform in the correct irreducible representation of this order. The above products satisfy the following quadratic relation,
\beal{es2806}
{V_{(I)}}^{ijk}_{~~~lmno} = {v_{(I)}}^{ijk}_{~~~lmno} + \frac{1}{2} {v_{(II)}}^{ijk}_{~~~lmno} = 0 ~.~
\eea
The above precisely is the quadratic relation we are looking for at this order.
\\

\item \underline{$-[2,2,0,0,0,0]_x s^5 t^6$ relations:} We need to consider several generator products with various symmetrizations and anti-symmetrizations of indices in order to construct the quadratic relation for this order. The first product to consider is the following,
\beal{es72810}
{v_{(III)}}^{ijklmn} = ( {A_{002}}^{ijk} - \frac{1}{2} u_2 B^{ijk}) {S_{012}}^{lmn} ~,~
\eea
where we recall that ${A_{011}}^{ijk} = {A_{002}}^{ijk} - \frac{1}{2} u_2 B^{ijk}$.
Above, we anti-symmetrize in the indices $[kl]$ and $[mn]$ to give
\beal{es72811}
{v_{(IV)}}^{ijklmn} = {v_{(III)}}^{ijklmn} - {v_{(III)}}^{ijlkmn} - {v_{(III)}}^{ijklnm} + {v_{(III)}}^{ijlknm}  ~,~
\eea
and further anti-symmetrize in the pairs of indices $[ij]$ and $[mn]$ to obtain,
\beal{es72812}
{v_{(V)}}^{ijklmn} = {v_{(IV)}}^{ijklmn} - {v_{(IV)}}^{mnklij} ~.~
\eea
The second generator product to consider is the following,
\beal{es2813}
{v_{(VI)}}^{ijklmn} = u_3 {A_{001}}^{ijk} {A_{001}}^{lmn} ~,~
\eea
which we anti-symmetrize in the indices $[kl]$ and $[mn]$ giving,
\beal{es72814}
{v_{(VII)}}^{ijklmn} = {v_{(VI)}}^{ijklmn} - {v_{(VI)}}^{ijlkmn} - {v_{(VI)}}^{ijklnm} + {v_{(VI)}}^{ijlknm} ~.~
\eea
A further symmetrization of the pair of indices $[ij]$ and $[kl]$ gives
\beal{es72815}
{v_{(VIII)}}^{ijklmn} = {v_{(VII)}}^{ijklmn} + {v_{(VII)}}^{klijmn} ~.`
\eea
From the above generator products, which correctly transform at this order, we can construct the following quadratic relation,
\beal{es72816}
{V_{(II)}}^{ijklmn} = {v_{(V)}}^{ijklmn} + {v_{(VIII)}}^{ijmnkl} = 0 ~.~
\eea
This is precisely we are looking for at this order.
\\

\item \underline{$-[3,0,1,0,0,0]_x s^5 t^6$ relations:} For the quadratic relation at this order, we need to consider the following generators products,
\beal{es72817}
{v_{(IX)}}^{ijk}_{~~~lmno} = {A_{002}}^{ipq} {S_{012}}^{rjk} \epsilon_{pqrlmno} ~,~
\eea
which we symmetrize in the indices $ijk$ as follows,
\beal{es72818}
{v_{(X)}}^{ijk}_{~~~lmno} = {v_{(IX)}}^{ijk}_{~~~lmno} + {v_{(IX)}}^{jki}_{~~~lmno} + {v_{(IX)}}^{kij}_{~~~lmno} ~.~
\eea
The second generator product which we need to consider is the following,
\beal{es72819}
{v_{(XI)}}^{ijk}_{~~~lmno} = u_2 {A_{001}}^{ipj} {A_{002}}^{qrk} \epsilon_{pqrlmno} ~,~
\eea
which we symmetrize in the indices $ijk$ as follows,
\beal{es72920}
{v_{(XII)}}^{ijk}_{~~~lmno} = {v_{(XI)}}^{ijk}_{~~~lmno} + {v_{(XI)}}^{jki}_{~~~lmno} + {v_{(XI)}}^{kij}_{~~~lmno} ~.~
\eea
From the above products, we construct the following quadratic relation,
\beal{es72921}
{V_{(III)}}^{ijk}_{~~~lmno} = {v_{(X)}}^{ijk}_{~~~lmno}  - \frac{1}{3} {v_{(XII)}}^{ijk}_{~~~lmno} = 0 ~.~
\eea
This is precisely the relation we are looking for at this order of the plethystic logarithm. 
\\

\end{itemize}

The final quadratic relation is of order $s^6 t^6$. It can be constructed by considering the following $SU(7)$ representation products
\beal{es72820}
\text{Sym}^{2} [3,0,0,0,0,0]_x &=& [6,0,0,0,0,0]_x + [2,2,0,0,0,0]_x ~,~
\nn\\
\text{Sym}^{2} [1,1,0,0,0,0]_x &=& [2,2,0,0,0,0]_x + [1,1,1,0,0,0]_x 
\nn\\
&&
+ [2,0,0,1,0,0]_x  + [0,0,2,0,0,0]_x ~,~
\nn\\
~[1,1,0,0,0,0]_x \times [3,0,0,0,0,0]_x &=&
[4,1,0,0,0,0]_x + [2,2,0,0,0,0]_x 
\nn\\
&&
+ [3,0,1,0,0,0]_x  + [1,1,1,0,0,0]_x ~,~
\nn\\
~[1,1,0,0,0,0]_x \times [1,1,0,0,0,0]_x &=& 
[2,2,0,0,0,0]_x + [3,0,1,0,0,0]_x 
\nn\\
&&
+ [0,3,0,0,0,0]_x + 2 [1,1,1,0,0,0]_x 
\nn\\
&&
+ [2,0,0,1,0,0]_x + [0,0,2,0,0,0]_x 
\nn\\
&&
+ [0,1,0,1,0,0]_x 
~.~
\eea
From the above, we can construct candidate products of generators of the vortex master space, in order to identify quadratic relations amongst them: 
\begin{itemize}
\item \underline{$-[2,2,0,0,0,0]_x s^6 t^6$ relations:} The final quadratic relation can be identified from the following generator product and its symmetrization and anti-symmetrization of indices,
\beal{es72821}
{z_{(I)}}^{ijklmn} = {S_{012}}^{ijk} {S_{012}}^{lmn} ~.~
\eea
We anti-symmetrize the above on the indices $[kl]$ and $[mn]$ 
\beal{es72822}
{z_{(II)}}^{ijklmn} = {z_{(I)}}^{ijklmn} - {z_{(I)}}^{ijlkmn} - {z_{(I)}}^{ijklnm} + {z_{(I)}}^{ijlknm} ~,~
\eea
and further symmetrize on the pairs of indices $[kl]$ and $[mn]$ to give
\beal{es72823}
Z^{ijklmn} = {z_{(II)}}^{ijklmn} + {z_{(II)}}^{ijmnkl} = 0 ~.~
\eea
The above quadratic relation vanishes non-trivially and by construction transforms in the representation of this order of the plethystic logarithm. The above is the quadratic relation corresponding to this order.
\\
\end{itemize}

\paragraph{Vortex moduli space:} Given the above explicit computation of the quadratic relations between generators, we can proceed in identifying the vortex moduli space for the $3$ $U(7)$ vortex theory. The vortex moduli space can be expressed as a $\mathbb{C}^{*}$ projection as follows,
\beal{es72823xx1}
&&
\wV_{3,7} = \wmaster_{3,7}/\{
B^{ijk} \simeq \lambda^{3} B^{ijk} ,
{A_{001}}^{ijk} \simeq \lambda^{3} {A_{001}}^{ijk} ,
\nn\\
&&
\hspace{3cm}
{A_{002}}^{ijk} \simeq \lambda^{3} {A_{002}}^{ijk} ,
{S_{012}}^{i j k} \simeq \lambda^{3} {S_{012}}^{i j k} 
\}
~.~
\eea
The master space of the vortex theory can be identified with the help of the quadratic relations as follows,
\beal{es72823xx2}
&&
\wmaster_{3,7} = \mathbb{C}[
u_2, u_3, B^{ijk}, {A_{001}}^{ijk}, {A_{002}}^{ijk}, {S_{012}}^{ijk}
]/\{
\nn\\
&&
\hspace{2cm}
H_{ij}^{~~k} = 0 ~,~
\nn\\
&&
\hspace{2cm}
{R_{(I)}}^{ij}_{~~klo} = 0 ~,~
{R_{(II)}}^{ij}_{~~klo} = 0 ~,~
{R_{(III)}}^{i}_{~jo} = 0 ~,~
\nn\\
&&
\hspace{2cm}
{O_{(I)}}^{ij}_{~~klo} = 0 ~,~
{O_{(II)}}^{ij}_{~~klo} = 0 ~,~
{O_{(III)}}^{ij}_{~~klo} = 0 ~,~
{O_{(IV)}}_{ijklmn o_1 o_2} = 0 ~,~
\nn\\
&&
\hspace{2cm}
{O_{(V)}}^{ijk}_{~~~lmno} = 0 ~,~
{O_{(VI)}}^{i}_{~jo} = 0 ~,~
\nn\\
&&
\hspace{2cm}
{P_{(I)}}^{ij}_{~~klo} = 0 ~,~
{P_{(II)}}^{ij}_{~~klo} = 0 ~,~
{P_{(III)}}^{ij}_{~~klo} = 0 ~,~
{P_{(IV)}}_{ijklmn o_1 o_2} = 0 ~,~
\nn\\
&&
\hspace{2cm}
{P_{(V)}}^{ijk}_{~~~lmno} = 0 ~,~
{P_{(VI)}}^{ijk}_{~~~lmno} = 0 ~,~
{P_{(VII)}}^{ijk}_{~~~lmno} = 0 ~,~
\nn\\
&&
\hspace{2cm}
{U_{(I)}}^{ij}_{~~klo} = 0 ~,~
{U_{(II)}}_{ijklmn o_1 o_2} = 0 ~,~
{U_{(III)}}^{ijk}_{~~~lmno} = 0 ~,~
{U_{(IV)}}^{ijk}_{~~~lmno} = 0 ~,~
\nn\\
&&
\hspace{2cm}
{U_{(V)}}^{ijklmn}= 0 ~,~
{U_{(VI)}}^{ijk}_{~~~lmno}= 0 ~,~
\nn\\
&&
\hspace{2cm}
{V_{(I)}}^{ijk}_{~~~lmno} = 0 ~,~
{V_{(II)}}^{ijklmn} = 0 ~,~
{V_{(III)}}^{ijk}_{~~~lmno} = 0 ~,~
\nn\\
&&
\hspace{2cm}
Z^{ijklmn} = 0
~~~
\}
~.~
\eea
\\

\subsection{$3$ $U(N)$ vortices on $\mathbb{C}$ \label{s3bN}}

In this section, we summarize the generalization of the $3$ vortex moduli space beyond $U(7)$. We begin with the Hilbert series of the master space for $3$ $U(N)$ vortices. It can be computed using the following Molien integral
\beal{esx1570}
g(t,s,x;\wmaster_{3,N}) =\oint \ud\mu_{SU(3)}
\PE\Big[
[0,1]_w [1,0,\dots,0]_{x} t + [1,1]_w s
\Big]~.~
\eea
The Hilbert series for the first few values of $N$ are as follows
\beal{esx1571}
g(t,t;\wmaster_{3,1}) &=&
\frac{1}{
(1 - t^2) (1 - t^3) (1 - t^6)
}
~~,
\nn\\
g(t,t;\wmaster_{3,2}) &=&
\frac{1 + 2 t^5 + 2 t^6 + t^{11}}{
(1 - t^2) (1 - t^3) (1 - t^4)^2 (1 - t^6)^2
}
~~,
\nn\\
g(t,t;\wmaster_{3,3}) &=&
\frac{
1
}{
(1 - t^2) (1 - t^3) (1 - t^4)^4 (1 - t^6)^3
}
\times
(1 + t^3 + 4 t^4 + 8 t^5 + 8 t^6 
\nn\\
&&
+ 4 t^7 + 9 t^8 + 13 t^9 + 14 t^{10} + 20 t^{11} + 14 t^{12} + 13 t^{13} + 9 t^{14} + 4 t^{15} + 8 t^{16} 
\nn\\
&&
+ 8 t^{17} + 4 t^{18} + t^{19} + t^{22})
~~,
\nn\\
g(t,t;\wmaster_{3,4}) &=&
\frac{
1
}{
(1 - t^2) (1 - t^3)^5 (1 - t^4)^6 (1 - t^6)^4
}
\times
(
1 + 14 t^4 + 20 t^5 
\nn\\
&&
+ 16 t^6 - 16 t^7 + 5 t^8 + 46 t^9 + 94 t^{10} + 16 t^{11} - 94 t^{12} - 156 t^{13} - 135 t^{14} 
\nn\\
&&
- 35 t^{15} - t^{16} + 65 t^{17} - 59 t^{18} - 161 t^{19} - 185 t^{20} + 125 t^{21} + 440 t^{22} 
\nn\\
&&
+ 440 t^{23} + 125 t^{24} - 185 t^{25} - 161 t^{26} - 59 t^{27} + 65 t^{28} - t^{29} - 35 t^{30} 
\nn\\
&&
- 135 t^{31} - 156 t^{32} - 94 t^{33} + 16 t^{34} + 94 t^{35} + 46 t^{36} + 5 t^{37} - 16 t^{38} 
\nn\\
&&
+ 16 t^{39} + 20 t^{40} + 14 t^{41} + t^{45}
)
~~,
\nn\\
g(t,t;\wmaster_{3,5}) &=&
\frac{1}{
(1 - t^2) (1 - t^3)^2 (1 - t^4)^8 (1 - t^6)^5
}
\times
(
1 - t + 9 t^3 + 23 t^4 
\nn\\
&&
+ 8 t^5 + 30 t^6 + 98 t^7 + 200 t^8 + 232 t^9 + 320 t^{10} + 482 t^{11} + 677 t^{12} 
\nn\\
&&
+ 806 t^{13} + 800 t^{14} + 1052 t^{15} + 988 t^{16} + 485 t^{17} + 18 t^{18} - 127 t^{19} 
\nn\\
&&
- 440 t^{20} - 970 t^{21} - 1728 t^{22} - 2074 t^{23} - 1778 t^{24} - 2074 t^{25} - 1728 t^{26} 
\nn\\
&&
- 970 t^{27} - 440 t^{28} - 127 t^{29} + 18 t^{30} + 485 t^{31} + 988 t^{32} + 1052 t^{33} 
\nn\\
&&
+ 800 t^{34} + 806 t^{35} + 677 t^{36} + 482 t^{37} + 320 t^{38} + 232 t^{39} + 200 t^{40} 
\nn\\
&&
+ 98 t^{41} + 30 t^{42} + 8 t^{43} + 23 t^{44} + 9 t^{45} - t^{47} + t^{48}
 )~~,
\eea
where for simplicity we have set all global $SU(N)$ fugacities to $x_i=1$ and have set the adjoint and fundamental fugacities to be the same $s=t$.

As a character expansion, the Hilbert series for the $3$ $U(N)$ vortex master space is
\beal{esx1572}
&&
g(t,s,x;\wmaster_{3,N}) =
\frac{1}{(1-s^2)(1-s^3)} 
\times
\nn\\
&& 
\sum_{n_0=0}^{\infty} \sum_{n_1=0}^{\infty} \sum_{n_2=0}^{\infty} \sum_{n_3=0}^{\infty}
\Big[
[n_1+n_2+3n_3, n_1+n_2, n_0, 0, \dots, 0]_x s^{n_1 + 2n_2 + 3n_3} t^{3 n_0 + 3n_1 + 3n_2 + 3n_3 }
\nn\\
&& 
\hspace{1cm}
+
[n_1+n_2, n_1+n_2+3n_3 +3, n_0, 0, \dots, 0]_x s^{n_1 + 2n_2 + 3n_3+3} t^{3 n_0 + 3n_1 + 3n_2 + 6n_3 + 6}
\Big]
~~,
\nn\\
\eea
where $[m_1,\dots,m_{N-1}]_x$ represents the character of an irreducible representation of the global $SU(N)$.

The plethystic logarithm takes the following form,
\beal{esx1573}
&&
\PL\Big[
g(t,s,x;\master_{3,N})
\Big]
=
s^2 + s^3 + [0,0,1,0,\dots,0]_x t^3
+ [1,1,0,0,0,0]_x s t^3
\nn\\
&&
\hspace{0.5cm}
+ [1,1,0,0,\dots,0]_x s^2 t^3
+ [3,0,0,0,\dots,0]_x s^3 t^3
\nn\\
&&
\hspace{0.5cm}
- [1,0,0,0,1,0,\dots,0]_x t^6
- (
[2,0,0,1,0,\dots,0]_x
+ [0,1,0,1,0,\dots,0]_x
\nn\\
&&
\hspace{0.5cm}
+ [1,0,0,0,1,0,\dots,0]_x
) s t^6
- (
[1,1,1,0,0,\dots,0]_x 
+ [0,0,2,0,0,\dots,0]_x
\nn\\
&&
\hspace{0.5cm}
+ [2,0,0,1,0,\dots,0]_x
+ [0,1,0,1,0,\dots,0]_x
+ [1,0,0,0,1,0,\dots,0]_x
) s^2 t^6
\nn\\
&&
\hspace{0.5cm}
+ (
[1,1,0,0,0,\dots,0]_x
+ [0,0,0,1,1,0,\dots,0]_x
) t^9
- (
[3,0,1,0,0,\dots,0]_x
\nn\\
&&
\hspace{0.5cm}
+2 [1,1,1,0,0,\dots,0]_x
+[0,0,2,0,0,\dots,0]_x
+2 [2,0,0,1,0,\dots,0]_x
\nn\\
&&
\hspace{0.5cm}
+ [0,1,0,1,0,0,\dots,0]_x
+ [1,0,0,0,1,0,\dots,0]_x
) s^3 t^6
+\dots
\nn\\
&&
\hspace{0.5cm}
- (
[0,0,2,0,0,\dots,0]_x
+ [1,1,1,0,0,\dots,0]_x
+ [2,0,0,1,0,\dots,0]_x
\nn\\
&&
\hspace{0.5cm}
+ [2,2,0,0,0,\dots,0]_x
+ [3,0,1,0,0,\dots,0]_x
) s^4 t^6 
+ \dots 
\nn\\
&&
\hspace{0.5cm}
- (
[1,1,1,0,\dots,0]_x
+ [2,2,0,0,\dots,0]_x
+ [3,0,1,0,\dots,0]_x
) s^5 t^6
+ \dots
\nn\\
&&
\hspace{0.5cm}
- [2,2,0,0,\dots,0]_x s^6 t^6
+ \dots ~~,
\eea
where we identify that the generators of the master space as,
\beal{esx1574}
s^2 &\ra& u_2=\Tr(\phi^2)
\nn\\
s^3 &\ra& u_3=\Tr(\phi^3)
\nn\\
~[0,0,1,0,0,\dots,0]_x t^3 &\ra&
{B}^{ijk} = \epsilon^{\alpha_1 \alpha_2 \alpha_3} Q_{\alpha_1}^{i} Q_{\alpha_2}^{j} Q_{\alpha_3}^{k}
\nn\\
~[1,1,0,0,0,\dots,0]_x s t^3 &\ra&
\left\{
\ba{l}
{A_{001}}^{i j k} =
\epsilon^{\alpha_1 \alpha_2 \alpha_3}
Q_{\alpha_1}^{i} Q_{\alpha_2}^{j} \phi_{\alpha_3}^{\beta} Q_{\beta}^{k}
\nn\\
\epsilon_{ijkm_1 \dots m_{N-3}} {A_{001}}^{ijk}=0 
\ea
\right.
\nn\\
~[1,1,0,0,0,\dots,0]_x s^2 t^3 &\ra&
\left\{
\ba{l}
{A_{002}}^{ijk} =
\epsilon^{\alpha_1 \alpha_2 \alpha_3}
Q_{\alpha_1}^{i} Q_{\alpha_2}^{j} \phi_{\alpha_3}^{\beta_1}\phi_{\beta_1}^{\beta_2} Q_{\beta_2}^{k}
\nn\\
\epsilon_{ijkm_1\dots m_{N-3}} {A_{002}}^{ijk} = - \frac{1}{3} u_2 \epsilon_{m_1\dots m_{N-3}rsu} {B}^{rsu}
\nn\\
{A_{011}}^{ijk} =
\epsilon^{\alpha_1 \alpha_2 \alpha_3}
Q_{\alpha_1}^{k} \phi_{\alpha_2}^{\beta_1} Q_{\beta_1}^{i} \phi_{\alpha_3}^{\beta_2} Q_{\beta_2}^{j}
\nn\\
\epsilon_{ijkm_1 \dots m_{N-3}} {A_{011}}^{ijk}= \frac{1}{6} u_2 \epsilon_{m_1 \dots m_{N-3}rsu} {B}^{rsu}
\nn\\
\ra {A_{002}}^{ijk} = {A_{011}}^{ijk} + \frac{1}{2} u_2 B^{ijk}
\ea\right.
\nn\\
~[3,0,0,0,0,\dots,0]_x s^3 t^3 &\ra&
\left\{\ba{l}
{A_{012}}^{ijk}
= 
\epsilon^{\alpha_1 \alpha_2 \alpha_3} 
Q_{\alpha_1}^{i}
\phi_{\alpha_2}^{\beta_1} Q_{\beta_1}^{j}
\phi_{\alpha_3}^{\beta_2} \phi_{\beta_2}^{\beta_3} Q_{\beta_3}^{k}
\nn\\
{S_{012}}^{ijk} =
{A_{012}}^{ijk}
+ {A_{012}}^{jki}
+ {A_{012}}^{kij}
\ea\right.
~~.
\nn\\
\eea

\paragraph{Quadratic relations:} The generalized quadratic relations amongst the generators of the $3$ $U(N)$ vortex master space are summarized in \tref{t3Na1} and \tref{t3Na2}.
\\

\paragraph{Vortex moduli space:} Given the above summary of the quadratic relations between generators, we can express the vortex moduli space for $3$ $U(N)$ vortices as the following $\mathbb{C}^{*}$ projection,
\beal{esx1574xx1}
&&
\wV_{3,N} = \wmaster_{3,N}/\{
B^{ijk} \simeq \lambda^{3} B^{ijk} ,
{A_{001}}^{ijk} \simeq \lambda^{3} {A_{001}}^{ijk} ,
\nn\\
&&
\hspace{3cm}
{A_{002}}^{ijk} \simeq \lambda^{3} {A_{002}}^{ijk} ,
{S_{012}}^{i j k} \simeq \lambda^{3} {S_{012}}^{i j k} 
\}
~.~
\eea
The master space of the vortex theory can be identified with the help of the quadratic relations as follows,
\beal{esx1574xx2}
&&
\wmaster_{3,N} = \mathbb{C}[
u_2, u_3, B^{ijk}, {A_{001}}^{ijk}, {A_{002}}^{ijk}, {S_{012}}^{ijk}
]/\{
\nn\\
&&
\hspace{2cm}
H_{i_1\dots i_{N-5}}^{~~k} = 0 ~,~
\nn\\
&&
\hspace{2cm}
{R_{(I)}}^{ij}_{~~k_1 \dots k_{N-4}} = 0 ~,~
\dots ~,~
{R_{(III)}}^{i}_{~j_1\dots j_{N-5}} = 0 ~,~
\nn\\
&&
\hspace{2cm}
{O_{(I)}}^{ij}_{~~k_{1}\dots k_{N-3}} = 0 ~,~
\dots ~,~
{O_{(VI)}}^{i}_{~j_1 \dots j_{N-5}} = 0 ~,~
\nn\\
&&
\hspace{2cm}
{P_{(I)}}^{ij}_{~~k_1 \dots k_{N-4}} = 0 ~,~
\dots ~,~
{P_{(VII)}}^{ijk}_{~~~m_1 \dots m_{N-3}} = 0 ~,~
\nn\\
&&
\hspace{2cm}
{U_{(I)}}^{ij}_{~~k_1 \dots k_{N-4}} = 0 ~,~
\dots ~,~
{U_{(VI)}}^{ijk}_{~~~m_1 \dots m_{N-3}}= 0 ~,~
\nn\\
&&
\hspace{2cm}
{V_{(I)}}^{ijk}_{~~~m_1 \dots m_{N-3}} = 0 ~,~
\dots ~,~
{V_{(III)}}^{ijk}_{~~~m_1 \dots m_{N-3}} = 0 ~,~
\nn\\
&&
\hspace{2cm}
Z^{ijklmn} = 0
~~~
\}
~.~
\eea
The dimension of $\wV_{3,N}$ is $3N-1$.
\\

\begin{table}[H]
\centering
\resizebox{\hsize}{!}{
\begin{tabular}{|c|l|}
\hline \hline
Order $t^6$ & Quadratic Relation
\\
\hline\hline
$-[1,0,0,0,1,0,\dots,0] t^6$ 
&
$H_{i_1 \dots i_{N-5}}^{~~~~~~~k} = \epsilon_{ijpqrsu} B^{pqr} B^{suk} = 0$
\\
\hline \hline
Order $st^6$ & Quadratic Relation
\\
\hline\hline
$-[0,1,0,1,0,\dots,0]_x s t^6$ & 
${R_{(I)}}^{ij}_{~~k_1\dots k_{N-4}} = {A_{001}}^{ijm} B^{pqr} \epsilon_{mpqrk_1 \dots k_{N-4}} =0 $
\\
\hline
$-[2,0,0,1,0,\dots,0]_x s t^6$ &
${R_{(II)}}^{ij}_{~~k_1 \dots k_{N-4}} = {A_{001}}^{imj} B^{pqr} \epsilon_{mpqrk_1 \dots k_{N-4}} = 0 $
\\
\hline
$-[1,0,0,0,1,0,\dots,0]_x s t^6$ &
${R_{(III)}}^{i}_{~j_1 \dots j_{N-5}} = {A_{001}}^{imn} B^{klp} \epsilon_{mnklpj_1 \dots j_{N-5}} =0 $
\\
\hline \hline
Order $s^2 t^6$ & Quadratic Relation
\\
\hline\hline
$-[0,1,0,1,0,\dots,0]_x s^2 t^6$ & 
${O_{(I)}}^{ij}_{~~k_1 \dots k_{N-3}} = {A_{002}}^{ijp} B^{qrs} \epsilon_{pqrsk_1 \dots k_{N-3}} = 0 $
\\
\hline
$- 2[2,0,0,1,0,\dots,0]_x s^2 t^6$ &
${O_{(II)}}^{ij}_{~~k_1 \dots k_{N-3}} = {A_{001}}^{mni} {A_{001}}^{pqj} \epsilon_{mnpqk_1 \dots k_{N-3}} = 0$
\\
&
${O_{(III)}}^{ij}_{~~k_1 \dots k_{N-3}} = {A_{002}}^{imj} B^{npq} \epsilon_{mnpqk_1 \dots k_{N-3}} = 0$
\\
\hline
$- [0,0,2,0,0,\dots,0]_x s^2 t^6$ &
${O_{(IV)}}_{ijklmn u_1 \dots u_{N-6} v_1\dots v_{N-6}} = {o_{(I)}}_{ijklmn u_1 \dots u_{N-6} v_1\dots v_{N-6}} - \frac{1}{9} {o_{(II)}}_{ijklmn u_1 \dots u_{N-6} v_1\dots v_{N-6}}$
\\
&
\hspace{0.5cm}
${o_{(I)}}_{ijklmn u_1\dots u_{N-6} v_1\dots v_{N-6}} = \epsilon_{pqsijk u_1\dots u_{N-6}} \epsilon_{uvrlmn v_1\dots v_{N-6}} {A_{001}}^{pqr} A_{001}^{uvs}$
\\
&
\hspace{0.5cm}
${o_{(II)}}_{ijklmn u_1 \dots u_{N-6} v_1 \dots v_{N-6}} = u_2 B^{pqs} B^{uvr} \epsilon_{pqsijk u_1 \dots u_{N-6}} \epsilon_{uvrlmn v_1 \dots v_{N-6}}$
\\
\hline
$- [1,1,1,0,0,\dots,0]_x s^2 t^6$ &
${O_{(V)}}^{ijk}_{~~~m_1 \dots m_{N-3}} = {o_{(III)}}^{ijk}_{~~~m_1 \dots m_{N-3}} - \frac{1}{3} {o_{(IV)}}^{ijk}_{~~~m_1 \dots m_{N-3}} = 0$
\\
&
\hspace{0.5cm}
${o_{(III)}}^{ijk}_{~~~m_1 \dots m_{N-3}} = {A_{001}}^{pqk} {A_{001}}^{ijr} \epsilon_{pqr m_1 \dots m_{N-3}} $
\\
&
\hspace{0.5cm}
${o_{(IV)}}^{ijk}_{~~~m_1 \dots m_{N-3}} = {A_{002}}^{ijk} B^{pqr} \epsilon_{pqr m_1 \dots m_{N-3}} $
\\
\hline
$- [1,0,0,0,1,0,\dots,0]_x s^2 t^6$ &
${O_{(VI)}}^{i}_{~j_1 \dots j_{N-5}} = {A_{002}}^{pqi} B^{lmn} \epsilon_{pqlmnj_1 - j_{N-5}} = 0$
\\
\hline\hline
Order $s^3 t^6$ & Quadratic Relation
\\
\hline\hline
$-[0,1,0,1,0,\dots,0]_x s^3 t^6$ & 
${P_{(I)}}^{ij}_{~~k_1 \dots k_{N-4}} = {A_{001}}^{ijm} {A_{002}}^{pqr} \epsilon_{mpqr k_1 \dots k_{N-4}} = 0$
\\
\hline
$-2 [2,0,0,1,0,\dots,0]_x s^3 t^6$ &
${P_{(II)}}^{ij}_{~~k_1 \dots k_{N-4}} = {A_{001}}^{mni} {A_{002}}^{pqj} \epsilon_{mnpq k_1 \dots k_{N-4}} = 0$
\\
&
${P_{(III)}}^{ij}_{~~k_1 \dots k_{N-4}} = {S_{012}}^{ijm} B^{npq} \epsilon_{mnpq k_1 \dots k_{N-4}} = 0$
\\
\hline
$-[0,0,2,0,0,\dots,0]_x s^3 t^6$
&
${P_{(IV)}}_{ijklmn o_1 \dots o_{N-6} w_1 \dots w_{N-6}} = {p_{(I)}}_{ijklmn o_1 \dots o_{N-6} w_1 \dots w_{N-6}} - \frac{1}{9} {p_{(II)}}_{ijklmn o_1 \dots o_{N-6} w_1 \dots w_{N-6}}$
\\
&
\hspace{0.5cm} 
${p_{(I)}}_{ijklmn o_1 \dots o_{N-6} w_1 \dots w_{N-6}} = {A_{001}}^{pqu} {A_{002}}^{rsv} \epsilon_{pqvijk o_1 \dots o_{N-6}} \epsilon_{rsulmn w_1 \dots w_{N-6}}$
\\
&
\hspace{0.5cm}
${p_{(II)}}_{ijklmn o_1 \dots o_{N-6} w_1 \dots w_{N-6}} = u_3 B^{pqu} B^{rsv} \epsilon_{pquijk o_1 \dots o_{N-6}} \epsilon_{rsvlmn w_1 \dots w_{N-6}}$
\\
\hline
$- 2 [1,1,1,0,0,\dots,0]_x s^3 t^6$ &
${P_{(V)}}^{ijk}_{~~~m_1 \dots m_{N-3}} = {p_{(III)}}^{ijk}_{~~~m_1 \dots m_{N-3}} - {p_{(IV)}}^{ijk}_{~~~m_1 \dots m_{N-3}} = 0$
\\
&
${P_{(VI)}}^{ijk}_{~~~m_1 \dots m_{N-3}} = {p_{(III)}}^{ijk}_{~~~m_1 \dots m_{N-3}} - \frac{1}{2} {p_{(V)}}^{ijk}_{~~~m_1 \dots m_{N-3}} = 0$
\\
&
\hspace{0.5cm}
${p_{(III)}}^{ijk}_{~~~m_1 \dots m_{N-3}} = {A_{001}}^{ijp} {A_{002}}^{qrk} \epsilon_{pqr m_1 \dots m_{N-3}}$
\\
&
\hspace{0.5cm}
${p_{(IV)}}^{ijk}_{~~~m_1 \dots m_{N-3}} = {A_{002}}^{ijp} {A_{001}}^{qrk} \epsilon_{pqr m_1 \dots m_{N-3}}$
\\
&
\hspace{0.5cm}
${p_{(V)}}^{ijk}_{~~~m_1 \dots m_{N-3}} = u_2 {A_{001}}^{ijp} B^{qrk} \epsilon_{pqr m_1 \dots m_{N-3}}$
\\
\hline
$-[3,0,1,0,0,\dots,0]_x s^3 t^6$ &
${P_{(VII)}}^{ijk}_{~~~m_1 \dots m_{N-3}} = {p_{(VI)}}^{ijk}_{~~~m_1 \dots m_{N-3}} + \frac{1}{3} {p_{(VII)}}^{ijk}_{~~~m_1 \dots m_{N-3}}$
\\
&
\hspace{0.5cm}
${p_{(VI)}}^{ijk}_{~~~m_1 \dots m_{N-3}}
=
 (
{A_{001}}^{ipj} {A_{002}}^{qrk} 
+  {A_{001}}^{jpk} {A_{002}}^{qri} 
+ {A_{001}}^{kpi} {A_{002}}^{qrj} 
)\epsilon_{pqr m_1 \dots m_{N-3}}$
\\
&
\hspace{0.5cm}
${p_{(VII)}}^{ijk}_{~~~m_1 \dots m_{N-3}}
=
{S_{012}}^{ijk} B^{pqr} \epsilon_{pqr m_1 \dots m_{N-3}}$
\\
\hline\hline
\end{tabular}
}
\caption{The quadratic relations for $3$ $U(N)$ vortices, for orders $t^6$ to $s^3 t^6$ of the Hilbert series. \label{t3Na1}}
\end{table}

\begin{table}[H]
\centering
\resizebox{\hsize}{!}{
\begin{tabular}{|c|l|}
\hline \hline
Order $t^6$ & Quadratic Relation
\\
\hline\hline
$-[2,0,0,1,0,\dots,0]_x s^4 t^6$ 
&
${U_{(I)}}^{ij}_{~~k_1\dots k_{N-4}} = {A_{002}}^{pqi} {A_{002}}^{mnj} \epsilon_{pqmn k_1 \dots k_{N-4}} = 0$
\\
\hline
$-[0,0,2,0,0,\dots,0]_x s^4 t^6$
&
${U_{(II)}}_{ijklmn u_1 \dots u_{N-6} v_1 \dots v_{N-6}} = {u_{(I)}}_{ijklmn u_1 \dots u_{N-6} v_1 \dots v_{N-6}}  -  \frac{1}{3} {u_{(II)}}_{ijklmn u_1 \dots u_{N-6} v_1 \dots v_{N-6}} = 0$
\\
&
\hspace{0.5cm}
${u_{(I)}}_{ijklmn u_1 \dots u_{N-6} v_1 \dots v_{N-6}} 
= 
{A_{002}}^{pqr} {A_{002}}^{uvs} \epsilon_{pqsijk u_1 \dots u_{N-6}} \epsilon_{uvrlmn v_1 \dots v_{N-6}}$
\\
&
\hspace{0.5cm}
${u_{(II)}}_{ijklmn u_1 \dots u_{N-6} v_1 \dots v_{N-6}}
=
u_2 u_2 B^{pqr} B^{uvw} \epsilon_{pqrijk u_1 \dots u_{N-6}} \epsilon_{uvwlmn v_1 \dots v_{N-6}}$
\\
\hline
$-2 [1,1,1,0,0,\dots,0]_x s^4 t^6$
&
${U_{(III)}}^{ijk}_{~~~m_1\dots m_{N-3}} =
{u_{(V)}}^{ijk}_{~~~m_1\dots m_{N-3}} - 2 {u_{(III)}}^{ijk}_{~~~m_1\dots m_{N-3}} + \frac{2}{9} {u_{(VI)}}^{ijk}_{~~~m_1\dots m_{N-3}} = 0$
\\
&
${U_{(IV)}}^{ijk}_{~~~m_1\dots m_{N-3}} =
{u_{(IV)}}^{ijk}_{~~~m_1\dots m_{N-3}} + \frac{1}{6} {u_{(VI)}}^{ijk}_{~~~m_1\dots m_{N-3}} = 0$
\\
&
\hspace{0.5cm}
${u_{(III)}}^{ijk}_{~~~m_1\dots m_{N-3}} =
{A_{002}}^{ijp} {A_{002}}^{qrk} \epsilon_{pqr m_1\dots m_{N-3}} $
\\
&
\hspace{0.5cm}
${u_{(IV)}}^{ijk}_{~~~m_1\dots m_{N-3}} =
{A_{001}}^{ijp} {S_{012}}^{qrk} \epsilon_{pqr m_1\dots m_{N-3}} $
\\
&
\hspace{0.5cm}
${u_{(V)}}^{ijk}_{~~~m_1\dots m_{N-3}} =
u_2 {A_{001}}^{ijk} {A_{001}}^{pqr} \epsilon_{pqr m_1\dots m_{N-3}} $
\\
&
\hspace{0.5cm}
${u_{(VI)}}^{ijk}_{~~~m_1\dots m_{N-3}} =
u_3 {A_{001}}^{ijk} B^{pqr} \epsilon_{pqr m_1\dots m_{N-3}}$
\\
\hline
$-[2,2,0,0,0,\dots,0]_x s^4 t^6$
&
${U_{(V)}}^{ijklmn} = {u_{(VIII)}}^{ijklmn} + {u_{(VIII)}}^{ijmnkl} = 0$
\\
&
\hspace{0.5cm}
${u_{(VII)}}^{ijklmn} = {A_{001}}^{ijk} {S_{012}}^{lmn}$
\\
&
\hspace{0.5cm}
${u_{(VIII)}}^{ijklmn} = {u_{(VII)}}^{ijklmn} - {u_{(VII)}}^{ijlkmn} -  {u_{(VII)}}^{ijklnm} + {u_{(VII)}}^{ijlknm} $
\\
\hline
$-[3,0,1,0,0,\dots,0]_x s^4 t^6$ &
${U_{(VI)}}^{ijk}_{~~~m_1 \dots m_{N-3}} = 
(
{A_{001}}^{ipq} {S_{012}}^{rjk}
+ {A_{001}}^{jpq} {S_{012}}^{rki}
+ {A_{001}}^{jpq} {S_{012}}^{rij} 
) \epsilon_{pqr m_1\dots m_{N-3}}
= 0$
\\
\hline \hline
Order $s^5 t^6$ & Quadratic Relation
\\
\hline \hline
$-[1,1,1,0,0,\dots,0]_x s^5 t^6$ &
${V_{(I)}}^{ijk}_{~~~m_1 \dots m_{N-3}} = {v_{(I)}}^{ijk}_{~~~m_1 \dots m_{N-3}} + \frac{1}{2} {v_{(II)}}^{ijk}_{~~~m_1 \dots m_{N-3}} = 0$
\\
&
\hspace{0.5cm}
${v_{(I)}}^{ijk}_{~~~m_1 \dots m_{N-3}} =
{A_{002}}^{ijp} {S_{012}}^{qrk}
\epsilon_{pqr m_1 \dots m_{N-3}}$
\\
&
\hspace{0.5cm}
${v_{(II)}}^{ijk}_{~~~m_1 \dots m_{N-3}} =
u_3 {A_{001}}^{ijp} {A_{001}}^{qrk}
\epsilon_{pqr m_1 \dots m_{N-3}}$
\\
\hline
$-[2,2,0,0,0,\dots,0]_x s^5 t^6$ &
${V_{(II)}}^{ijklmn} = {v_{(V)}}^{ijklmn} + {v_{(VIII)}}^{ijmnkl} = 0$
\\
&
\hspace{0.5cm}
${v_{(III)}}^{ijklmn} = ( {A_{002}}^{ijk} - \frac{1}{2} u_2 B^{ijk}) {S_{012}}^{lmn}$
\\
&
\hspace{0.5cm}
${v_{(IV)}}^{ijklmn} = {v_{(III)}}^{ijklmn} - {v_{(III)}}^{ijlkmn} - {v_{(III)}}^{ijklnm} + {v_{(III)}}^{ijlknm} $
\\
&
\hspace{0.5cm}
${v_{(V)}}^{ijklmn} = {v_{(IV)}}^{ijklmn} - {v_{(IV)}}^{mnklij}$
\\
&
\hspace{0.5cm}
${v_{(VI)}}^{ijklmn} = u_3 {A_{001}}^{ijk} {A_{001}}^{lmn}$
\\
&
\hspace{0.5cm}
${v_{(VII)}}^{ijklmn} = {v_{(VI)}}^{ijklmn} - {v_{(VI)}}^{ijlkmn} - {v_{(VI)}}^{ijklnm} + {v_{(VI)}}^{ijlknm}$
\\
&
\hspace{0.5cm}
${v_{(VIII)}}^{ijklmn} = {v_{(VII)}}^{ijklmn} + {v_{(VII)}}^{klijmn}$
\\
\hline
$-[3,0,1,0,0,\dots,0]_x s^5 t^6$ &
${V_{(III)}}^{ijk}_{~~~m_1\dots m_{N-3}} = {v_{(X)}}^{ijk}_{~~~m_1 \dots m_{N-3}}  - \frac{1}{3} {v_{(XII)}}^{ijk}_{~~~m_1 \dots m_{N-3}} = 0$
\\
&
\hspace{0.5cm}
${v_{(IX)}}^{ijk}_{~~~m_1\dots m_{N-3}} = {A_{002}}^{ipq} {S_{012}}^{rjk} \epsilon_{pqr m_1\dots m_{N-3}}$
\\
&
\hspace{0.5cm}
${v_{(X)}}^{ijk}_{~~~m_1\dots m_{N-3}} = {v_{(IX)}}^{ijk}_{~~~m_1\dots m_{N-3}} + {v_{(IX)}}^{jki}_{~~~m_1\dots m_{N-3}} + {v_{(IX)}}^{kij}_{~~~m_1\dots m_{N-3}}$
\\
&
\hspace{0.5cm}
${v_{(XI)}}^{ijk}_{~~~m_1\dots m_{N-3}} = u_2 {A_{001}}^{ipj} {A_{002}}^{qrk} \epsilon_{pqr m_1\dots m_{N-3}}$
\\
&
\hspace{0.5cm}
${v_{(XII)}}^{ijk}_{~~~m_1\dots m_{N-3}} = {v_{(XI)}}^{ijk}_{~~~m_1\dots m_{N-3}} + {v_{(XI)}}^{jki}_{~~~m_1\dots m_{N-3}} + {v_{(XI)}}^{kij}_{~~~lmno}$
\\
\hline \hline
Order $s^6 t^6$ & Quadratic Relation
\\
\hline \hline
$-[2,2,0,0,0,0]_x s^6 t^6$ &
$Z^{ijklmn} = {z_{(II)}}^{ijklmn} + {z_{(II)}}^{ijmnkl} = 0$
\\
&
\hspace{0.5cm}
${z_{(I)}}^{ijklmn} = {S_{012}}^{ijk} {S_{012}}^{lmn}$
\\
&
\hspace{0.5cm}
${z_{(II)}}^{ijklmn} = {z_{(I)}}^{ijklmn} - {z_{(I)}}^{ijlkmn} - {z_{(I)}}^{ijklnm} + {z_{(I)}}^{ijlknm}$
\\
\hline \hline
\end{tabular}
}
\caption{The quadratic relations for $3$ $U(N)$ vortices, for orders $s^4 t^6$ to $s^6 t^6$ of the Hilbert series. \label{t3Na2}}
\end{table}

\section{$4$ $U(N)$ vortices on $\mathbb{C}$ \label{s4}}

\begin{figure}[H]
\begin{center}
\resizebox{0.6\hsize}{!}{
\includegraphics[trim=0cm 0cm 0cm 0cm,totalheight=16 cm]{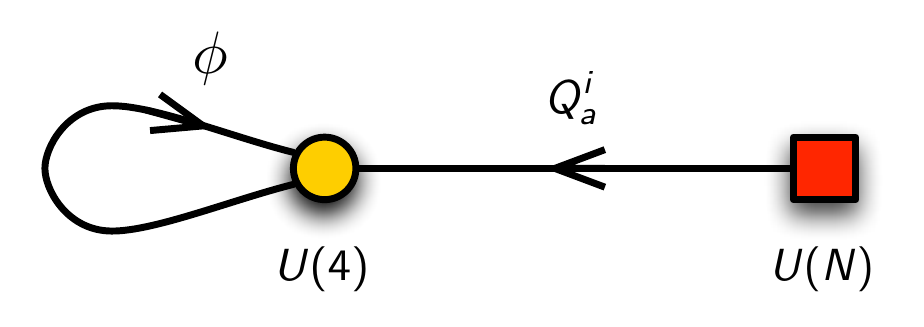}
}  \caption{Quiver diagram of the 4 $U(N)$ vortex theory. \label{fquiverk4}}
 \end{center}
 \end{figure}

\paragraph{The moduli space.} The moduli space of $4$ $U(N)$ vortices $\V_{4,N}$ is a partial $\mathbb{C}^{*}$ quotient of the master space $\master_{4,N}$. The generators $x_1,\dots,x_d$ of $\master_{4,N}$ can be considered as coordinates for the $\mathbb{C}^{*}$ projection which takes the form
\beal{es1200aa1}
(x_1,\dots,x_d) \simeq (\lambda^{w_1} x_1,\dots,\lambda^{w_d} x_d) ~.~
\eea
Above, $\lambda$ is the $\mathbb{C}^{*}$ parameter and $w_1,\dots,w_d$ are respectively the $U(1)$ weights for the coordinates $x_1,\dots,x_d$. Accordingly, the vortex master space is $\mathbb{C}^{*}$ projected as follows,
\beal{es1200aa2}
\V_{4,N} = \master_{4,N} / \{
x_1 \simeq \lambda^{w_1}, 
\dots,
x_d \simeq \lambda^{w_d}
\}~.~
\eea
\\

\paragraph{The Molien integral and Hilbert series.} The Hilbert series of the $4$ $U(N)$ vortex master space is given by the following Molien integral,
\beal{es1200aa3}
g(t,s,x;\master_{4,N}) = 
\oint \ud\mu_{SU(4)} ~ \PE\Big[
[0,0,1]_w [1,0,\dots,0]_x t + [1,0,1]_w s
\Big]~,~
\eea
where $Q_{\alpha}^{i}$ transforms in $[0,0,1]_w [1,0,\dots,0]_x t$ and $\phi$ transforms in $[1,0,1]_w s$. $\ud\mu_{SU(4)}$ is the $SU(4)$ Haar measure.
\\

\paragraph{Center of mass contribution.} The integrand in \eref{es1200aa3} for the master space Hilbert series can be rewritten as follows,
\beal{es1200aa4}
&&
\PE\Big[
[0,0,1]_w [1,0,\dots,0]_x t + [1,0,1]_w s
\Big]
=
\nn\\
&& \hspace{0.5cm}
\frac{1}{1-s}
\PE\Big[
[0,0,1]_w [1,0,\dots,0]_x t +
(
w_1 w_3 + w_1 w_2 w_3^{-1} + w_1^{-1} w_2 w_3 + w_1^2 w_2^{-1} 
\nn\\
&&\hspace{2.5cm}
+ w_2^{-1} w_3^{2}  + w_1^{-1} w_2^{2} w_3^{-1}
+ 2
+ w_1 w_2^{-2} w_3 + w_2 w_3^{-2} + w_1^{-2} w_2
+ w_1 w_2^{-1} w_3^{-1} 
\nn\\
&&\hspace{2.5cm}
+ w_1^{-1} w_2^{-1} w_3
+ w_1^{-1} w_3^{-1}
) s
\Big]~,~
\eea
where the character of the adjoint of $SU(4)$ is given by 
\beal{es1200aa5}
[1,0,1]_w 
&=&
w_1 w_3 + w_1 w_2 w_3^{-1} + w_1^{-1} w_2 w_3 + w_1^2 w_2^{-1} 
+ w_2^{-1} w_3^{2}  + w_1^{-1} w_2^{2} w_3^{-1}
+ 3
\nn\\
&&
+ w_1 w_2^{-2} w_3 + w_2 w_3^{-2} + w_1^{-2} w_2
+ w_1 w_2^{-1} w_3^{-1} 
+ w_1^{-1} w_2^{-1} w_3
+ w_1^{-1} w_3^{-1}
~.~
\nn\\
\eea
The $\frac{1}{1-s}$ prefactor in \eref{es1200aa5} does not interfere with the Molien integral and also is independent of the $\mathbb{C}^{*}$ projection of the vortex master space. It refers to the center of mass position of the $4$ vortices. 
\\

\subsection{$4$ $U(1)$ vortices on $\mathbb{C}$ \label{s4b1}}

The Hilbert series of the $4$ $U(1)$ vortex master space is
\beal{es1201}
g(t,s,x;\wmaster_{4,1})
=
\frac{1}{(1 - s^2) (1 - s^3) (1 - s^4) (1 - s^6 t^4)}~~.
\eea
The generalization of the master space Hilbert series for any $k$ $U(1)$ vortices is 
\beal{es1201b1}
g(t,s,x;\wmaster_{k,1})
=
\frac{1}{ (1 - s^{k(k+1)/2} t^{k}) \prod_{i=2}^{k} (1 - s^i) }~~.
\eea
The vortex master space for $k$  $U(1)$ vortices is
\beal{es1201b2}
\wV_{k,1} = \mathbb{C}^{k-1} ~.~
\eea
\\

\subsection{$4$ $U(2)$ vortices on $\mathbb{C}$ \label{s4b2}}

 The Hilbert series for the $4$ $SU(2)$ vortex moduli space is given by the following Molien integral,
\beal{es1210}
g(t,s,x;\wmaster_{4,2})
&=&
\oint
\ud\mu_{SU(4)}
\PE\Big[
[0,0,1]_w [1]_x t +
[1,0,1]_w s
\Big]~~.
\eea
When solved, the above integral gives the Hilbert series 
\beal{es1211}
g(t,s,x;\wmaster_{4,2})
&=&
\frac{1}{
(1 - s^2) (1 - s^3) (1 - s^4)
(1 - s^2 t^4) 
(1 - s^3 t^4)^2 (1 - s^6 t^4)^2
}
\times
\nn\\
&&
(1 + s^3 t^4 + 4 s^4 t^4 + 3 s^5 t^4 + 3 s^6 t^4 + 3 s^8 t^8 + 3 s^9 t^8 + 4 s^{10} t^8 
\nn\\
&&
+ s^{11} t^8 + s^{14} t^{12})
~~,
\eea
where we have set for simplicity the global $SU(2)$ fugacity to $x=1$. The base manifold is a non-complete intersection of dimension $9$. As a character expansion the Hilbert series is
\beal{es1212}
&&
g (t,s,x;\wmaster_{4,2}) =
\frac{1}{
(1 - s^2) (1 - s^3) (1 - s^4)
(1 - s^2 t^4) (1 - s^4 t^4)
}
\nn\\
&&\hspace{3cm}
\times
\sum_{n_3=0}^{\infty} \sum_{n_6=0}^{\infty} 
\Big[
[2 n_3 + 4 n_6]_x s^{3 n_3 + 6 n_6} t^{4 n_3 + 4 n_6}
\nn\\
&&\hspace{5cm}
+ [2 n_3 + 4 n_6 + 2]_x s^{3 n_3 + 6 n_6 + 4} t^{4 n_3 + 4 n_6 + 4} 
\nn\\
&&\hspace{5cm}
+ [2 n_3 + 4 n_6 + 2]_x s^{3 n_3 + 6 n_6 + 5} t^{4 n_3 + 4 n_6 + 4} 
\nn\\
&&\hspace{5cm}
+ [2 n_3 + 4 n_6 + 4]_x s^{3 n_3 + 6 n_6 + 9} t^{4 n_3 + 4 n_6 + 8} 
\Big]~~.
\eea
The plethystic logarithm of the Hilbert series is
\beal{es1213}
\PL\Big[
g(t,s,x;\wmaster_{4,2})
\Big]
&=&
s^2+s^3+s^4
+ [2]_x s^3 t^4
+ [2]_x s^4 t^4
+ [2]_x s^5 t^4
+ [4]_x s^6 t^4
\nn\\
&&
- s^6 t^8
- [2]_x s^8 t^8
- (2+[2]_x+[4]_x) s^8 t^8
- (1+2[2]_x + [4]_x)s^9 t^8
\nn\\
&&
- (1+[2]_x+2[4]_x) s^{10} t^8
- ([2]_x + [4]_x) s^{11} t^8
\nn\\
&&
- (1+ [4]_x) s^{12} t^8
+ \dots~~.
\eea
\\

\section{Summary of Hilbert series for vortices \label{apphs}}

\subsection{Unrefined Hilbert series \label{sunref}}

We summarize in this section the Hilbert series of the vortex master spaces in unrefined form. This means, we set the fugacities corresponding to the global symmetry $SU(N)$ to $x_i=1$. The only remaining fugacities are $t$, corresponding to the remaining $U(1)$ R-symmetry and the fugacity $s$ corresponding to the $U(1)$ residual gauge symmetry. The unrefined Hilbert series take the following form:
\\

\paragraph{1 Vortex:}

\beal{eappx10}
g(t,s;\wmaster_{1,1}) &= &
\frac{1}{1-t}
\eea

\beal{eappx11}
g(t,s;\wmaster_{1,2}) &= &
\frac{1}{(1- t)^2} 
\eea

\beal{eappx12}
g(t,s;\wmaster_{1,3}) &= &
\frac{1}{(1- t)^3}~~,
\eea
\\

\paragraph{2 Vortices:}

\beal{eappx20}
g(t,s;\wmaster_{2,1}) = 
\frac{1}{(1 - s^2) (1 - s t^2)}
\eea

\beal{eappx21}
g(t,s;\wmaster_{2,2}) = 
\frac{1 + s t^2}{(1 - s^2) (1 - t^2) (1 - s t^2)^2}
\eea

\beal{eappx22}
g(t,s;\wmaster_{2,3}) = 
\frac{1 + 3 s t^2 - 3 s t^4 - s^2 t^6}{(1 - s^2) (1 - t^2)^3 (1 - s t^2)^3}
\eea

\beal{eappx23}
g(t,s;\wmaster_{2,4}) &=& 
\frac{
1
}{
(1 - s^2) (1 - t^2)^5 (1 - s t^2)^5
}\times 
(1 + t^2 + 5 s t^2 - 10 s t^4 - 5 s^2 t^4 + s t^6 
\nn\\
&&
\hspace{1cm}
- s^3 t^6 + 5 s^2 t^8 + 10 s^3 t^8 - 5 s^3 t^{10} - s^4 t^{10} - s^4 t^{12})
\eea

\beal{eappx24}
g(t,s;\wmaster_{2,5}) &=&
\frac{
1
}{
(1 - s^2) (1 - t^2)^7 (1 - s t^2)^7
}\times
(
1 + 3 t^2 + 8 s t^2 + t^4 - 21 s t^4 - 14 s^2 t^4 
\nn\\
&&
\hspace{1cm}
- 7 s t^6 -  7 s^2 t^6 + 51 s^2 t^8 + 70 s^3 t^8 + 5 s^4 t^8 - 5 s^2 t^{10} -  70 s^3 t^{10} 
\nn\\
&&
\hspace{1cm}
- 51 s^4 t^{10} + 7 s^4 t^{12} + 7 s^5 t^{12} + 14 s^4 t^{14} +  21 s^5 t^{14} - s^6 t^{14} 
\nn\\
&&
\hspace{1cm}
- 8 s^5 t^{16} - 3 s^6 t^{16} - s^6 t^{18}
 )
\eea
\\

\paragraph{3 Vortices:}

\beal{eappx30}
g(t,s;\wmaster_{3,1}) = 
\frac{1}{(1-s^2)(1-s^3)(1-s^3 t^3)}
\eea

\beal{eappx31}
g(t,s;\wmaster_{3,2}) = 
\frac{
1 + 2 s^2 t^3 + 2 s^3 t^3 + s^5 t^6
}{
(1 - s^2) (1 - s^3) (1 - t^3) (1 - s t^3)^2 (1 -  s^3 t^3)^2
}
\eea

\beal{eappx32}
g(t,s;\wmaster_{3,3}) &=& 
\frac{
1
}{
(1 - s^2) (1 - s^3) (1 - t^3) (1 - s t^3)^4 (1 - s^3 t^3)^4
}\times
(
1 + 4 s t^3 + 8 s^2 t^3 
\nn\\
&&
\hspace{1cm}
+ 6 s^3 t^3 + s^2 t^6 + 5 s^3 t^6 + 6 s^4 t^6 + 3 s^5 t^6 - 6 s^6 t^6 - s^4 t^9 - 12 s^5 t^9 
\nn\\
&&
\hspace{1cm}
- 15 s^6 t^9 - 15 s^7 t^9 - 12 s^8 t^9 - s^9 t^9 - 6 s^7 t^{12} +  3 s^8 t^{12} + 6 s^9 t^{12}
\nn\\
&&
\hspace{1cm} 
+ 5 s^{10} t^{12} + s^{11} t^{12} + 6 s^{10} t^{15} +  8 s^{11} t^{15} + 4 s^{12} t^{15} + s^{13} t^{18}
)
\eea

\beal{eappx33}
g(t,s;\wmaster_{3,4}) &=&
\frac{
1
}{
(1 - s^2) (1 - s^3) (1 - t^3)^4 (1 - s t^3)^6 (1 - s^3 t^3)^6
}\times
(
1 + 14 s t^3 + 20 s^2 t^3 
\nn\\
&&
\hspace{1cm}
+ 14 s^3 t^3 - 16 s t^6 + 5 s^2 t^6 +  46 s^3 t^6 + 62 s^4 t^6 + 20 s^5 t^6 - 21 s^6 t^6 
\nn\\
&&
\hspace{1cm}
+ 4 s t^9 -  44 s^2 t^9 - 104 s^3 t^9 - 124 s^4 t^9 - 166 s^5 t^9 - 152 s^6 t^9 
\nn\\
&&
\hspace{1cm}
-  190 s^7 t^9 - 100 s^8 t^9 - 4 s^9 t^9 + 21 s^2 t^{12} + 25 s^3 t^{12} + 15 s^4 t^{12} 
\nn\\
&&
\hspace{1cm}
+ 153 s^5 t^{12} + 135 s^6 t^{12} + 101 s^7 t^{12} +  110 s^8 t^{12} + 85 s^9 t^{12} 
\nn\\
&&
\hspace{1cm}
+ 139 s^{10} t^{12} + 61 s^{11} t^{12} +  10 s^{12} t^{12} + 14 s^3 t^{15} + 34 s^4 t^{15}  - 20 s^5 t^{15} 
\nn\\
&&
\hspace{1cm}
+  156 s^6 t^{15} + 396 s^7 t^{15} + 264 s^8 t^{15} + 248 s^9 t^{15} +  256 s^{10} t^{15} 
\nn\\
&&
\hspace{1cm}
+ 292 s^{11} t^{15} + 116 s^{12} t^{15} - 26 s^{13} t^{15} -  2 s^{14} t^{15} + s^4 t^{18} + s^5 t^{18} 
\nn\\
&&
\hspace{1cm}
- 109 s^6 t^{18} - 275 s^7 t^{18} -  218 s^8 t^{18} - 460 s^9 t^{18} - 788 s^{10} t^{18} 
\nn\\
&&
\hspace{1cm}
- 788 s^{11} t^{18} -  460 s^{12} t^{18} - 218 s^{13} t^{18} - 275 s^{14} t^{18} - 109 s^{15} t^{18} 
\nn\\
&&
\hspace{1cm}
+  s^{16} t^{18} + s^{17} t^{18} - 2 s^7 t^{21} - 26 s^8 t^{21} + 116 s^9 t^{21} +  292 s^{10} t^{21} 
\nn\\
&&
\hspace{1cm}
+ 256 s^{11} t^{21} + 248 s^{12} t^{21} + 264 s^{13} t^{21} +  396 s^{14} t^{21} + 156 s^{15} t^{21} 
\nn\\
&&
\hspace{1cm}
- 20 s^{16} t^{21} + 34 s^{17} t^{21} +  14 s^{18} t^{21} + 10 s^9 t^{24} + 61 s^{10} t^{24} 
\nn\\
&&
\hspace{1cm}
+ 139 s^{11} t^{24} +  85 s^{12} t^{24} + 110 s^{13} t^{24} + 101 s^{14} t^{24} + 135 s^{15} t^{24} 
\nn\\
&&
\hspace{1cm}
+  153 s^{16} t^{24} + 15 s^{17} t^{24} + 25 s^{18} t^{24} + 21 s^{19} t^{24} -  4 s^{12} t^{27} - 100 s^{13} t^{27} 
\nn\\
&&
\hspace{1cm}
- 190 s^{14} t^{27} - 152 s^{15} t^{27} -  166 s^{16} t^{27} - 124 s^{17} t^{27} - 104 s^{18} t^{27} 
\nn\\
&&
\hspace{1cm}
- 44 s^{19} t^{27} +  4 s^{20} t^{27} - 21 s^{15} t^{30} + 20 s^{16} t^{30} + 62 s^{17} t^{30} +  46 s^{18} t^{30} 
\nn\\
&&
\hspace{1cm}
+ 5 s^{19} t^{30} - 16 s^{20} t^{30} + 14 s^{18} t^{33} +  20 s^{19} t^{33} + 14 s^{20} t^{33} + s^{21} t^{36}
)
\nn\\
\eea

\subsection{Highest Weight Hilbert Series \label{shish}}

We have in the sections above computed the Hilbert series for vortex master spaces in order to characterize their algebraic structure as well as to identify their generators for the $\mathbb{C}^{*}$ projection to the vortex moduli space. The Hilbert series were refined such that one had the following collection of fugacities,
\beal{es5000}
s &\ra& \phi_{\alpha\beta} 
\nn\\
t &\ra& Q_{\alpha}^{i}
\nn\\
~[n_1,\dots,n_{N-1}]_x &\ra& SU(N) ~~\text{global symmetry}
~~.
\eea

Let us summarize the character expansions for the master space Hilbert series for $1,2,3$ $U(N)$ vortices,
\beal{es5001}
g(t,s,x;\wmaster_{1,N})
&=&
\sum_{n_0=0}^{\infty} [n_0,0,\dots,0]_x t^{n_0}~,~
\nn\\
g(t,s,x;\wmaster_{2,N})
&=&
\sum_{n_0=0}^{\infty}
\sum_{n_1=0}^{\infty}
[2n_1,n_0,0,\dots,0]_x s^{n_1} t^{2(n_0+n_1)}~,~
\nn\\
g(t,s,x;\wmaster_{3,N}) 
&=&
\frac{1}{(1-s^2)(1-s^3)} 
\times
\nn\\
&& 
\hspace{-2.5cm}
\sum_{n_0=0}^{\infty} \sum_{n_1=0}^{\infty} \sum_{n_2=0}^{\infty} \sum_{n_3=0}^{\infty}
\Big[
[n_1+n_2+3n_3, n_1+n_2, n_0, 0, \dots, 0] s^{n_1 + 2n_2 + 3n_3} t^{3 n_0 + 3n_1 + 3n_2 + 3n_3 }
\nn\\
&& 
\hspace{-2.5cm}
+
[n_1+n_2, n_1+n_2+3n_3 +3, n_0, 0, \dots, 0] s^{n_1 + 2n_2 + 3n_3+3} t^{3 n_0 + 3n_1 + 3n_2 + 6n_3 + 6}
\Big]~.~
\eea
\\

\paragraph{Highest weight Hilbert series.} We now use a more compact form of writing characters of irreducible representations in a Hilbert series \cite{hananykalveks}. Given that characters of the form $[n_1,\dots,n_r]$ are written in terms of highest weight Dynkin labels, we introduce for each of the $r$ labels its own fugacity $\mu_i$ such that
\beal{es5010}
[n_1,n_2,\dots,n_r]_G \ra 
\prod_{i=1}^{r} \mu_i^{n_i}  = \mu_1^{n_1} \mu_2^{n_2} \dots \mu_r^{n_r}~,~
\eea
for a group $G$ of rank $r$. Effectively, the above map replaces a character with a product of fugacities carrying as exponents the highest weight Dynkin labels of the irreducible representation of the group $G$.

There are various motivations for introducing the above map. One of them is the ability to write character expansions of Hilbert series given by infinite concatenated sums as compact rational functions. For the character expansion in \eref{es5001} corresponding to the Hilbert series of $1,2,3$ $U(N)$ vortex master spaces, the highest weight forms are as follows
\beal{es5011}
g(t,s,x;\wmaster_{1,N}) &\ra&
\frac{1}{1- \mu_1 t}
~,~
\nn\\
g(t,s,x;\wmaster_{2,N}) &\ra&
\frac{1}{
(1 - s^2) 
(1 - \mu_2 t^2) 
(1 - \mu_1^2 s t^2)
}
~,~
\nn\\
g(t,s,x;\wmaster_{3,N}) &\ra&
\frac{
1 + \mu_1 \mu_2 s^2 t^3 + \mu_1^2 \mu_2^2 s^4 t^6
}{
(1-s^2)
(1-s^3)
(1 - \mu_3 t^3) 
(1 - \mu_1 \mu_2 s t^3) 
(1 - \mu_1^3 s^3 t^3) 
(1 - \mu_2^3 s^3 t^6)
}~.~
\nn\\
\eea
\\

\section{Conclusions}

With this work, we have classified fully for the first time the moduli spaces for supersymmetric gauge theories of up to $3$ $U(N)$ vortices. This was done by describing the vortex moduli space as a partially weighted projective space coming from a $\mathbb{C}^{*}$ projection of the vortex master space. In our classification, we have given the full algebraic structure of the vortex master spaces by identifying all of their generators and quadratic relations formed amongst the generators. The information given by the residual $U(1)$ gauge charges carried by the generators allows us to describe the projection into the full vortex moduli space.

Our results for $2$ vortices agree with previous results in \cite{Hanany:2003hp,Eto:2010aj}. In view of our complete analysis of moduli spaces for $3$ vortices and the preliminary computations we have presented here for $4$ vortices, it would be interesting to generalize our results to higher number of vortices. With our current methods and tools, this task seems to be a challenge at this moment.

We have seen with the computation of the Hilbert series for vortex master spaces that its expression as an infinite expansion in terms of characters of $SU(N)$ is an increasingly difficult problem to solve. The most evident example is the Hilbert series of the master space for $4$ $U(2)$ vortices which for the purpose of our argument we present again as follows,
\beal{eend1}
&&
g (t,s,x;\wmaster_{4,2}) =
\frac{1}{
(1 - s^2) (1 - s^3) (1 - s^4)
(1 - s^2 t^4) (1 - s^4 t^4)
}
\nn\\
&&\hspace{3cm}
\times
\sum_{n_3=0}^{\infty} \sum_{n_6=0}^{\infty} 
\Big[
[2 n_3 + 4 n_6]_x s^{3 n_3 + 6 n_6} t^{4 n_3 + 4 n_6}
\nn\\
&&\hspace{5cm}
+ [2 n_3 + 4 n_6 + 2]_x s^{3 n_3 + 6 n_6 + 4} t^{4 n_3 + 4 n_6 + 4} 
\nn\\
&&\hspace{5cm}
+ [2 n_3 + 4 n_6 + 2]_x s^{3 n_3 + 6 n_6 + 5} t^{4 n_3 + 4 n_6 + 4} 
\nn\\
&&\hspace{5cm}
+ [2 n_3 + 4 n_6 + 4]_x s^{3 n_3 + 6 n_6 + 9} t^{4 n_3 + 4 n_6 + 8} 
\Big]~~.
\eea
As discussed in section \sref{shish}, we use highest weight fugacities. For the Hilbert series of the master space of $4$ $U(2)$ vortices above, the new fugacity is $\mu$ for $SU(2)$ such that the representations are mapped to
\beal{eend2}
[n]_x \rightarrow \mu^{n} ~.~
\eea
The above map dramatically simplifies the expression in \eref{eend1} to 
\beal{eend3}
g (t,s,x;\wmaster_{4,2}) 
&\rightarrow&
\frac{1}{(1 - s^2) (1 - s^3) (1 - s^4) (1 - s^2 t^4) (1 - s^4 t^4)}
\nn\\
&&
\times
\frac{
1 + \mu^2 s^4 t^4 + \mu^2 s^5 t^4 + \mu^4 s^9 t^8
}{
(1 - \mu^2 s^3 t^4) (1 - \mu^4 s^6 t^4)
}
~.~
\eea
The expression becomes a rational function, which of great interest has a palindrome as its numerator. This is not the actual Hilbert series itself. This can be seen by comparing the so called highest weight Hilbert series in \eref{eend3} with the actual unrefined Hilbert series 
\beal{eend4}
g (t,s;\wmaster_{4,2})  &=&
\frac{
1
}{
(1 - s^2) (1 - s^3) (1 - s^4) (1 - s^2 t^4) (1 - s^4 t^4)
(1 - s^3 t^4)^2 (1 - s^6 t^4)^2
}
\nn\\
&&
\times
(
1 + s^3 t^4 + 3 s^4 t^4 + 3 s^5 t^4 + 3 s^6 t^4 - s^7 t^8 - s^8 t^8 + s^{10} t^8 + s^{11} t^8 
\nn\\
&&
- 3 s^{12} t^{12} - 3 s^{13} t^{12} - 3 s^{14} t^{12} - s^{15} t^{12} - s^{18} t^{16}
)
~.~
\eea
From here, it can be seen that the map in \eref{eend2} transforms a palindromic Hilbert series of a non-compact Calabi-Yau space to another different rational function with a palindromic numerator. It is of great interest to explore the meaning of the function in \eref{eend3} in comparison to the original vortex Hilbert series in \eref{eend4}.
\\

\section*{Acknowledgements}

A. H. and R.-K. S. gratefully acknowledge hospitality at the Simons Center for Geometry and Physics, Stony Brook University where some of the research for this paper was performed. We thank also Giuseppe Torri for input during the early stages of this work. R.-K. S. would also like to thank Kazushi Ueda and Sungsoo Kim for useful discussions. 
\\

\bibliographystyle{JHEP}
\bibliography{mybib}


\end{document}

%% file: pref.tex
\newcommand{\be}{\begin{equation}}
\newcommand{\ee}{\end{equation}}
\newcommand{\beq}{\begin{equation}}
\newcommand{\beql}[1]{\begin{equation}\label{#1}}
\newcommand{\eeq}{\end{equation}}
\newcommand{\ba}{\begin{array}}
\newcommand{\ea}{\end{array}}
\newcommand{\bea}{\begin{eqnarray}}
\newcommand{\beal}[1]{\begin{eqnarray}\label{#1}}
\newcommand{\eea}{\end{eqnarray}}
\newcommand{\ben}{\begin{enumerate}}
\newcommand{\een}{\end{enumerate}}
\newcommand{\bean}{\begin{eqnarray*}}
\newcommand{\eean}{\end{eqnarray*}}
\newcommand{\eref}[1]{(\ref{#1})}
\newcommand{\sref}[1]{\S\ref{#1}}
\newcommand{\tref}[1]{Table~\ref{#1}}
\newcommand{\nn}{\nonumber}

\newcommand{\fref}[1]{Figure \ref{#1}}
\newcommand{\btab}[1]{\begin{tabular}{#1}}
\newcommand{\etab}{\end{tabular}}

\newcommand{\master}{\mathcal{F}^\flat}

\def \Black {\pdfsetcolor{0 0 0 1}}

\newcommand{\comment}[1]{}

\newcommand{\ud}{\mathrm{d}}

\newcommand{\PE}{\mathrm{PE}}
\newcommand{\PL}{\mathrm{PL}}

\newcommand{\qed}{\nobreak \ifvmode \relax \else
      \ifdim\lastskip<1.5em \hskip-\lastskip
      \hskip1.5em plus0em minus0.5em \fi \nobreak
      \vrule height0.75em width0.5em depth0.25em\fi}